\documentclass[fleqn,usenatbib]{mnras}

\usepackage{newtxtext,newtxmath}
\usepackage[T1]{fontenc}
\usepackage{ae,aecompl}
\usepackage[dvipsnames]{xcolor}

\usepackage{graphicx}	
\usepackage{multirow}
\usepackage{multicol}
\usepackage{makecell}
\usepackage{array}
\usepackage[section]{placeins}
\usepackage{hyperref}
\usepackage{pdflscape}
\usepackage{adjustbox}

\newcommand{\kev}{\mbox{keV}}

\newcommand{\msun}{\mbox{$M_\odot$}}
\newcommand{\Msun}{\mbox{$M_\odot$}}

\newcommand{\hst}{\emph{HST}}
\newcommand{\chandra}{\emph{Chandra}}
\newcommand{\gaia}{\emph{Gaia}}

\newcommand{\ergs}{\mbox{${\rm erg}~{\rm s}^{-1}$}}

\newcommand{\ergsh}{\mbox{${\rm erg}~{\rm s}^{-1}~{\rm Hz}^{-1}$}}

\newcommand{\bv}{\mbox{$B\!-\!V$}} 
\newcommand{\B}{\mbox{$B_{435}$}}
\newcommand{\R}{\mbox{$R_{625}$}}
\newcommand{\V}{\mbox{$V_{606}$}}
\newcommand{\I}{\mbox{$I_{814}$}}

\newcommand{\ha}{\mbox{H$\alpha$}}
\newcommand{\hr}{\mbox{$\ha\!-\!\R$}}
\newcommand{\UV}{\mbox{$UV_{275}$}}
\newcommand{\U}{\mbox{$U_{336}$}}
\newcommand{\uvu}{\mbox{$\UV\!-\!\U$}}

\newcommand{\ub}{\mbox{$\U\!-\!\B$}}
\newcommand{\vi}{\mbox{$\V\!-\!\I$}}
\newcommand{\rerr}{\mbox{$r_{\rm err}$}}
\newcommand{\ignore}[1]{}

\newcommand{\cory}[1]{{\color{black}{#1}}}

\title[Exotica in M4]{Exotica in the Globular Cluster M4, Studied with Chandra, HST, and the VLA}

\author[P.M. Lugger et al.]{
Phyllis M. Lugger,$^{1}$\thanks{E-mail: lugger@iu.edu (PML)}
Haldan N. Cohn,$^{1}$
Craig O. Heinke,$^{2}$
Jiaqi Zhao,$^{2}$
Yue Zhao,$^{3}$ \newauthor and
Jay Anderson$^{4}$ \\ ~ \\
$^{1}$Department of Astronomy, Indiana University, 727 E. Third St.,
Bloomington, IN 47405, USA\\
$^{2}$Department of Physics, University of Alberta, Edmonton, AB T6G
2G7, Canada\\
$^{3}$Department of Physics and Astronomy, University of Southampton, University Road, Southampton SO17 1BJ, UK \\
$^{4}$Space Telescope Science Institute, 3700 San Martin Dr., Baltimore, MD 21218, USA}

\date{Accepted 2023 June 13. Received 2023 Jun 11; in original form 2023 February 14}

\pubyear{2022}

\usepackage{placeins}
\usepackage{multicol}
\usepackage{balance}

\newcommand{\urlwofont}[1]{\urlstyle{same}\url{#1}} 

\newcommand{\slow}{S_5}
\newcommand{\shigh}{S_7}
\newcommand{\mujy}{\mathrm{\mu Jy}}

\defcitealias{Bahramian20}{B20}
\defcitealias{Shishkovsky20}{S20}

\begin{document}
\label{firstpage}
\pagerange{\pageref{firstpage}--\pageref{lastpage}}
\maketitle

\begin{abstract}

\noindent Using the Hubble Ultraviolet Globular Cluster Survey (HUGS) and additional \hst\ archival data, we have carried out a search for optical counterparts to the low-luminosity \chandra\ X-ray sources in the globular cluster M4 (NGC~6121). We have also searched for optical or X-ray counterparts to radio sources detected by the VLA. We find 24 new confident optical counterparts to \chandra\ sources for a total of 40, including the 16 previously identified. Of the 24 new identifications, 18 are stellar coronal X-ray sources (active binaries, ABs), the majority located along the binary sequence in a \vi\ colour-magnitude diagram and generally showing an \ha\ excess. In addition to confirming the previously detected cataclysmic variable (CV, CX4), we identify one confident new CV (CX76), and two candidates (CX81 and CX101). One MSP is known in M4 (CX12), and another strong candidate has been suggested (CX1); we identify some possible MSP candidates among optical and radio sources, such as VLA20, which appears to have a white dwarf counterpart. One X-ray source with a sub-subgiant optical counterpart and a flat radio spectrum (CX8, VLA31) is particularly mysterious. 
The radial distribution of X-ray sources suggests a relaxed population of average mass $\sim1.2 - 1.5\,\Msun$. Comparing the numbers of ABs, MSPs, and CVs in M4 with other clusters indicates that AB numbers are proportional to cluster mass (primordial population), MSPs to stellar encounter rate (dynamically formed population), while CVs seem to be produced both primordially and dynamically.  
\\

\end{abstract}




\section{Introduction}

\begin{table*}
\caption{Comparison of the properties of M4 with a sample of nearby, well-studied clusters. Values for solar distance ($d_\odot$), central density ($\rho_0$), and mass ($M$) are from \citet{Baumgardt21}, interaction rates ($\Gamma$) are from \citet{Bahramian13}, X-ray emissivities ($\xi_X = L_X/M$) are from \citet{Heinke20}. Limiting $L_X$ values are from \citet{Bahramian20} for M4 and NGC~6397, from \citet{Heinke05} for 47~Tuc, from \citet{Cool13} for $\omega$\,Cen, and from \citet{Cohn21} for NGC~6752. The value for $\omega$\,Cen applies to the centre of this large cluster; the sensitivity falls off with radial offset.}
\label{t:Cluster_properties_Chandra_obs}
\begin{center}
\begin{tabular}{lcccrccc}
\hline
Cluster & 
$d_\odot$ & 
$\log \rho_0$ &
$\log M$ &
$\Gamma$~~ &
$\xi_X$ &
Exposure& 
log\,$L_{X,\mathrm{lim}}$ \\
& [kpc] & [$\msun\,\mathrm{pc}^{-3}$] &
[\msun] & & [$10^{27}~\mathrm{erg~s}^{-1}\msun^{-1}$] 
& ~[ks] &
[$\mathrm{erg~s}^{-1}$] \\
\hline
47 Tuc        & 4.5 & 4.4 & 6.0 & 1000  &  5.8 & 540 &   29.9  \\
$\omega$\,Cen & 5.4 & 3.3 & 6.6 &   90  &  0.9 & 291 &   30.1  \\
M4            & 1.9 & 4.2 & 4.9 &   27  &  3.6 & 119 &   29.5  \\
NGC 6397      & 2.5 & 6.4 & 5.0 &   84  & 15.0 & 325 &   29.0  \\
NGC 6752      & 4.1 & 5.9 & 5.4 &  401  &  4.4 & 346 &   29.5  \\
\hline
\end{tabular}
\end{center}
\end{table*}

The nearest globular cluster, M4 (NGC~6121), has long been known to host a substantial population of low-luminosity X-ray sources ($L_X \lesssim 10^{34}~\ergs$; \citealt{Bassa04b,Verbunt01}). M4 has a moderate central density, thus providing an important basis of comparison with the well-studied nearby clusters $\omega$\,Cen \citep{Cool13,Henleywillis18}, which has a low central density, 47~Tuc \citep{Heinke05,RiveraSandoval18}, which has a moderate central density, and NGC~6397 \citep{Cohn10} and NGC~6752 \citep{Lugger17,Cohn21}, which are core collapsed and thus have extremely high central densities. The \citet{Bahramian20} catalogue of \chandra\ sources in 38 globular clusters, which is based on deep ACIS (Advanced CCD Imaging Spectrometer) imaging, lists 161 sources in the entire ACIS field of M4\@. Of these, 52 sources are within the central 1.7\,arcmin radius region, which corresponds to the half-width of the \hst\ ACS/WFC\footnote{Advanced Camera for Surveys/Wide Field Channel} field of view. The class of low-luminosity X-ray sources includes cataclysmic variables (CVs), magnetically active binaries (ABs), millisecond pulsars (MSPs), and (potentially) black hole (BH) and/or neutron star X-ray binaries \citep{van_den_Berg20}. 

In order to build on previous efforts to identify and characterise the X-ray source population of M4, we employed photometry from the Hubble Ultraviolet Globular Cluster Survey \citep[HUGS,][]{Piotto15,Nardiello18} and additional archival \hst\ data. 
The \citet{Bahramian20} \chandra\ source catalogue and the HUGS photometry database both supersede those used in the previous comprehensive study of \chandra\ sources and their optical counterparts in M4 by \citet{Bassa04b,Bassa05}. 

\subsection{Background}

Hard binaries play a critical role in cluster dynamical evolution, since they initially support cluster cores against collapse \citep{Kremer20} and later halt deep collapse, leading to core bounce and subsequent oscillations \citep{Hut92,Vesperini10}. The central question addressed by this study is how compact binary X-ray source populations are shaped by primordial formation and dynamical evolutionary processes. This issue has been the subject of considerable research, as recently reviewed by \citet{Cheng18}, \citet{Belloni19}, and \citet{Heinke20}. Globular clusters have long been known to be $\sim100$ times overabundant in low-mass X-ray binaries (LMXBs, with $L_X \gtrsim 10^{35}~\mathrm{erg~s}^{-1}$) relative to the field, pointing to a dynamical origin of these neutron-star-containing binaries in dense clusters \citep{Clark75}. Subsequent studies showed that the total number of X-ray sources in a globular cluster scales with the density-dependent interaction rate 
$\Gamma \propto \int_r (\rho^2/v) dV \propto \rho_0^2 r_c^3/v_0$, where $\rho_0$ is central density, $r_c$ is the core radius, and  $v_0$ is the central velocity dispersion \citep[the second expression approximates the interactions as all occurring in a constant-density cluster core;][] {Johnston96,Heinke03,Pooley03,Pooley06,Bahramian13}.
However, it has been observed that the low-density populations of open clusters have X-ray emissivities (X-ray luminosity per unit mass) that are higher than those of all but the highest density globular clusters \citep{Verbunt00,Ge15,van_den_Berg20}. For the open clusters NGC 6791 and M67, 
the number of CVs per unit mass is higher than in the field, which is consistent with the emissivity result. The origin of this is not certain, but the high evaporative mass loss of open clusters, combined with mass segregation, is a plausible explanation -- the remaining mass in the clusters is concentrated toward high-mass objects, especially binaries \citep{Verbunt00}. It appears that a complex interaction between binary formation and destruction determines the relative numbers of CVs and ABs in a globular cluster, and the luminosity functions of each of these groups \citep{Ivanova06,Belloni19,Heinke20}.

Previous \chandra\ and \hst\ studies of the X-ray emitting binary populations in other nearby globular clusters include those of NGC~6397 \citep{Grindlay01,Taylor01,Bogdanov10,Cohn10}, NGC~6752 \citep{Pooley02,Thomson12,Lugger17,Cohn21}, 47~Tuc \citep{Grindlay01b,Edmonds03a,Edmonds03b,Heinke05,Bhattacharya17,Cheng18,RiveraSandoval18}, and $\omega$\,Cen \citep{Haggard09,Cool13,Henleywillis18}. A compilation of the properties of these clusters and a summary of the \chandra\ observations are given in Table~\ref{t:Cluster_properties_Chandra_obs}.

\subsection{CVs in Globular Clusters} 

{Cataclysmic variables 
are 
semi-detached binary systems in which a white dwarf primary accretes material from a 
main-sequence secondary that overflows its Roche lobe. The term `cataclysmic' refers to the very large amplitude luminosity variations that may be observed in these systems.} CVs are the most numerous type of accreting X-ray source, and have been clearly identified in a number of globular clusters \citep[e.g.][]{Cool95,Pooley02,Edmonds03a,Knigge03,Cohn10,Thomson12,Cool13,Lugger17,RiveraSandoval18,Cohn21}. Identification of an X-ray source's optical counterpart as a likely CV generally has been claimed using one or more of: blue colours, photometric H$\alpha$ excess, optical variability, or a CV-type emission-line spectrum \citep{Knigge12}, along with a proper motion consistent with the cluster. A majority of the strong candidate CVs in clusters have been 
{identified using the trio of \chandra\ X-ray emission, blue colours, and \ha\ excess.}
{LMXBs, and MSPs with nondegenerate companions (redbacks), can appear similar to CVs, but both classes are significantly rarer than CVs (e.g. in 47 Tuc, there are 43 CVs, 6 quiescent LMXBs, and 3 redbacks; \citealt{RiveraSandoval18,Heinke05,Miller-Jones15,Ridolfi21}).}

The production of CVs in globular clusters is complex, as some CVs are produced through dynamical channels, such as exchange interactions with primordial binaries, and possibly tidal captures, while others are of primordial origin  \citep[e.g.][]{Davies97,Ivanova06,Shara06,Belloni19}. 
Dynamical production of CVs in dense clusters 
is supported by the correlation between the stellar encounter rates $\Gamma$ and numbers of cluster sources  
\citep{Pooley06,Heinke06,Bahramian13}. 
For a sufficiently large CV population in a cluster, the spatial distribution and luminosity function of the CVs provide information on their ages and dynamical state. 

\begin{table}
\caption{CV numbers in M4 and a sample of nearby, well-studied clusters. `Candidate' includes `Confirmed,' where confirmation is usually by a detection of excess \ha\ emission.}
\label{t:CVs_in_clusters}
\begin{center}
\begin{tabular}{lrrrr}
\hline
Cluster & 
\multicolumn{3}{c}{$N_\mathrm{CV}$} \\
\cline{2-4}%
& Predicted & Candidate & Confirmed & Refs \\
\hline
47 Tuc          &  36 & 44 & 29 & 1,2 \\
$\omega$\,Cen   & 144 & 27
&  6 & 3,4 \\
M4              &   4 &  4 &  1 & 5,6 \\ 
NGC 6397        &   4 & 15 & 11 & 7 \\
NGC 6752        &  11 & 18 &  7 & 8,9,10 \\ 
\hline \\
\multicolumn{5}{p{0.9\columnwidth}}
{\emph{Notes.}1: \citet{Edmonds03a}, 2: \citet{RiveraSandoval18},
3: \citet{Cool13}, 4: \citet{Henleywillis18}, 5: \citet{Bassa04b, Bassa05}, 6: this work, 7: \citet{Cohn10}, 8: \citet{Pooley03}, 9: \citet{Lugger17}, 10: \citet{Cohn21}}
\end{tabular}
\end{center}
\end{table}

Theoretically, lower-density clusters are expected to have more primordially-formed CVs than dynamically-formed CVs \citep{Davies97,Belloni19}, and there is observational support for this prediction \citep{Kong06,Haggard09,Cheng18,Belloni19}. It has also been suggested that many primordial soft binaries in clusters have been destroyed by dynamical interactions, thus preventing them from becoming X-ray-emitting close binaries  \citep{Ivanova05,Fregeau09,Belloni19}. In high-density clusters, formation of new CVs by dynamical processes may dominate over primordial CV destruction, leading to an excess number of CVs relative to the observed X-ray emissivity of the field CV population. 

Two recent studies have investigated the dependence of X-ray emissivity on cluster properties such as central density, mass, and interaction rate \citep{Cheng18,Heinke20}. The latter study found that for globular clusters with central densities above $10^{4}\,\msun\,\mathrm{pc}^{-3}$, 
X-ray emissivity and central density correlate. 
This supports the prediction that dynamical formation of CVs is dominant over primordial formation in these denser clusters.

Table~\ref{t:CVs_in_clusters} provides a comparison of the number of CV candidates in a cluster to the number predicted for its mass (Table~\ref{t:Cluster_properties_Chandra_obs}), based on the number of CVs per unit stellar mass in the solar vicinity. 
{The identification of CVs depends on both X-ray and optical sensitivity in a complex way (a full analysis is outside the scope of this paper), so the numbers of confirmed CVs should be taken as a lower limit on the true CV numbers.}
The local density of CVs with $L_X \ge 10^{30}~\ergs$ is $\sim 2\times10^{-6}~\mathrm{pc}^{-3}$ \citep{Pretorius12}. Taken with the local mass density of $0.05~\msun~\mathrm{pc}^{-3}$ \citep{Chabrier01}, this gives a field CV number per unit mass of $\sim 4\times10^{-5}~\msun^{-1}$.
%
Table~\ref{t:CVs_in_clusters} shows there is an excess of CV candidates  in the high-density globular clusters NGC~6397 and NGC~6752, relative to the number predicted from the field density {(or observed in the moderate-density cluster M4)},  especially for NGC~6397. This may be a result of dynamical interactions in the collapsed cores. 
{As M4 is nearby, with (now) rather deep X-ray and moderately deep optical observations, we plan to use it as a baseline for future comparisons among clusters.}

Bright CVs should be younger on average, as CVs often start with (relatively) massive companions, and the companion's optical luminosity and mass loss rate both decay with time due to mass loss to the white dwarf \citep{Hellier01}. Studies of NGC~6397 \citep{Cohn10}, NGC~6752 \citep{Lugger17,Cohn21},  47~Tuc \citep{RiveraSandoval18}, and  $\omega$\,Cen \citep{Cool13} have revealed significant populations of X-ray and optically faint CVs, which are expected to have orbital periods generally below 2 hours, based on predictions from binary evolution \citep{Ivanova06,Belloni19}. These simulations predict two (or more) times as many detectable CVs below the 2\,--\,3~h period gap as above it. 

Observations of local CVs by \citet{Pala20} find 3\,--\,8 times more CVs below the period gap, and suggest that $M_R\approx9$ roughly divides CVs at the period gap. 
In 47~Tuc, there are about $3\times$ more CVs below the period gap than above it \citep{RiveraSandoval18}. In $\omega$\,Cen, the numbers of CVs below and above 
the gap are comparable \citep{Cool13}, but this result is based on shallower \chandra\ observations (see Table~\ref{t:Cluster_properties_Chandra_obs}), 
{which would have missed most of the faint 47~Tuc CVs.
(Deeper \chandra\ observations of $\omega$\,Cen have been taken, and optical follow-up is now in progress.)} For the core-collapsed clusters NGC~6397 \citep{Cohn10} and NGC~6752 \citep{Lugger17,Cohn21}, for which the \chandra\ data reach well below  $10^{30}~\ergs$, there are comparable numbers of CVs below and above the period gap. 
Possible explanations for this relative lack of faint CVs in dense, core collapsed clusters include the destruction or ejection of old (faint) CVs through dynamical encounters, or by a recent burst of dynamical CV formation (making more bright CVs), in an extreme-density phase of core collapse. 

\subsection{ABs in Globular Clusters}

The most abundant class of low-luminosity X-ray source in globular clusters is magnetically active binaries that have enhanced chromospheric and coronal activity, usually due to tidal locking in a close system \citep{van_den_Berg20}. Classes of ABs include RS CVn stars (which contain at least one subgiant or giant), BY Dra stars (which contain two main sequence stars), W UMa stars (where  one, or both, stars fills its Roche lobe), and sub-subgiants (which lie below the subgiant branch on the CMD). ABs generally have lower X-ray luminosities than CVs; 
for instance in 47 Tuc, \citet{Heinke05} found an average $L_X\sim10^{30}~\ergs$ for a sample of 60 ABs and $L_X\sim10^{31}~\ergs$ for a sample of 22 CVs.
In NGC~6397, for which a limiting $L_X \approx 10^{29}~\ergs$ has been reached, a population of 42 ABs has been detected within the half-light radius \citep{Cohn10}. Four clusters, 47~Tuc \citep{Heinke05,Bhattacharya17}, M4 \citep{Bassa04b,Pooley16,Bahramian20}, M22 \citep{Bahramian20}, and NGC~6752 \citep{Lugger17,Cohn21} have been observed to a depth of $L_X < 10^{30}~\mathrm{erg~s}^{-1}$, allowing the detection of significant AB populations. 

ABs are expected to be an {essentially primordial population in clusters, having evolved from the short-period end of the binary period distribution \citep{Verbunt02,van_den_Berg20}.} This suggests that AB population size should scale with cluster mass, with the proviso that there is variation in the overall binary fraction among clusters. This would imply that for clusters with  large AB populations (and without large populations of dynamically formed CVs and LMXBs), X-ray emissivity should be independent of such cluster properties as central density. However, the X-ray emissivities of low-density ($\rho \lesssim 10^2 \msun\,\mathrm{pc}^{-3}$) open clusters are substantially higher than those of all but the densest globular clusters \citep{Verbunt00,Heinke20}. Thus, it appears that dynamical interactions may be destroying ABs in globular clusters. This study, along with a comparison to clusters of different masses and densities, will provide a measure of the dependence of the AB population size on cluster properties and thus a measure of the relative importance of primordial and dynamical AB formation 
and destruction 
channels.

\subsection{The Importance of M4} 

M4 makes a compelling target for comparison with other nearby, well-studied clusters, given its proximity (both reducing the effect of image crowding and allowing the imaging to go deeper), the moderate density of its core ($\rho_0 \sim 10^4~\msun~\rm{pc}^{-3}$) leading to a moderate stellar interaction rate $\Gamma$, 
the presence of an MSP \citep{Lyne88},  
and the existence of 38 MAVERIC\footnote{Milky Way ATCA and VLA Exploration of Radio sources In Clusters} radio sources, some of which are likely to be MSPs and/or quiescent BH X-ray binaries \citep{Shishkovsky20,Zhao21}. The presence of a confirmed MSP in M4 indicates that dynamical processes are active in its core, which has implications for the CV and AB populations, as discussed above. 


Previous \hst\ studies to find optical counterparts to \chandra\ sources in M4 are \citet{Bassa04b,Bassa05} and \citet{Pooley15,Pooley16}. Using 
25~ks of \chandra\ ACIS 
imaging, 
and archival \hst\ WFPC2 imaging, \citet{Bassa04b,Bassa05} detected 31 sources and obtained optical identifications for 
20 
of these. These optical counterparts include two CV candidates, the brighter of which (CX1) now appears to be a neutron star binary \citep[likely a redback MSP;][]{Kaluzny12,Pooley16}, one confirmed MSP, one AGN \citep[CX2,][]{Bassa05}, one foreground star, and 12 AB candidates. 
Early results from an analysis of the total of 119~ks of \chandra\ exposure 
indicate a rich population of X-ray sources of various classes \citep{Pooley15,Pooley16}. We have made an additional 28 identifications using the \citet{Bahramian20} source catalogue and the HUGS photometry database, beyond those found by \citet{Bassa04b}, 
as discussed in Section~\ref{sec:source-identification}.

\section{Data}
\subsection{UV and Optical Data}

The broad-band UV and optical data come from the Hubble UV Globular Cluster Survey \citep[HUGS;][]{Piotto15,Nardiello18}. This provides 5-band imaging and photometry of M4 in F275W (\UV), F336W (\U), F435W (\B), F606W (\V), and F814W (\I). The former three bands were obtained with the WFC3\footnote{Wide Field Camera 3}, the latter two with the ACS/WFC. {\citet{Legnardi23} have recently corrected the HUGS magnitudes for differential reddening. We employed these corrected magnitudes, which were kindly provided by M.~Legnardi (priv.\ comm.). This corrected photometry uses the KS2 method 1 approach \citep{Nardiello18}.} These data were supplemented by relatively shallow narrow-band F658N (\ha) and 
broad-band F625W (\R) ACS/WFC imaging from programme GO-10120 (PI: S. Anderson). These \ha\ and \R\ frames were reduced using DAOPHOT aperture photometry. Based on these data, (\UV, \uvu), (\U, \ub), and (\V, \vi) CMDs were constructed. In addition an (\hr, \vi) colour-colour diagram was generated.

{Inclusion of the differential reddening correction makes a substantial improvement in the tightness of the fiducial sequences in the (\V, \vi) CMD. The effect on bluer CMDs is much less evident, owing to the natural tendency of the \UV--\U--\B\ filter triplet to produce broad main sequences that provide evidence for the presence of multiple populations. Nonetheless, we now use the differential-reddening-corrected magnitudes for all of the CMDs. }

\begin{table}
\caption{UV and optical data used in this study}
\label{t:UV_optical_data}
\begin{center}
\begin{tabular}{lclcr} 
\hline
Programme & Observation date range & Inst. & Filter & Exp. \\
          &                        &       &        & \multicolumn{1}{c}{(s)}  \\
\hline 
GO-13297 & 2014-07-05 to 2015-02-17 & WFC3 & F275W & 3086 \\
GO-13297 & 2014-07-05 to 2015-02-17 & WFC3 & F336W & 1200 \\
GO-13297 & 2014-07-05 to 2015-02-17 & WFC3 & F438W & 131  \\
GO-10227 & 2006-03-05               & ACS  & F606W & 110  \\
GO-10775 & 2006-03-05               & ACS  & F814W & 120  \\
GO-10120 & 2004-07-26               & ACS  & F625W & 15$^a$   \\
GO-10120 & 2004-07-26               & ACS  & F658N & 340$^a$  \\
\hline \\
\multicolumn{5}{p{0.95\columnwidth}}{\emph{Note.}$^a$We chose to use to use the single `short' F625W and `long' F658N exposures, as these are contiguous in time, which minimises \hr\ colour offsets that are due to source variability.} \\
\end{tabular}
\end{center}
\end{table}

\subsection{VLA Data}
\cory{M4 is one of 25 GCs observed by the {\it Karl G. Jansky Very Large Array} (VLA) for the {\it Milky way ATCA and VLA Exploration of Radio sources In Clusters} (MAVERIC) survey (project codes VLA/13B-014, VLA/15A-100;  \citealt{Shishkovsky20}) with a total on-source time of $6.9\,$h. The observations were taken in the most-extended A configuration, using the C-band receiver. Details of calibration and data reduction processes are described in \citet{Shishkovsky20}. The data were imaged in low ($5~\mathrm{GHz}$) and high frequency ($7.2~\mathrm{GHz}$) sub-bands with beam sizes of $1\farcs18 \times 0\farcs08$ and $0\farcs84\times 0\farcs63$, and RMSs of $2.3\,\mathrm{\mu Jy~beam^{-1}}$ and $2.1\,\mathrm{\mu Jy~beam^{-1}}$, respectively. The catalogue reports positions, low- and high-frequency flux densities ($\slow$ and $\shigh$), and a radio spectral index $\alpha$ defined as $S_\nu \propto \nu^\alpha$. In M4, there are 37 radio sources that are detected at the 5\,$\sigma$ level.}

\subsection{Chandra data}
\citet{Bahramian20} 
used deep \chandra\ ACIS imaging to produce a comprehensive source catalogue of 161 sources in M4, which we use. 
They
labelled each X-ray source with a `CXOU\_J' designation that is based on its celestial coordinates. In order to provide a convenient source numbering system, we have extended the `CX' sequence numbering system used by \citet{Bassa04b}, 
ordering the sources by descending $L_X~\mathrm{(0.5-10~keV)}$, as given by \citet{Bahramian20}. 
We break `ties' by numbering the sources in order of increasing RA\@. 
Table~\ref{t:source_numbering} gives the correspondence between the `CX' and `CXOU\_J' numbering. \citet{Bahramian20} include marginal sources in their catalogue, clearly identified; we search these error circles as well, but many of them are empty, suggesting that most of these marginal sources are not real.

\begin{table*}
\centering
\caption{Source Numbering Convention}
\label{t:source_numbering}
\begin{tabular}{rcrcrc}
\hline
CX & CXOU\_J & ~~~~CX & CXOU\_J & ~~~~CX & CXOU\_J \\
\hline
1	&	162334.13-263134.7	&	61	&	162344.00-262945.9	&	115	&	162335.50-263542.7	\\
3	&	162338.08-263138.0	&	62	&	162330.06-263205.7	&	116	&	162319.83-263315.4	\\
4	&	162334.33-263039.2	&	63	&	162342.32-263037.9	&	117	&	162339.78-262937.1	\\
6	&	162338.10-262922.2	&	64	&	162340.70-262939.3	&	118	&	162351.15-262918.8	\\
7	&	162345.89-262854.9	&	65	&	162340.16-262925.2	&	119	&	162337.22-262840.2	\\
8	&	162331.46-263057.9	&	66	&	162323.65-263318.7	&	120	&	162331.62-263034.5	\\
10	&	162335.03-263119.2	&	67	&	162323.88-263448.7	&	121	&	162349.73-263152.5	\\
11	&	162332.38-263045.4	&	68	&	162352.28-263159.0	&	122	&	162347.45-262858.6	\\
12	&	162338.20-263154.1	&	69	&	162317.26-263330.1	&	123	&	162314.82-263251.7	\\
13	&	162334.31-263202.0	&	70	&	162329.23-262949.5	&	124	&	162328.60-263056.3	\\
14	&	162326.00-263354.4	&	71	&	162322.17-262714.6	&	125	&	162335.99-263124.7	\\
15	&	162336.78-263144.3	&	72	&	162335.07-263204.2	&	126	&	162352.78-262849.8	\\
16	&	162333.68-263417.1	&	73	&	162340.21-262926.5	&	127	&	162334.38-262837.8	\\
17	&	162335.98-263101.6	&	74	&	162333.18-263109.3	&	128	&	162342.00-262925.1	\\
18	&	162345.77-263116.8	&	75	&	162331.30-263148.6	&	129	&	162343.91-262751.7	\\
19	&	162328.95-262951.1	&	76	&	162341.67-263115.5	&	130	&	162335.92-263240.1	\\
20	&	162336.89-263139.4	&	77	&	162341.58-262937.7	&	131	&	162333.58-263151.5	\\
21	&	162334.67-263204.4	&	78	&	162338.81-263456.3	&	132	&	162338.75-263303.9	\\
22	&	162333.36-263145.4	&	79	&	162348.74-263206.0	&	133	&	162330.27-263356.1	\\
24	&	162342.09-263136.6	&	80	&	162335.81-263132.5	&	134	&	162336.34-263245.0	\\
25	&	162333.51-263229.9	&	81	&	162331.83-263156.6	&	135	&	162328.61-262701.6	\\
26	&	162338.89-263148.2	&	82	&	162334.82-263159.9	&	136	&	162345.60-262758.9	\\
27	&	162333.28-263157.6	&	83	&	162337.00-263133.6	&	137	&	162342.06-262921.9	\\
28	&	162334.97-263224.3	&	84	&	162336.64-263143.5	&	138	&	162347.77-263050.1	\\
30	&	162328.40-263022.4	&	85	&	162336.45-263030.5	&	139	&	162332.31-262633.6	\\
32	&	162335.21-263525.8	&	86	&	162341.47-263205.4	&	140	&	162347.89-262817.7	\\
33	&	162334.27-262956.0	&	87	&	162335.79-263137.4	&	141	&	162341.35-263347.6	\\
34	&	162349.77-263323.5	&	88	&	162339.70-263120.2	&	142	&	162350.68-262749.1	\\
35	&	162352.35-263229.9	&	89	&	162335.37-263450.6	&	143	&	162333.77-262630.0	\\
36	&	162346.42-263115.8	&	90	&	162324.29-263013.8	&	144	&	162337.08-263208.6	\\
37	&	162346.38-263114.7	&	91	&	162333.76-263405.8	&	145	&	162345.52-262730.4	\\
38	&	162337.79-263518.9	&	92	&	162344.21-262645.5	&	146	&	162331.80-262647.0	\\
39	&	162336.09-262736.9	&	93	&	162346.83-262936.0	&	147	&	162328.34-263320.6	\\
40	&	162334.62-262717.8	&	94	&	162328.54-263134.2	&	148	&	162340.36-263332.7	\\
41	&	162352.09-263214.6	&	95	&	162338.48-262952.5	&	149	&	162347.48-262803.5	\\
42	&	162335.50-262707.6	&	96	&	162317.60-262842.1	&	150	&	162340.15-262646.3	\\
43	&	162326.61-262658.8	&	97	&	162339.18-262955.3	&	151	&	162344.34-263154.0	\\
44	&	162352.12-263217.6	&	98	&	162322.79-263155.0	&	152	&	162316.75-263256.0	\\
45	&	162351.94-262833.1	&	99	&	162351.40-263324.8	&	153	&	162334.48-262704.2	\\
46	&	162338.08-262858.8	&	100	&	162330.38-263049.6	&	154	&	162332.75-262815.9	\\
47	&	162352.19-263306.5	&	101	&	162334.54-263110.8	&	155	&	162329.97-262919.5	\\
48	&	162342.90-262726.8	&	102	&	162339.82-263126.6	&	156	&	162331.10-263300.5	\\
49	&	162352.63-262950.4	&	103	&	162319.14-262850.2	&	157	&	162325.66-262930.5	\\
50	&	162350.04-262901.6	&	104	&	162334.77-263143.8	&	158	&	162325.72-262817.9	\\
51	&	162352.07-262937.7	&	105	&	162327.01-263249.5	&	159	&	162340.36-263013.3	\\
52	&	162322.56-263340.2	&	106	&	162323.86-263028.6	&	160	&	162328.61-263543.0	\\
53	&	162335.67-262709.7	&	107	&	162319.20-263206.4	&	161	&	162337.21-263522.6	\\
54	&	162329.58-262839.7	&	108	&	162320.00-262740.8	&	162	&	162350.99-262940.4	\\
55	&	162333.85-263200.3	&	109	&	162320.17-262751.2	&	163	&	162346.66-263120.0	\\
56	&	162339.82-263557.4	&	110	&	162333.28-263113.7	&	164	&	162318.25-263116.0	\\
57	&	162330.27-263045.1	&	111	&	162319.45-262943.6	&	165	&	162325.42-263102.8	\\
58	&	162345.02-263030.0	&	112	&	162328.03-263528.6	&	166	&	162336.79-263202.7	\\
59	&	162323.37-263211.5	&	113	&	162351.74-263316.0	&	167	&	162340.44-263205.5	\\
60	&	162336.29-263553.8	&	114	&	162320.56-263217.6	&	168 &	162331.54-263057.6	\\
\hline
\end{tabular}
\end{table*}

\section{Astrometry}\label{astrometry}

{Given the multi-wavelength datasets used in this study, which were collected at different epochs, it is important to establish a common astrometric reference frame. Since the HUGS positional catalogue is referred to the astrometric system of the \emph{Gaia} Data Release 1 Catalogue \citep[see][]{Nardiello18}, we adopted the HUGS object positions as our fundamental reference system. We next computed a boresight correction of the \citet{Bahramian20} \chandra\ source positions to the HUGS/\emph{Gaia} frame based on the offsets of the sources CX1, CX3, CX4, CX10, CX11, CX12, and CX15 from our proposed HUGS counterparts.\footnote{The correspondence between the CX source numbering and the \citet{Bahramian20} source designations is given in Table~\ref{t:source_numbering}.} These 7 sources are among the brightest in M4; we eliminated CX8 and CX20 from this list, given their larger than typical offsets from the proposed HUGS counterpart positions. The resulting boresight correction in angular displacement is $\Delta\alpha\cos\delta = -0.099\,\mathrm{arcsec}$ and $\Delta\delta = -0.131\,\mathrm{arcsec}$. Interestingly, these boresight offsets are nearly precisely what is expected based on the \emph{Gaia}-based proper motion of M4 over the 7.3 year interval between the 2007.7 \chandra\ observations and the 2015.0 HUGS epoch. 
\citet{Vasiliev19} gives proper motion components\footnote{The proper motion nomenclature convention used by \citet{Vasiliev19} is: $\overline{\mu_\alpha}\equiv[\mathrm{d}\alpha/\mathrm{d}t]\cos\delta$,  $\overline{\mu_\delta}\equiv\mathrm{d}\delta/\mathrm{d}t$.} for M4 of $\overline{\mu_\alpha} = -12.490\,\mathrm{mas\,yr}^{-1}$ and $\overline{\mu_\delta} = -19.001\,\mathrm{mas\,yr}^{-1}$. These produce shifts of $\Delta\alpha\cos\delta = -0.091\,\mathrm{arcsec}$ and $\Delta\delta = -0.139\,\mathrm{arcsec}$, agreeing within 0.01\,arcsec with our offset.

As a check on the radio--optical--X-ray astrometric alignment, we considered the millisecond pulsar PSR B1620-26, which is detected in the radio, optical, and X-ray. The various measurements of its position in each of the wavelength bands are compared in Table~\ref{t:msp_position}. The radio position has been measured both by pulsar timing \citep{Thorsett99} and by a direct VLA detection \citep{Shishkovsky20}. The former measurement has an epoch of 1992.28, and thus must be advanced in proper motion for comparison with the HUGS/\emph{Gaia} frame. We note that the proper motion components given by \citet{Thorsett99} have a fairly large uncertainty, 8 per cent in RA and 20 per cent in Decl. \citet{Vasiliev19} have computed much higher accuracy mean proper motion components for M4, based on \emph{Gaia} measurements. We note that the timing position, advanced to epoch 2015.0 
using the 
\emph{Gaia} 
proper motion, agrees within 
0.03 arcsec with the VLA, HST, and \chandra\ positions. Thus, we conclude that there is good astrometric alignment of the radio, optical, and X-ray data with the \emph{Gaia} system.

\begin{table}
{    
    \caption{Measurements of the position of PSR B1620-26}
    \label{t:msp_position}
    \begin{center}
    \begin{tabular}{lccp{48pt}<{\raggedright}}
    \hline
    Type & RA-Decl & Epoch & Notes \\
    \hline\hline
    Timing$^a$   & 16:23:38.199 $-$26:31:54.34 & 2015.0 & PSR B2629-26 \\      
    Timing$^b$   & 16:23:38.201 $-$26:31:54.20 & 2015.0 & PSR B2629-26 \\      
    VLA$^c$      & 16:23:38.201	$-$26:31:54.24 & 2014.5 & M4-VLA9 \\
    HST$^d$      & 16:23:38.201	$-$26:31:54.22 & 2015.0 & R0006434 \\
    \chandra$^e$ & 16:23:38.199	$-$26:31:54.23 & 2015.0 & CX12 \\
    \hline
    \multicolumn{4}{p{0.95\columnwidth}}{\emph{Notes.}$^a$PSR position from \citet{Thorsett99}, advanced to 2015.0 using approximate proper motion from \citet{Thorsett99}.} \\
    \multicolumn{4}{p{0.95\columnwidth}}{$^b$PSR position from \citet{Thorsett99}, advanced to 2015.0 using \emph{Gaia} mean cluster proper motion from \citet{Vasiliev19}.} \\
    \multicolumn{4}{l}{$^c$Source position from \citet{Shishkovsky20}.} \\
    \multicolumn{4}{l}{$^d$Source position from \citet{Nardiello18}.} \\
    \multicolumn{4}{p{0.95\columnwidth}}{$^e$Source position from \citet{Bahramian20}, boresight corrected to HUGS/\emph{Gaia} frame.}
    \end{tabular}
    \end{center}
}
\end{table}
}

\section{Variability} \label{Variability}

\subsection{Comparison to M4 Core Project\label{M4_core_project}}

\begin{figure*}
  \centering
  \includegraphics[width=0.725\textwidth]{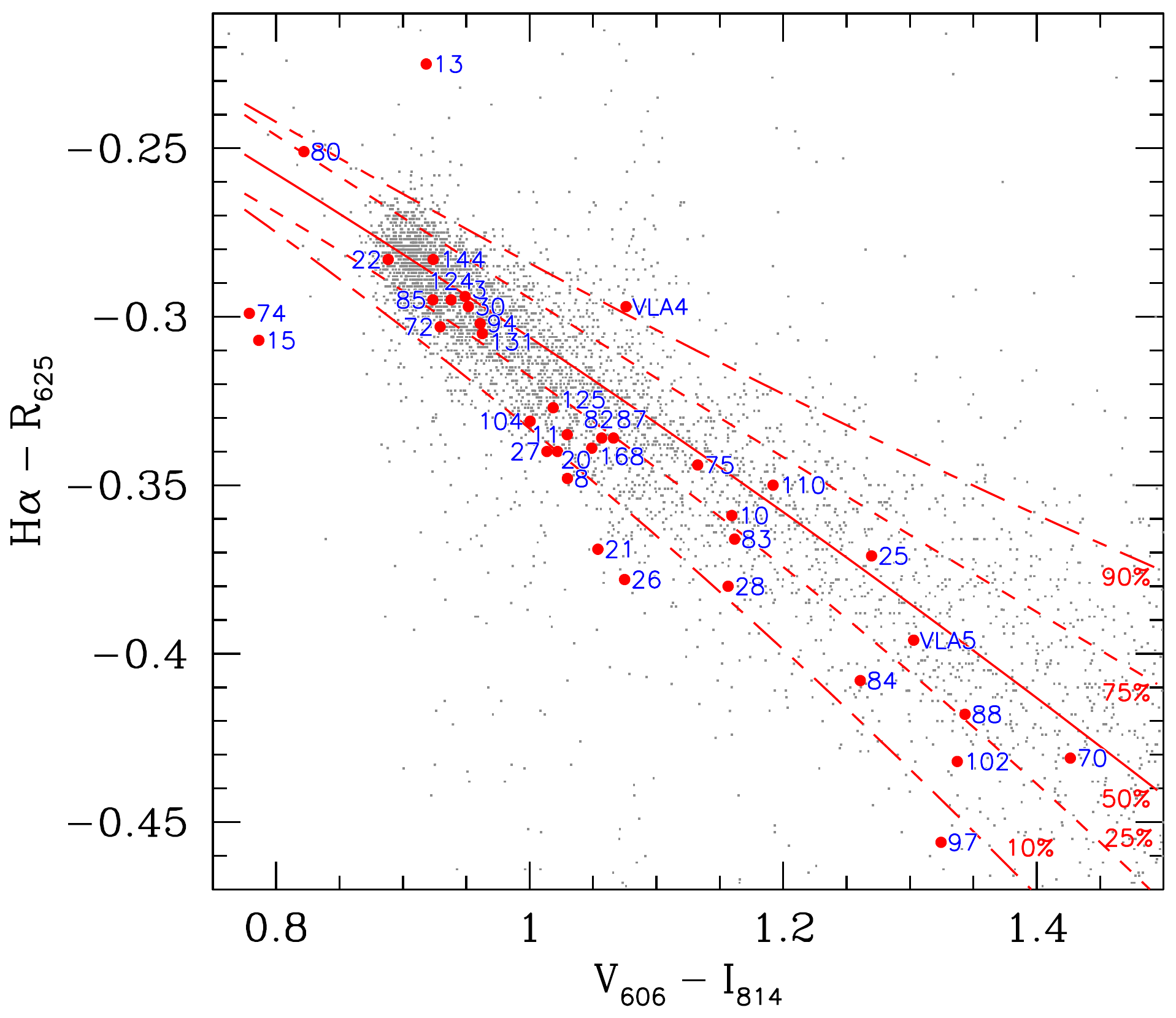}
  \caption{ \hr\ index versus \vi\ broad-band colour. 
  The solid curve indicates the median of the \hr\ distribution, the dashed curves indicate the upper and lower quartiles, and the long-short dashed curves indicate the upper and lower deciles. All curves are second-order polynomial fits. Most of the sources fall to the \ha-excess side of the median. The sources CX4 and CX101 (large \ha\ excess) and VLA20 (large \ha\ deficit) fall outside the limits of the \hr\ axis.}
  \label{f:color-color_diagram}
\end{figure*}

As part of the M4 Core Project, \citet{Nascimbeni14} 
detected 38 variable stars in the core of M4, using deep \hst\ WFC3 imaging with the filters F467M and F775W. They provide light curves for each of these variables, demonstrating that 19 of the 38 are eclipsing binaries. \citet{Nascimbeni14} found that 9 of the 38 variables correspond to \chandra\ sources from the original \citet{Bassa04b} list. We find that an additional 7 of the variables correspond to \citet{Bahramian20} sources from the extended CX source list. Table~\ref{t:cross-IDs} provides the cross-IDs for the 16 variable star--X-ray source matches. It can be seen that the counterpart types -- nearly all AB -- inferred from the broad-band photometric properties in the present study are consistent with the binary nature of these objects, as demonstrated by their light curves from \citet{Nascimbeni14}. 
A number 
of the objects in Table~\ref{t:cross-IDs} show strong evidence of \ha\ emission (see Fig.~
\ref{f:color-color_diagram}), consistent with an active binary nature. Several of the contact eclipsing binaries demonstrate evidence of strange colors in our CMDs (see below), likely due to large variability.

\subsection{Variability from HUGS photometry}

\begin{figure*}
  \centering
  \includegraphics[width=0.7\textwidth]{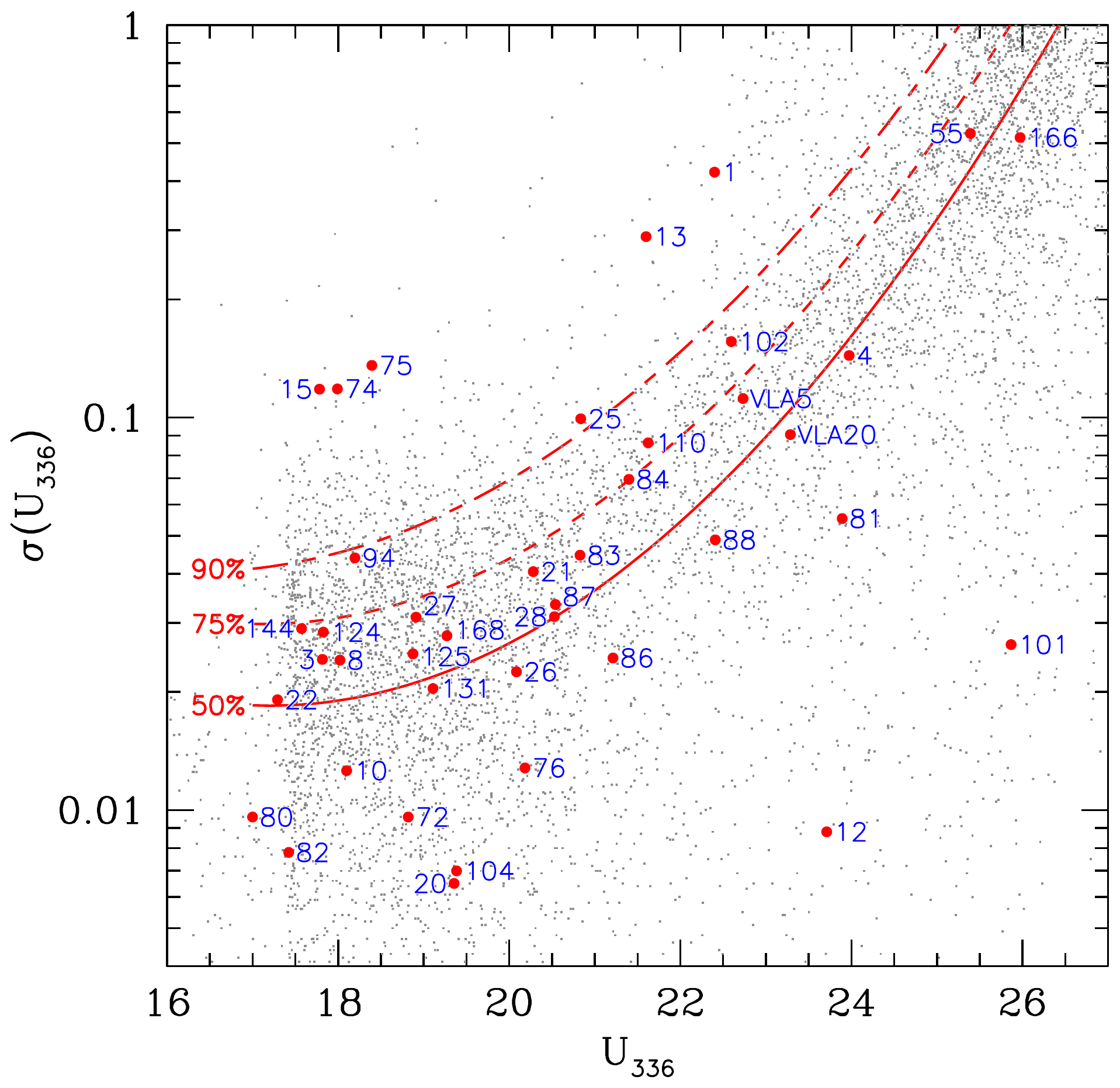}
  \caption{RMS of the $\le$ four HUGS \U\ magnitude measurements, vs.\ \U\ magnitude. This plot provides a measure of variability, particularly for counterparts with a short variability timescale. The counterparts to CX1, CX13, CX15, CX74, and CX75 show the strongest variability signal. All five of these significantly variable counterparts are known to have short binary periods of $\lesssim 7$\,h: CX1 \citep{Kaluzny12}, and CX13, CX15, CX74, and CX75 \citep{Nascimbeni14}.}
  \label{f:m4_sigma_u_plot}
\end{figure*}

Each of the HUGS magnitudes has an associated rms ($\sigma$), which measures the dispersion over the multiple magnitude measurements. This value may be used as a measure of variability by plotting it against magnitude as in Fig.~\ref{f:m4_sigma_u_plot}, for \U. In this figure, curves representing the median, the 75th percentile, and the 90th percentile are plotted. Of the variables in Table~\ref{t:cross-IDs}, it can be seen that CX13, CX15, CX25, CX74, CX75, and CX94 register as variables in Fig.~\ref{f:m4_sigma_u_plot}. 
The $\sigma$--magnitude plot provides an imperfect measure of variability, as 
it is based on a small number of magnitude measurements ($\leq 4)$. Thus, the light curve sampling is generally incomplete, particularly for the longer period variables. We note that CX1, CX13, CX15, CX74, and CX75, which show the strongest variability signal in Fig.~\ref{f:m4_sigma_u_plot}, have periods of $\le 0.3$~day. In contrast, CX25 ($P=1.9$~day) and CX94 ($P=5.9$~day) show a weaker variability signal. In any case, this plot does provide a useful indication of counterparts which display variability, such as CX1, which is known to have a 0.26 day sinusoidal period. 

\begin{table}
\centering
\caption{Cross-IDs with \citet{Nascimbeni14}}
\label{t:cross-IDs}
\begin{tabular}{rclllc}
\hline
CX & ID$^a$ & Type$^b$ & Notes$^c$ & Present Type & \ha?\,$^d$ \\
\hline
3	&	3236	&	EB?	&	BSEQ/TO; CX3	&	AB	&	n	\\
8	&	7864	&	UNK	&	Above BSEQ; CX8	&	AB/??	&	y	\\
11	&	8081	&	UNK	&	BSEQ; CX11 	&	AB	&	n	\\
13	&	3407	&	cEB	&	NM; CX13   	&	AB	&	n	\\
15	&	3401	&	cEB	&	TO; CX15   	&	AB	&	y	\\
20	&	3627	&	EB 	&	MS/BSEQ; CX20	&	AB	&	y	\\
21	&	3153	&	EB 	&	MS/BSEQ; CX21	&	AB	&	y	\\
25	&	2316	&	EB 	&	BSEQ; CX25 	&	AB	&	n	\\
28	&	2108	&	dEB	&	BSEQ; CX28 	&	AB?	&	n	\\
72	&	2992	&	EB?	&	MS/BSEQ    	&	AB?	&	n	\\
74	&	6807	&	cEB	&	BSEQ       	&	AB?	&	y	\\
75	&	5430	&	cEB	&	MS         	&	AB?	&	n	\\
84	&	3487	&	EB 	&	BSEQ       	&	AB	&	n	\\
86	&	1001	&	EB?	&	BSEQ       	&	AB	&	--	\\
88	&	3575	&	EB 	&	BSEQ       	&	AB	&	n	\\
94	&	7202	&	dEB	&	BSEQ       	&	AB	&	n	\\
\hline
\multicolumn{6}{l}{\makecell[{{p{0.9\columnwidth}}}]{$^a$ID \# from \citet{Nascimbeni14}.}}\\
\multicolumn{6}{l}{\makecell[{{p{0.9\columnwidth}}}]{$^b$Binary type from \citet{Nascimbeni14}. EB = eclipsing binary; c = contact; d = detached; UNK = unknown.}}\\
\multicolumn{6}{l}{\makecell[{{p{0.9\columnwidth}}}]{$^c$Notes from \citet{Nascimbeni14}. BSEQ = binary sequence; NM = non-member.}}\\
\multicolumn{6}{l}{\makecell[{{p{0.9\columnwidth}}}]{$^d$\ha\ excess confirmed in 
the colour-colour diagram (Fig.~\ref{f:color-color_diagram}).}}\\
\end{tabular}
\end{table}

\begin{figure*}
  \centering
  \includegraphics[width=0.47\textwidth]{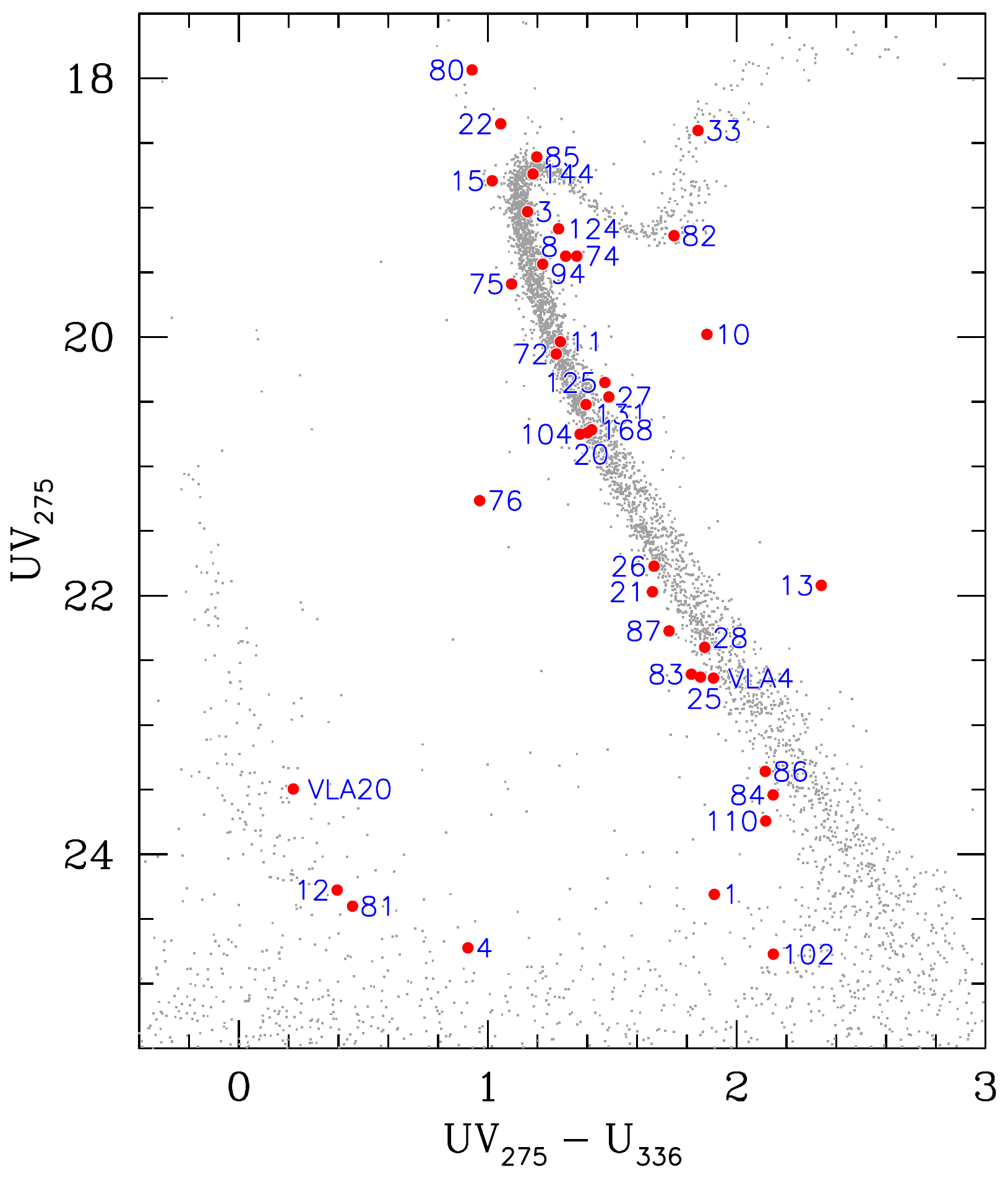}
  \includegraphics[width=0.47\textwidth]{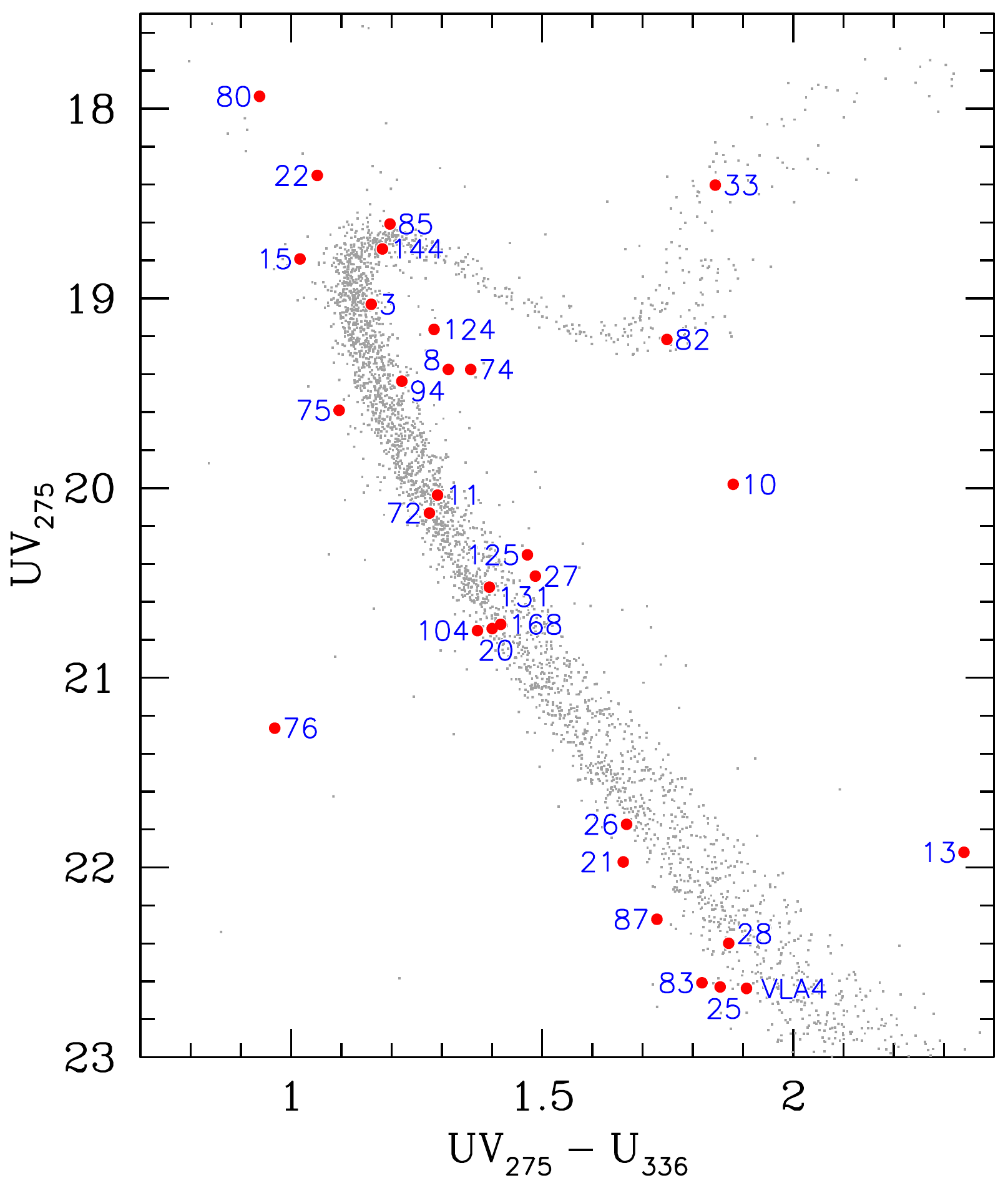}
  \caption{Left panel: Colour-magnitude diagram, based on HUGS photometry in \UV\ and \U, illustrating the location of candidate counterparts to \chandra\ X-ray sources and to VLA sources without a \chandra\ counterpart. Right panel: Zoom of CMD to show more detail near the MS. CX1 has been interpreted as a neutron star binary \citep{Kaluzny12,Pooley16}. CX12 is the binary counterpart to the millisecond pulsar B1620-26 \citep{Sigurdsson03}, which is detected as MAVERIC radio source M4-VLA9. The secondary in this system is a low-mass white dwarf. CX4, CX81, and VLA20 have a similar white-dwarf-like presentation to that of CX12. CX4 and CX81 are faint CV candidates. VLA20 is an MSP candidate. CX76 is a bright CV candidate.  
  }
  \label{f:m4_cmd1_hugs}
\end{figure*}

\begin{figure*}
  \centering
  \includegraphics[width=0.475\textwidth]{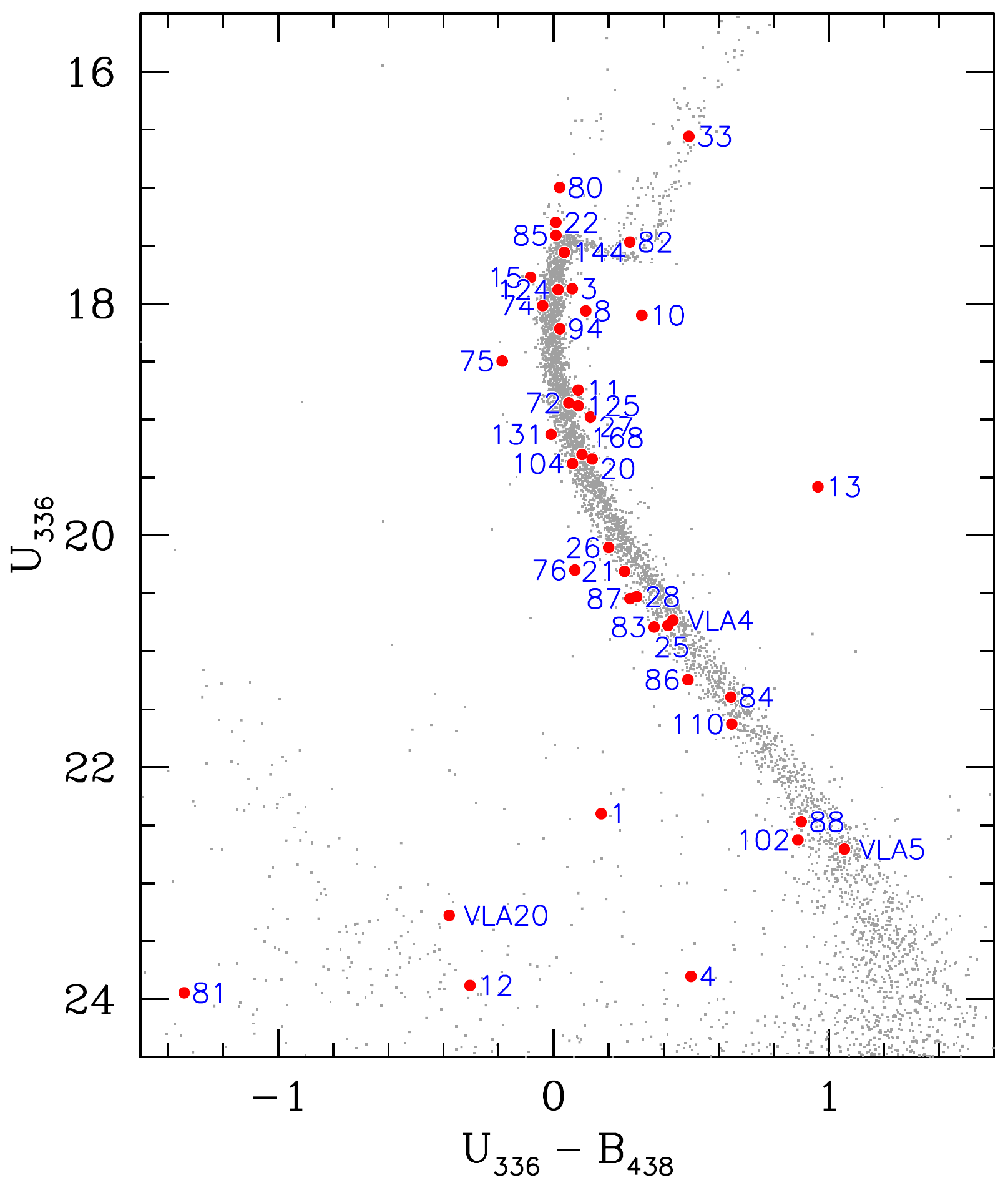}
  \includegraphics[width=0.475\textwidth]{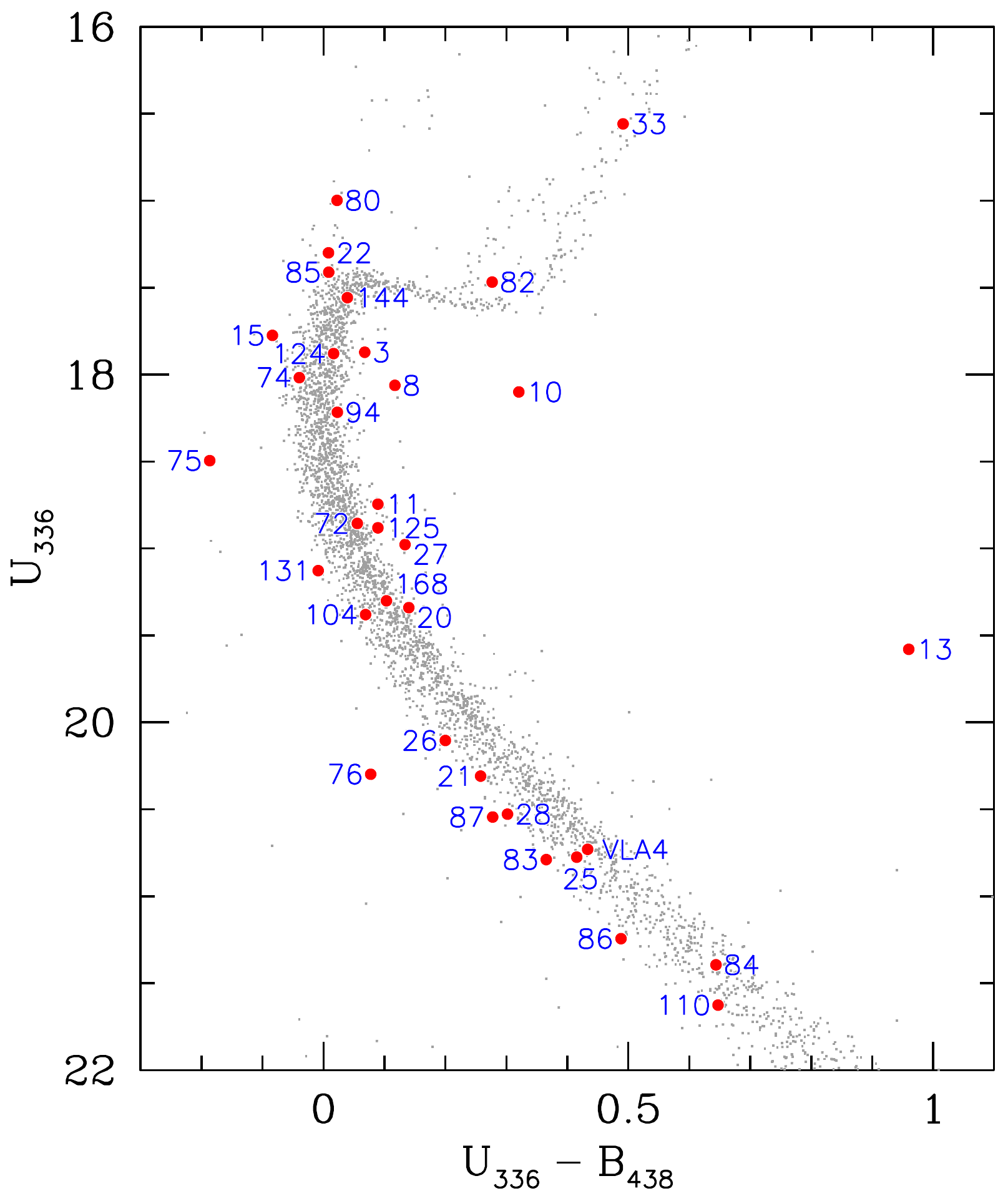}
  \caption{Left panel: Colour-magnitude diagram, based on HUGS photometry in \U\ and \B, showing the same counterparts as in Fig.~\ref{f:m4_cmd1_hugs}. Right panel: Zoom of CMD. The counterparts to CX15 and CX75 are consistent with being W~UMa-type contact binaries.}
  \label{f:m4_cmd2_hugs}
\end{figure*}

\begin{figure*}
  \centering
  \includegraphics[width=0.475\textwidth]{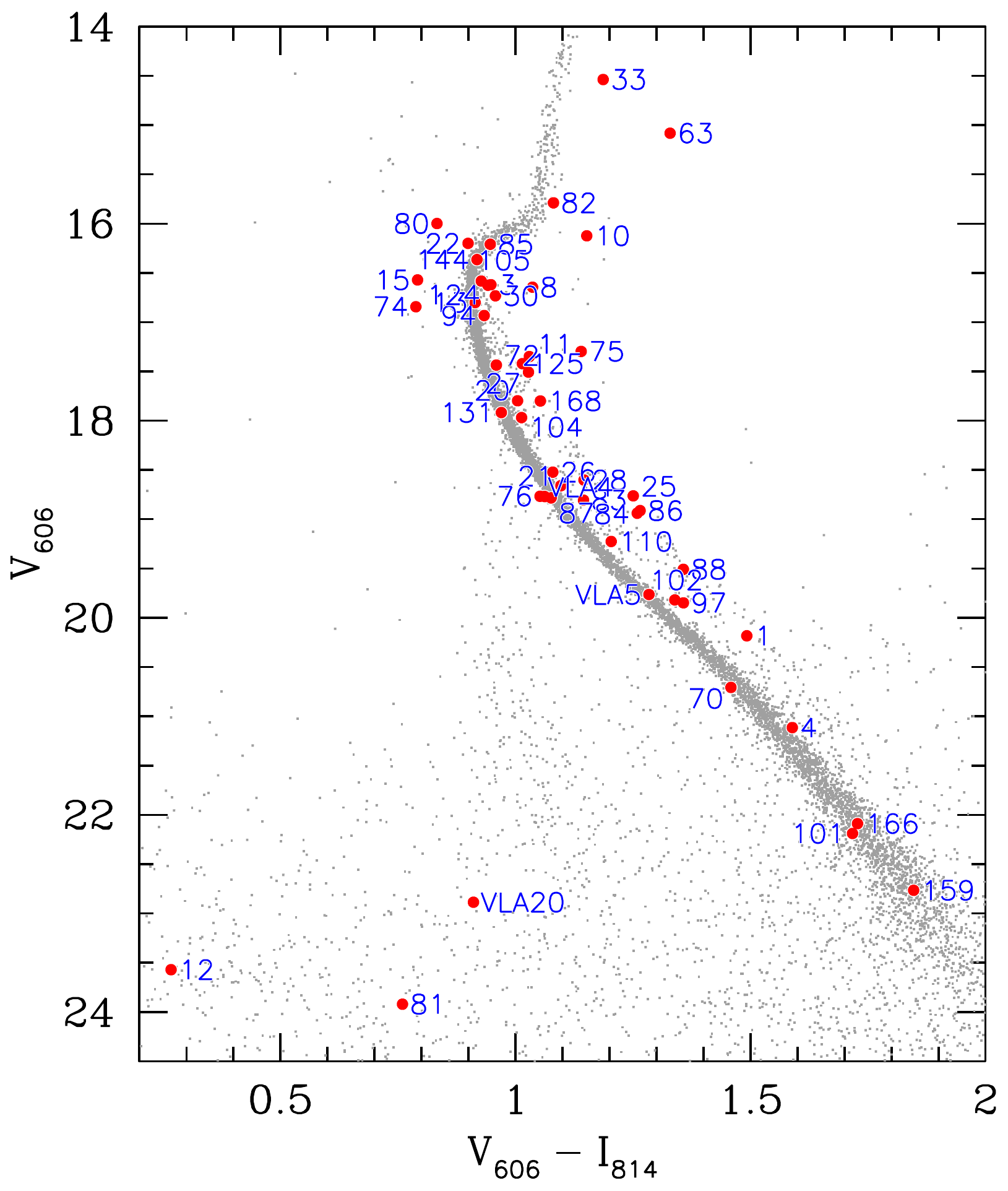}
  \includegraphics[width=0.475\textwidth]{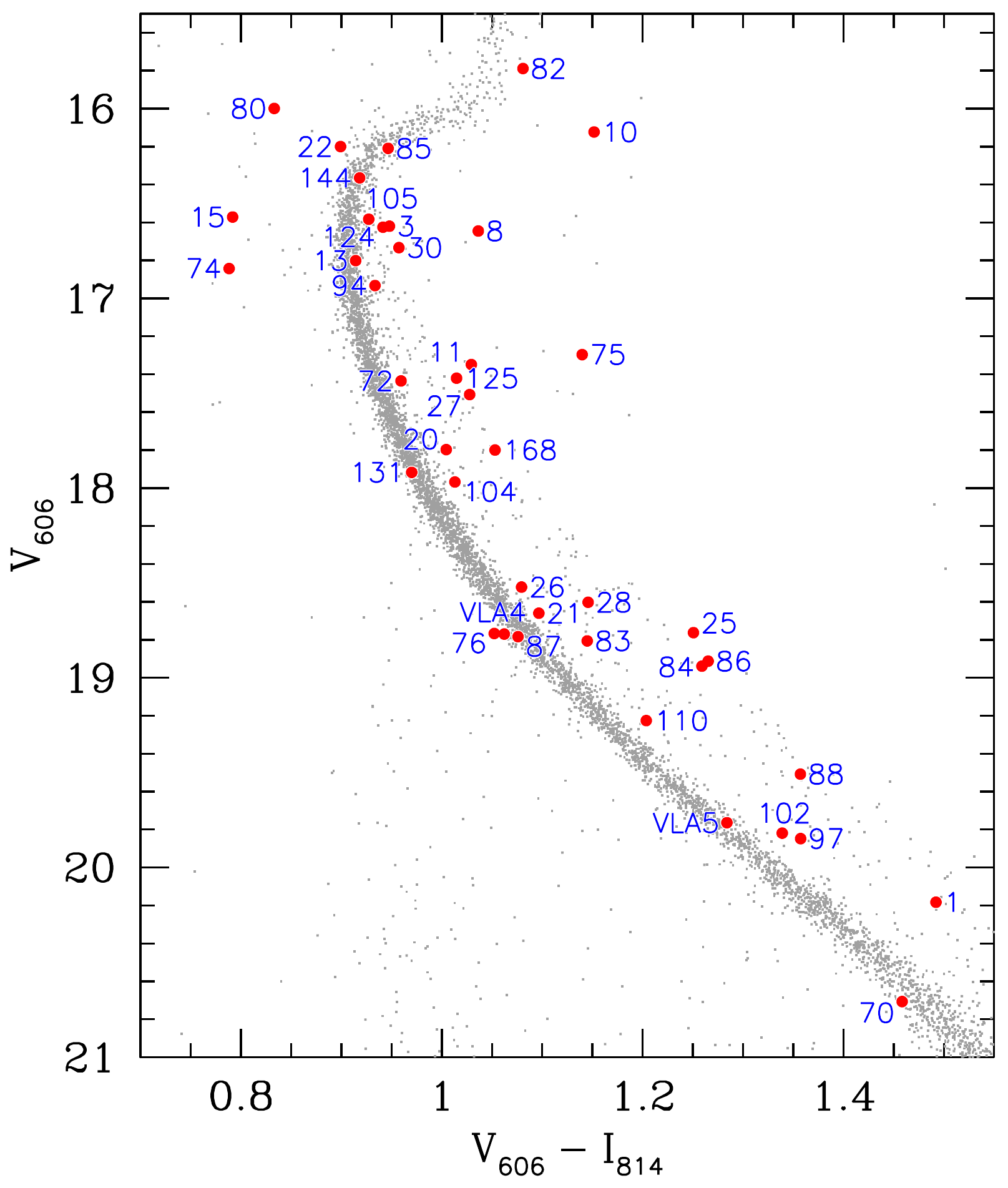}
  \caption{Left panel: Colour-magnitude diagram, based on HUGS photometry in \V\ and \I, showing the same counterparts as in Fig.~\ref{f:m4_cmd1_hugs}. Right panel: Zoom of CMD. There are some counterparts that are only detected in these filters, owing to the larger size of the ACS/WFC field relative to the WFC3/UVIS field. Most of the counterparts that lie to the right side of the fiducial sequences are consistent with being ABs.
  }
  \label{f:m4_cmd3_hugs}
\end{figure*}


\newcommand{\note}{\nointerlineskip\hangindent 1.5ex \hangafter 1}
\begin{landscape}
\begin{table}
\centering
\caption{Optical Counterpart Summary}
\label{t:counterparts}
\begin{adjustbox}{max width=\linewidth}
\begin{tabular}{rccccccp{48pt}<{\centering}p{36pt}<{\centering}cp{144pt}<{\raggedright}}
\hline
Source$^a$ &
RA, Dec (J2000)$^b$ &
$\rerr~('')^c$ &
$r~(')^d$ &
$L_X$~(0.5--10\,keV)$^e$ &
HUGS \# &
Offset$^f$&
Type$^g$ &
Bassa Type$^h$ &
PM$^i$ &
Notes \\
\hline
	1	&	16:23:34.125	$-$26:31:34.90	&	0.30	&	0.25	&	1.54E+32	&	R0007667	&	0.14	&	MSP?/ qLMXB?	&	CV	&	0.00	&	\note red in \vi, quite blue in \uvu\ and \ub, UV-variable; PM likely affected by brighter neighbor	\\
	3	&	16:23:38.074	$-$26:31:38.18	&	0.32	&	0.65	&	2.89E+31	&	R0001235	&	0.08	&	AB	&	AB	&	96.40	&	\note slightly red in all CMDs	\\
	4	&	16:23:34.320	$-$26:30:39.31	&	0.34	&	0.91	&	1.99E+31	&	R0010636	&	0.05	&	CV	&	CV/AB	&	96.60	&	\note WD seq in \uvu, blue in \ub, MS in \vi, large \ha-excess	\\
	8	&	16:23:31.460	$-$26:30:58.01	&	0.35	&	1.02	&	5.1E+30	&	R0001877	&	0.51	&	SSG/??	&	AB/V52	&	97.50	&	\note SSG in all CMDs, \ha-excess	\\
	10	&	16:23:35.031	$-$26:31:19.38	&	0.32	&	0.23	&	1.63E+31	&	R0001536	&	0.04	&	AB/SSG	&	AB	&	97.90	&	\note SSG in all CMDs	\\
	11	&	16:23:32.395	$-$26:30:45.75	&	0.38	&	1.01	&	3.11E+30	&	R0002030	&	0.21	&	AB	&	--	&	97.10	&	\note on MS in \uvu, red in \ub\ and \vi	\\
	12	&	16:23:38.199	$-$26:31:54.23	&	0.36	&	0.76	&	2.58E+30	&	R0006434	&	0.07	&	MSP	&	MSP	&	--	&	\note on WD seq in all CMDs	\\
	13	&	16:23:34.310	$-$26:32:02.20	&	0.35	&	0.53	&	3.02E+30	&	R0000806	&	0.15	&	AB	&	AB/V49	&	--	&	\note slightly red in \vi\ CMD; PM nonmember, large PM affects other CMD locations	\\
	15	&	16:23:36.781	$-$26:31:44.49	&	0.38	&	0.40	&	3.66E+30	&	R0001106	&	0.18	&	AB	&	AB/V48	&	98.00	&	\note BS in all CMDs, \ha-excess; eclipsing binary 
 \\
	17	&	16:23:35.977	$-$26:31:01.84	&	0.42	&	0.54	&	2.24E+30	&	--	&	--	&	--	&	--	&	--	&	\note empty error circle near very bright RG	\\
	19	&	16:23:28.915	$-$26:29:51.08	&	0.50	&	2.20	&	2.48E+30	&	--	&	--	&	--	&	--	&	--	&	\note out of UVIS FoV; undetected object just outside of error circle in ACS FoV	\\
	20	&	16:23:36.884	$-$26:31:39.61	&	0.39	&	0.39	&	2.00E+30	&	R0007345	&	0.24	&	AB	&	AB	&	97.80	&	\note MS in UV CMDs, red in \vi, \ha-excess \\
	21	&	16:23:34.658	$-$26:32:04.61	&	0.38	&	0.54	&	1.06E+30	&	R0005745	&	0.02	&	AB	&	--	&	98.10	&	\note slightly blue in \uvu, slightly red in \vi, \ha-excess	\\
	22	&	16:23:33.358	$-$26:31:45.64	&	0.38	&	0.47	&	1.89E+30	&	R0001090	&	0.14	&	AB?/BS	&	AB	&	96.70	&	\note BS in all CMDs	\\
	24	&	16:23:42.088	$-$26:31:37.06	&	0.55	&	1.54	&	4.65E+29	&	R0007270	&	0.54	&	Fg?	&	Amb.	&	--	&	\note not detected in \UV, \U, or \B, extremely red in \vi	\\
	25	&	16:23:33.492	$-$26:32:30.14	&	0.38	&	1.03	&	1.15E+30	&	R0004510	&	0.26	&	AB	&	AB	&	96.30	&	\note slightly blue in \uvu, on MS in \ub, red in \vi	\\
	26	&	16:23:38.882	$-$26:31:48.22	&	0.49	&	0.86	&	1.15E+30	&	R0006718	&	0.37	&	AB	&	AB	&	96.70	&	\note 
 MS in \uvu\ and \ub, red in \vi, \ha-excess \\
	27	&	16:23:33.272	$-$26:31:57.94	&	0.43	&	0.60	&	6.45E+29	&	R0000883	&	0.29	&	AB	&	AB	&	97.10	&	\note red in all CMDs, \ha-excess	\\
	28	&	16:23:34.972	$-$26:32:24.53	&	0.40	&	0.86	&	1.60E+30	&	R0004784	&	0.25	&	AB	&	AB	&	97.70	&	\note slightly blue in \ub, red in \vi	\\
	30	&	16:23:28.378	$-$26:30:22.33	&	0.53	&	1.92	&	2.11E+30	&	R0002281	&	0.10	&	AB	&	--	&	--	&	\note out of UVIS FoV; red in \vi	\\
	33	&	16:23:34.268	$-$26:29:56.22	&	0.35	&	1.62	&	2.24E+31	&	R0002524	&	0.11	&	AB	&	V56/Amb.	&	97.30	&	\note red in \vi; RG in other  CMDs	\\
	55	&	16:23:33.835	$-$26:32:00.57	&	0.38	&	0.55	&	3.13E+30	&	--	&	--	&	--	&	--	&	--	&	\note empty error circle	\\
	57	&	16:23:30.237	$-$26:30:45.25	&	0.48	&	1.36	&	2.79E+30	&	--	&	--	&	--	&	--	&	--	&	\note empty error circle	\\
	62	&	16:23:30.047	$-$26:32:05.53	&	0.50	&	1.28	&	2.11E+30	&	--	&	--	&	--	&	--	&	--	&	\note empty error circle	\\
	63	&	16:23:42.332	$-$26:30:38.47	&	0.49	&	1.83	&	2.00E+30	&	R0002125	&	0.12	&	Fg?	&	--	&	--	&	\note out of UVIS FoV; to red of RGB in \vi; \emph{Gaia}	PM indicates nonmember \\
	70	&	16:23:29.214	$-$26:29:49.45	&	0.62	&	2.18	&	1.40E+30	&	R0012571	&	0.39	&	AGN	&	--	&	--	&	\note out of UVIS FoV; 
 MS in \vi; actual counterpart is likely a radio-bright, optically undetected AGN = M4-VLA1	\\
	72	&	16:23:35.069	$-$26:32:04.34	&	0.36	&	0.53	&	1.32E+30	&	R0000770	&	0.16	&	AB	&	--	&	97.20	&	\note MS in UV CMDs, red in \vi; EB? in \citet{Nascimbeni14}	\\
%
%
\hline
\end{tabular}
\end{adjustbox}
\end{table}
\end{landscape}
\begin{landscape}
\begin{table}
\centering
\contcaption{Optical Counterpart Summary}
\begin{adjustbox}{min width=\textwidth}
\begin{tabular}{rccccccp{48pt}<{\centering}p{36pt}<{\centering}cp{144pt}<{\raggedright}}
\hline
Source$^a$ &
RA, Dec (J2000)$^b$ &
$\rerr~('')^c$ &
$r~(')^d$ &
$L_X$~(0.5--10\,keV)$^e$ &
HUGS \# &
Offset$^f$&
Type$^g$ &
Bassa Type$^h$ &
PM$^i$ &
Notes \\
\hline
	74	&	16:23:33.184	$-$26:31:09.55	&	0.44	&	0.60	&	1.20E+30	&	R0001710	&	0.36	&	AB	&	--	&	97.40	&	\note red in \uvu, MS in \ub, blue in \vi, \ha-excess; cEB in \citet{Nascimbeni14}	\\
	75	&	16:23:31.308	$-$26:31:48.79	&	0.41	&	0.91	&	1.17E+30	&	R0001028	&	0.22	&	AB	&	--	&	98.00	&	\note slightly blue in \uvu, blue in \ub, very red in \vi	\\
	76	&	16:23:41.661	$-$26:31:15.83	&	0.46	&	1.47	&	1.17E+30	&	R0008632	&	0.52	&	CV	&	--	&	98.00	&	\note blue in all CMDs	\\
	80	&	16:23:35.811	$-$26:31:32.69	&	0.44	&	0.13	&	1.07E+30	&	R0001322	&	0.15	&	BS	&	--	&	97.90	&	\note BS in all CMDs	\\
	81	&	16:23:31.758	$-$26:31:56.54	&	0.52	&	0.86	&	1.05E+30	&	R0006301	&	0.09	&	CV?	&	--	&	--	&	\note on WD seq in all CMDs	\\
	82	&	16:23:34.817	$-$26:32:00.12	&	0.39	&	0.46	&	1.05E+30	&	R0000841	&	0.03	&	RS	&	--	&	97.30	&	\note RS in \vi, at SG/RG juncture in other CMDs	\\
	83	&	16:23:36.993	$-$26:31:33.70	&	0.44	&	0.40	&	9.83E+29	&	R0007525	&	0.50	&	AB	&	--	&	96.60	&	\note blue in \uvu, red in \vi	\\
	84	&	16:23:36.638	$-$26:31:43.76	&	0.47	&	0.37	&	9.49E+29	&	R0006962	&	0.13	&	AB	&	--	&	97.40	&	\note blue in \uvu, red in \vi	\\
	85	&	16:23:36.444	$-$26:30:30.62	&	0.54	&	1.07	&	9.28E+29	&	R0002209	&	0.36	&	SG?	&	--	&	97.30	&	\note SG in all CMDs	\\
	86	&	16:23:41.478	$-$26:32:05.69	&	0.47	&	1.50	&	8.66E+29	&	R0005592	&	0.39	&	AB	&	--	&	97.80	&	\note slightly blue in \uvu\ and \ub, red in \vi	\\
	87	&	16:23:35.779	$-$26:31:37.51	&	0.42	&	0.15	&	8.61E+29	&	R0007423	&	0.35	&	AB?	&	--	&	97.90	&	\note slightly blue in \uvu\ and \ub, on MS in \vi	\\
	88	&	16:23:39.710	$-$26:31:20.49	&	0.55	&	1.02	&	8.30E+29	&	R0008392	&	0.49	&	AB	&	--	&	96.50	&	\note not detected in \UV, on MS in \ub, red in \vi	\\
	94	&	16:23:28.529	$-$26:31:34.41	&	0.47	&	1.49	&	6.58E+29	&	R0001298	&	0.36	&	AB	&	--	&	97.20	&	\note slightly red in UV CMDs, red in \vi\	\\
	95	&	16:23:38.506	$-$26:29:52.71	&	0.86	&	1.82	&	6.14E+29	&	--	&	--	&	--	&	--	&	--	&	\note poor \chandra\ detection; out of UVIS FoV; empty error circle in ACS FoV	\\
	97	&	16:23:39.157	$-$26:29:55.42	&	0.58	&	1.85	&	5.64E+29	&	R0012464	&	0.38	&	AB	&	--	&	--	&	\note out of UVIS FoV;  red in \vi	\\
	100	&	16:23:30.374	$-$26:30:49.75	&	0.82	&	1.30	&	4.75E+29	&	--	&	--	&	--	&	--	&	--	&	\note marginal \chandra\ detection; empty error circle	\\
	101	&	16:23:34.516	$-$26:31:10.72	&	0.55	&	0.40	&	4.51E+29	&	R0008999	&	0.92	&	CV?/AB?	&	--	&	96.30	&	\note out of UVIS FoV; MS in \vi; large \ha\ excess	\\
	102	&	16:23:39.836	$-$26:31:26.77	&	0.50	&	1.04	&	4.46E+29	&	R0007856	&	0.30	&	AB	&	--	&	98.10	&	\note slightly blue in \uvu\ and \ub,  red in \vi	\\
	104	&	16:23:34.747	$-$26:31:43.93	&	0.43	&	0.21	&	3.92E+29	&	R0007155	&	0.26	&	AB	&	--	&	98.20	&	\note slightly blue in \uvu,  red in \vi	\\
	105	&	16:23:26.986	$-$26:32:49.43	&	0.53	&	2.24	&	3.72E+29	&	R0000215	&	0.60	&	AB	&	--	&	--	&	\note out of UVIS FoV; slightly red in \vi	\\
	110	&	16:23:33.315	$-$26:31:13.82	&	0.51	&	0.54	&	2.93E+29	&	R0008707	&	0.48	&	AB	&	--	&	98.10	&	\note blue in \uvu, MS in \ub, red in \vi	\\
	120	&	16:23:31.609	$-$26:30:34.45	&	0.57	&	1.26	&	1.72E+29	&	--	&	--	&	--	&	--	&	--	&	\note empty error circle	\\
	124	&	16:23:28.588	$-$26:30:56.40	&	0.60	&	1.60	&	1.47E+29	&	R0001915	&	0.82	&	AB	&	--	&	97.70	&	\note red in \uvu, MS in \ub, red in \vi	\\
	125	&	16:23:35.985	$-$26:31:24.86	&	0.54	&	0.22	&	1.44E+29	&	R0001440	&	0.23	&	AB	&	--	&	97.90	&	\note red in all CMDs	\\
	130	&	16:23:35.919	$-$26:32:40.13	&	0.54	&	1.14	&	1.07E+29	&	--	&	--	&	--	&	--	&	--	&	\note poor \chandra\ detection; empty error circle	\\
	131	&	16:23:33.595	$-$26:31:51.73	&	0.54	&	0.48	&	9.55E+28	&	R0006557	&	0.51	&	MS?	&	--	&	98.10	&	\note poor \chandra\ detection; very slightly blue in \ub	\\
	132	&	16:23:38.748	$-$26:33:03.68	&	0.48	&	1.71	&	9.29E+28	&	--	&	--	&	--	&	--	&	--	&	\note out of UVIS FoV; empty error circle in ACS FoV	\\
	134	&	16:23:36.365	$-$26:32:44.85	&	0.47	&	1.23	&	8.50E+28	&	--	&	--	&	--	&	--	&	--	&	\note empty error circle	\\
%
%
\hline
\end{tabular}
\end{adjustbox}
\end{table}
\end{landscape}
\begin{landscape}
\begin{table}
\centering
\contcaption{Optical Counterpart Summary}
\begin{adjustbox}{min width=\textwidth}
\begin{tabular}{rccccccp{48pt}<{\centering}p{36pt}<{\centering}cp{144pt}<{\raggedright}}
\hline
Source$^a$ &
RA, Dec (J2000)$^b$ &
$\rerr~('')^c$ &
$r~(')^d$ &
$L_X$~(0.5--10\,keV)$^e$ &
HUGS \# &
Offset$^f$&
Type$^g$ &
Bassa Type$^h$ &
PM$^i$ &
Notes \\
\hline
	144	&	16:23:37.092	$-$26:32:09.13	&	0.72	&	0.73	&	5.68E+28	&	R0000720	&	0.84	&	??	&	--	&	97.90	&	\note poor \chandra\ detection -- likely false positive;  MSTO/SG juncture in all CMDs	\\
	151	&	16:23:44.332	$-$26:31:54.25	&	0.90	&	2.07	&	5.03E+28	&	--	&	--	&	--	&	--	&	--	&	\note poor \chandra\ detection; out of ACS FoV; on edge of UVIS FoV	\\
	156	&	16:23:31.124	$-$26:33:00.66	&	0.67	&	1.73	&	4.64E+28	&	--	&	--	&	--	&	--	&	--	&	\note poor \chandra\ detection; out of UVIS FoV; empty error circle in ACS FoV	\\
	159	&	16:23:40.324	$-$26:30:13.28	&	0.82	&	1.75	&	4.41E+28	&	R0011785	&	0.22	&	MS?	&	--	&	--	&	\note poor \chandra\ detection; out of UVIS FoV; MS in \vi	\\
	166	&	16:23:36.809	$-$26:32:02.75	&	0.50	&	0.61	&	3.46E+28	&	R0005945	&	0.68	&	MS	&	--	&	95.20	&	\note Not detected in \UV\ or \U; MS in \vi	\\
	167	&	16:23:40.450	$-$26:32:05.76	&	1.00	&	1.29	&	3.16E+28	&	--	&	--	&	--	&	--	&	--	&	\note poor \chandra\ detection; empty error circle	\\
	168	&	16:23:31.529	$-$26:30:57.74	&	0.35	&	1.02	&	2.5E+30	&	R0009481	&	0.10	&	AB	&	--	&	96.80	&	\note on MS in \uvu\ and \ub, red in \vi	\\
\hline
\multicolumn{11}{p{8.75in}}{\emph{Notes.} $^a$Extension of \citet{Bassa04b} numbering system (sources CX1--CX31) to include the full set of 161 sources reported by \citet{Bahramian20}. Sources are numbered in order of descending X-ray luminosity. Source CX2 has been replaced by sources CX36 and CX37 (which lie outside of the HUGS FoV). Sources CX5 and CX9 from \citet{Bassa04b} have been replaced by CX33. Source CX23 has been replaced by sources CX65 and CX73 (which lie outside of the HUGS FoV). Sources CX29 and CX31 have been dropped, since these were not detected by \citet{Bahramian20}, as discussed in the text.}\\
\multicolumn{11}{l}{$^b$Source position from \citet{Bahramian20}, advanced to 2015.0 using \emph{Gaia} mean cluster proper motion from \citet{Vasiliev19}.}\\
\multicolumn{11}{l}{$^c$95 per cent confidence X-ray error circle radius in arcsec, computed using the prescription of \citet{Hong05}.}\\
\multicolumn{11}{l}{$^d$Projected distance from cluster centre in arcmin.}\\
\multicolumn{11}{l}{$^e$X-ray luminosity in \ergs, based on a BXA power-law fit from \citet{Bahramian20}.}\\
\multicolumn{11}{l}{$^f$Offset of counterpart from X-ray source position in units of $r_\mathrm{err}$.}\\
\multicolumn{11}{p{8.75in}}{$^g$Counterpart type: CV = cataclysmic variable; AB = active binary; RG = red giant; BS = blue stragger; RS = red straggler; SSG = sub-subgiant; MS = main sequence; Fg = foreground; AGN = active galactic nucleus; ? indicates less certain classification. When more than one type is listed, the first type is considered to be the most likely classification.}\\
\multicolumn{11}{p{8.75in}}{$^h$\citet{Bassa04b} counterpart type. In addition to the types above, Amb. = ambiguous and V objects are variable stars from \citet{Kaluzny97} and \citet{Mochejska02}.} \\
\multicolumn{11}{l}{$^i$Probability of cluster membership of counterpart in per cent, from HUGS database.}\\
\end{tabular}
\end{adjustbox}
\end{table}
\end{landscape}

\begin{table*}
    \centering
    {
    \caption{Positions and photometry for HUGS counterparts to \chandra\ and VLA sources}
    \label{t:positions_photometry}
    \begin{tabular}{lccccccccccc}
    \hline
    CX &
    HUGS ID &
    $X^a$ &
    $Y^a$ &
    RA, Decl (J2000)$^b$ &
    \UV &
    \U &
    \B &
    \V &
    \I &
    \R &
    \ha \\
    \hline
%
1	&	R0007667	&	5366.41	&	4952.09	&	16:23:34.128~~$-$26:31:34.92	&	24.31	&	22.40	&	22.23	&	20.18	&	18.69	&	--	&	--	\\
3	&	R0001235	&	4025.24	&	4869.56	&	16:23:38.076~~$-$26:31:38.19	&	19.03	&	17.87	&	17.80	&	16.62	&	15.67	&	16.34	&	16.05	\\
4	&	R0010636	&	5301.51	&	6360.67	&	16:23:34.319~~$-$26:30:39.30	&	24.72	&	23.80	&	23.30	&	21.11	&	19.53	&	20.51	&	19.68	\\
8	&	R0001877	&	6273.97	&	5882.37	&	16:23:31.456~~$-$26:30:58.18	&	19.38	&	18.06	&	17.94	&	16.65	&	15.61	&	16.33	&	15.98	\\
10	&	R0001536	&	5059.89	&	5345.77	&	16:23:35.030~~$-$26:31:19.38	&	19.98	&	18.10	&	17.78	&	16.12	&	14.97	&	15.88	&	15.53	\\
11	&	R0002030	&	5956.44	&	6195.56	&	16:23:32.391~~$-$26:30:45.81	&	20.04	&	18.75	&	18.66	&	17.35	&	16.32	&	17.14	&	16.81	\\
12	&	R0006434	&	3982.89	&	4463.69	&	16:23:38.201~~$-$26:31:54.22	&	24.28	&	23.88	&	24.18	&	23.57	&	23.30	&	--	&	--	\\
13	&	R0000806	&	5303.18	&	4261.53	&	16:23:34.314~~$-$26:32:02.20	&	21.92	&	19.58	&	18.62	&	16.80	&	15.89	&	16.62	&	16.40	\\
15	&	R0001106	&	4465.71	&	4711.67	&	16:23:36.779~~$-$26:31:44.43	&	18.79	&	17.77	&	17.86	&	16.57	&	15.78	&	16.27	&	15.97	\\
20	&	R0007345	&	4428.86	&	4835.50	&	16:23:36.888~~$-$26:31:39.53	&	20.74	&	19.34	&	19.20	&	17.80	&	16.79	&	17.51	&	17.16	\\
21	&	R0005745	&	5186.22	&	4200.54	&	16:23:34.658~~$-$26:32:04.61	&	21.97	&	20.31	&	20.05	&	18.66	&	17.56	&	18.30	&	17.93	\\
22	&	R0001090	&	5626.65	&	4680.86	&	16:23:33.362~~$-$26:31:45.63	&	18.35	&	17.30	&	17.29	&	16.20	&	15.30	&	15.94	&	15.65	\\
24	&	R0007270	&	2657.62	&	4892.17	&	16:23:42.101~~$-$26:31:37.29	&	27.28	&	--	&	--	&	22.48	&	19.67	&	--	&	--	\\
25	&	R0004510	&	5579.87	&	3553.87	&	16:23:33.499~~$-$26:32:30.15	&	22.63	&	20.78	&	20.36	&	18.76	&	17.51	&	18.35	&	17.98	\\
26	&	R0006718	&	3753.62	&	4611.80	&	16:23:38.875~~$-$26:31:48.37	&	21.77	&	20.10	&	19.90	&	18.52	&	17.44	&	18.30	&	17.93	\\
27	&	R0000883	&	5659.32	&	4367.21	&	16:23:33.265~~$-$26:31:58.02	&	20.46	&	18.98	&	18.84	&	17.51	&	16.48	&	17.17	&	16.83	\\
28	&	R0004784	&	5077.11	&	3695.88	&	16:23:34.980~~$-$26:32:24.54	&	22.40	&	20.53	&	20.23	&	18.60	&	17.46	&	18.34	&	17.96	\\
30	&	R0002281	&	7320.34	&	6791.21	&	16:23:28.377~~$-$26:30:22.28	&	--	&	--	&	--	&	16.73	&	15.78	&	16.52	&	16.22	\\
33	&	R0002524	&	5319.11	&	7450.74	&	16:23:34.267~~$-$26:29:56.25	&	18.40	&	16.56	&	16.07	&	14.54	&	13.35	&	--	&	--	\\
63	&	R0002125	&	2579.41	&	6382.86	&	16:23:42.330~~$-$26:30:38.42	&	--	&	--	&	--	&	15.08	&	13.75	&	--	&	--	\\
70	&	R0012571	&	7041.22	&	7619.21	&	16:23:29.199~~$-$26:29:49.58	&	--	&	--	&	--	&	20.71	&	19.25	&	20.41	&	19.98	\\
72	&	R0000770	&	5047.56	&	4206.41	&	16:23:35.066~~$-$26:32:04.38	&	20.13	&	18.86	&	18.80	&	17.44	&	16.48	&	17.16	&	16.86	\\
74	&	R0001710	&	5687.16	&	5598.71	&	16:23:33.184~~$-$26:31:09.39	&	19.38	&	18.02	&	18.06	&	16.84	&	16.05	&	16.55	&	16.25	\\
75	&	R0001028	&	6325.24	&	4598.88	&	16:23:31.305~~$-$26:31:48.87	&	19.59	&	18.49	&	18.68	&	17.30	&	16.16	&	16.79	&	16.45	\\
76	&	R0008632	&	2806.13	&	5429.89	&	16:23:41.664~~$-$26:31:16.06	&	21.27	&	20.30	&	20.22	&	18.77	&	17.71	&	--	&	--	\\
80	&	R0001322	&	4796.12	&	5008.74	&	16:23:35.806~~$-$26:31:32.69	&	17.94	&	17.00	&	16.98	&	16.00	&	15.17	&	15.77	&	15.52	\\
81	&	R0006301	&	6172.05	&	4405.76	&	16:23:31.756~~$-$26:31:56.50	&	24.40	&	23.94	&	25.29	&	23.92	&	23.16	&	--	&	--	\\
82	&	R0000841	&	5132.58	&	4314.08	&	16:23:34.816~~$-$26:32:00.12	&	19.22	&	17.47	&	17.19	&	15.79	&	14.71	&	15.42	&	15.08	\\
83	&	R0007525	&	4388.66	&	4979.83	&	16:23:37.006~~$-$26:31:33.83	&	22.61	&	20.79	&	20.42	&	18.81	&	17.66	&	18.53	&	18.16	\\
84	&	R0006962	&	4513.97	&	4730.06	&	16:23:36.637~~$-$26:31:43.70	&	23.54	&	21.39	&	20.75	&	18.94	&	17.68	&	18.63	&	18.23	\\
85	&	R0002209	&	4580.37	&	6585.40	&	16:23:36.441~~$-$26:30:30.42	&	18.61	&	17.41	&	17.40	&	16.21	&	15.26	&	15.92	&	15.62	\\
86	&	R0005592	&	2873.37	&	4175.75	&	16:23:41.467~~$-$26:32:05.59	&	23.36	&	21.24	&	20.76	&	18.91	&	17.65	&	--	&	--	\\
87	&	R0007423	&	4808.10	&	4889.28	&	16:23:35.771~~$-$26:31:37.41	&	22.27	&	20.54	&	20.27	&	18.78	&	17.71	&	18.50	&	18.16	\\
88	&	R0008392	&	3475.57	&	5321.43	&	16:23:39.693~~$-$26:31:20.34	&	26.38	&	22.47	&	21.57	&	19.51	&	18.15	&	19.14	&	18.72	\\
94	&	R0001298	&	7272.47	&	4966.17	&	16:23:28.517~~$-$26:31:34.35	&	19.44	&	18.22	&	18.19	&	16.93	&	16.00	&	16.74	&	16.44	\\
97	&	R0012464	&	3659.59	&	7476.91	&	16:23:39.150~~$-$26:29:55.22	&	--	&	--	&	--	&	19.85	&	18.49	&	19.46	&	19.01	\\
101	&	R0008999	&	5235.59	&	5552.54	&	16:23:34.513~~$-$26:31:11.21	&	26.05	&	26.76	&	--	&	22.19	&	20.47	&	22.12	&	21.19	\\
102	&	R0007856	&	3424.26	&	5156.22	&	16:23:39.845~~$-$26:31:26.87	&	24.77	&	22.62	&	21.74	&	19.82	&	18.48	&	19.42	&	18.99	\\
104	&	R0007155	&	5153.58	&	4722.83	&	16:23:34.754~~$-$26:31:43.98	&	20.75	&	19.38	&	19.31	&	17.97	&	16.95	&	17.72	&	17.39	\\
105	&	R0000215	&	7787.65	&	3058.40	&	16:23:26.999~~$-$26:32:49.69	&	--	&	--	&	--	&	16.58	&	15.66	&	--	&	--	\\
110	&	R0008707	&	5642.62	&	5492.74	&	16:23:33.315~~$-$26:31:13.57	&	23.74	&	21.63	&	20.98	&	19.23	&	18.02	&	18.98	&	18.63	\\
124	&	R0001915	&	7247.12	&	5939.41	&	16:23:28.592~~$-$26:30:55.92	&	19.16	&	17.88	&	17.86	&	16.62	&	15.68	&	16.43	&	16.13	\\
125	&	R0001440	&	4736.47	&	5204.31	&	16:23:35.982~~$-$26:31:24.97	&	20.35	&	18.88	&	18.79	&	17.42	&	16.41	&	17.18	&	16.86	\\
131	&	R0006557	&	5551.92	&	4531.87	&	16:23:33.582~~$-$26:31:51.52	&	20.52	&	19.13	&	19.14	&	17.92	&	16.95	&	17.62	&	17.32	\\
144	&	R0000720	&	4369.60	&	4097.26	&	16:23:37.062~~$-$26:32:08.69	&	18.74	&	17.56	&	17.52	&	16.37	&	15.45	&	16.10	&	15.81	\\
159	&	R0011785	&	3257.97	&	7016.21	&	16:23:40.333~~$-$26:30:13.41	&	--	&	--	&	--	&	22.77	&	20.92	&	22.45	&	21.74	\\
166	&	R0005945	&	4457.42	&	4239.40	&	16:23:36.804~~$-$26:32:03.08	&	27.91	&	25.56	&	24.38	&	22.09	&	20.36	&	21.72	&	21.19	\\
168	&	R0009481	&	6249.93	&	5893.21	&	16:23:31.527~~$-$26:30:57.75	&	20.72	&	19.30	&	19.20	&	17.80	&	16.75	&	17.52	&	17.18	\\
VLA4	&	R0009986	&	7293.82	&	6044.15	&	16:23:28.455~~$-$26:30:51.78	&	22.64	&	20.73	&	20.30	&	18.77	&	17.71	&	18.52	&	18.22	\\
VLA5	&	R0004648	&	4339.81	&	3718.55	&	16:23:37.150~~$-$26:32:23.64	&	26.04	&	22.70	&	21.65	&	19.76	&	18.48	&	19.44	&	19.04	\\
VLA20	&	R0005914	&	3728.37	&	4316.65	&	16:23:38.950~~$-$26:32:00.02	&	23.49	&	23.28	&	23.66	&	22.89	&	21.98	&	22.43	&	22.42	\\
\hline
\multicolumn{12}{l}{\emph{Notes.} $^a$HUGS star position in frame coordinate system defined by \citet{Nardiello18}.}\\
\multicolumn{12}{l}{$^b$Celestial coordinates are for an observation epoch of 2015.0 \citep[see][]{Nardiello18}.}\\
\end{tabular}
}
\end{table*}

\section{\emph{Chandra} Source Identification}
\label{sec:source-identification}

As in \citet{Cohn10}, \citet{Lugger17}, and \citet{Cohn21}, we classified objects found in \chandra\ error circles based on their colour-magnitude diagram (CMD) location. In those studies, CVs were distinguished by their blue colours, relative to the main sequence (MS), and their significant \hr\ excesses. ABs were distinguished by their red colours and modest \ha\ excesses. Although deep \ha\ and comparison \R\ imaging is not available for the present study, we performed aperture photometry on the shallow images in these two bands that are available in the \hst\ archive (see Table~\ref{t:UV_optical_data}). Thus, our classifications here are based on CMDs that are constructed from the 5-band HUGS photometry, augmented by 
an (\hr, \vi) colour-colour diagram. In this study, CVs are defined as objects with significant \uvu\ and/or \ub\ excesses (along with evidence of cluster membership, such as a consistent proper motion, \hr\ excess, or location on the main sequence in redder filters), while ABs are 
identified as generally lying 
to the right of the main sequence or red giant branch (RGB) in \vi\ colour. \hr\ excesses, where measurable, are taken as additional evidence supporting the CV and AB classifications.  {Calculating the \ha\ photometric equivalent width (EW) following \citet{DeMarchi10}, we see that all \ha-excess counterparts except CX4 and CX101 (likely CVs) have \ha\ EW $<$5 \AA, as generally observed for chromospherically active binaries \citep{Beccari14,Pallanca17}.} 
Identifications of stars by \citet{Nascimbeni14} as eclipsing binaries constitute strong evidence for active binaries (see Section~\ref{M4_core_project}).

Finding charts in the \V\ band for the 61 \chandra\ sources that lie within the HUGS field of view are shown in Appendix~B, which is included in the supplementary online material. Given its status as the nearest globular cluster, M4 has a very dispersed appearance on the sky. As a consequence, in many cases there is a unique object in each \chandra\ error circle, which makes it a likely counterpart provided that it deviates from the fiducial sequences in the CMDs. Many of the \citet{Bahramian20} sources lie outside of the HUGS field of view (FoV) and cannot be classified based on their optical properties (except for a few that are bright enough for detection by \gaia, see Section~\ref{sec:gaia-chandra}). In a few cases, the source lies within the HUGS FoV, but the error circle is empty. 

Table~\ref{t:counterparts} summarises the result of the classification procedure. Only those sources that lie within the HUGS FoV are listed. Classifications are given for 44 confident sources, which includes 16 X-ray sources previously classified by \citet{Bassa04b}. We were able to recover 
most 
of their 31 X-ray sources. 
CX29 and CX31, 
the two lowest luminosity sources in their sample,  may represent false positive detections, as the total exposure of 119~ks analysed by \citet{Bahramian20} is nearly 5 times longer than the 25~ks subset of this exposure analysed by \citet{Bassa04b}. 
Inspection of all three \chandra\ observations does not support the reality of 
CX29 and CX31, so we 
removed them from our `CX' list. 
We also could not confirm the reality of CX5 and CX9 as two separate sources. \citet{Bassa04b} identify these as overlapping sources, separately identified using the {\tt wavdetect} algorithm, but inspection of all three \chandra\ images suggests the X-ray point-spread function is consistent with other sources at this off-axis angle (consistent with \citealt{Bahramian20}). We retain CX5/CX9 as a single source, here labeled CX33.
\textcolor{black}{However, we identify CX8 as  consisting of two separate sources, with a fainter X-ray source (here labeled CX168) located $\sim$1" NE of the brighter source; both have secure optical counterparts. We estimate the $L_X$ of CX168 to be 0.5 that of the brighter object, CX8, based on the numbers of photons in the core of each source, and scale the \citet{Bahramian20} $L_X$ values.}

Additional changes to the CX labelling of the list of 31 sources from \citet{Bassa04b} are given in the footnotes to Table~\ref{t:counterparts}. \citet{Bahramian20} provide a quality flag for each source, indicating whether the detection is `poor,' `marginal,' or `confident,' where poor detections have a false detection probability of $\ge1$\,per cent. We have indicated the marginal and poor detections in the notes column of Table~\ref{t:counterparts}. {Six of these have empty error circles, and 3 have stars with no unusual properties (likely chance coincidences); most likely, few or none of these are real X-ray sources.}

Broad-band CMDs that show the objects listed in Table~\ref{t:counterparts} are presented in Figs.~\ref{f:m4_cmd1_hugs}--\ref{f:m4_cmd3_hugs}. 
The (\hr, \vi) colour-colour diagram is presented in Fig.~\ref{f:color-color_diagram}. 
Table~\ref{t:positions_photometry} gives the positions and photometry for each of the counterparts listed in Table~\ref{t:counterparts} and plotted in Figs.~\ref{f:m4_cmd1_hugs}--\ref{f:color-color_diagram}.

\subsection{Compact Object Binary Candidates}

Several of the proposed counterparts show evidence of being a binary containing either a white dwarf (WD) or neutron star (NS).

\citet{Bassa04b} identified a V=17.37 star as the counterpart of CX1, which they suggested was a CV. However, they also noted the poor astrometric match (about 2 $\sigma$ away). \citet{Kaluzny12} used the HST analysis of \citet{Anderson08} to instead identify another, fainter star (V=20.65) as the likely counterpart to CX1, bolstered by the clear sinusoidal variation of the fainter star in both 
ground-based and HST observations, {as well as variation in the {\it Chandra} X-ray lightcurve}. The sinusoidal variation suggests 
a 0.2628 (preferred), or 0.5256 (for ellipsoidal variation),  day orbital period. \citet{Kaluzny12} argued for a neutron star (and likely radio pulsar) nature of the primary, based on 
the 
X-ray/optical flux ratio, and the lack of a strong UV excess {and long-term optical or X-ray variations, as would be} expected from an X-ray bright CV.
We find that CX1 shows strong evidence for variability in \U, based on the rms of its HUGS magnitude (see Fig.~\ref{f:m4_sigma_u_plot}), consistent with the optical variability found by \citet{Kaluzny12}. 
We do not detect CX1 in the radio (Section~\ref{sec:individual_matches_with_VLA_sources}), 
{but as the RMS of our VLA images only reaches the mean flux (extrapolated to 5 GHz, assuming a spectral slope of $-1.9$) of the best-fit lognormal luminosity distribution for MSPs of \citet{Bagchi11}, we cannot constrain the presence of a pulsar. The evidence from \citet{Kaluzny12} indicates that CX1 is probably an MSP.
} 

CX4 was classified as a likely CV by \citet{Bassa04b}, based on its high X-ray/optical flux ratio (as its CMD location on the main sequence in \vi\ does not point to a definitive classification). We find CX4 to be quite blue in both UV CMDs, reaching the WD sequence in \uvu, and to have a proper motion consistent with the cluster, confirming its classification as a CV. Furthermore, it has a strong \hr\ excess. 
Its main sequence position in \vi\  allows us to estimate the mass of the companion star (which must provide most of the optical light) as $\sim$0.5 \Msun\  (using an isochrone of \citealt{Dotter07}, assuming an age of 12.8 Gyr from \citealt{Hansen04}, and taking $(m-M)_V=12.8$ from  \citealt{Harris96}), which suggests an orbital period of roughly 4.5 hours \citep{Howell01}. 

CX12 is a well-studied MSP with a white dwarf companion (and also a planet-mass companion in a wide orbit; \citealt{Thorsett99,Sigurdsson03}). It lies on the WD sequence in all three CMDs.  Its radio properties are discussed in Section~\ref{sec:individual_matches_with_VLA_sources}, and its X-ray properties in Appendix~\ref{sec:M4A}.

CX76 is a newly identified X-ray source in \citet{Bahramian20}. We here identify it with a moderately bright optical counterpart that is marginally blue compared to the main sequence in \vi, and progressively bluer in bluer filters. Its CMD locations suggest that it is a CV with a low-mass donor star (we may estimate $\sim$0.65 \Msun, using the \citealt{Dotter07} isochrone as above), 
with an orbital period of order 5.5 hours.

CX81 lies on the WD sequence in all broad-band CMDs.  It is too faint to be detected in \hr. The HUGS data do not provide a proper motion estimate (as it is only barely detected), but comparison of the 2008 and 2014 images suggests it moves with the cluster stars, and thus is a cluster member.  Given its very blue \vi\ colour, we infer this is likely to be a CV below the period gap, though an MSP with a WD companion is possible (however, in Section~\ref{sec:X-ray_CMD} below, we show that its X-ray colours and luminosity do not match other cluster MSPs). 

 CX101 has a very faint candidate counterpart, which lies on the main sequence in \vi, but appears to show a very strong \ha\ excess, 
 $\sim$\,0.4 magnitudes.  Such an excess is larger than typical for chromospherically active stars (e.g. \citealt{Beccari14}), making this star a candidate CV. However, the star is near the detection limit, and we do not fully trust the photometry here; if the excess is not real, this star may not be the true counterpart.


\subsection{Chromospherically active binary candidates}\label{sec:chromospherically_AB_candidates}

ABs often differ from average cluster stars in several ways; lying on the binary sequence above the MS (this sequence includes objects that lie up to  0.75 mag above the MS in the (\V, \vi) CMDs, which represents the equal-mass limit); showing an \hr\ excess 
(lower quartile in Fig
.~\ref{f:color-color_diagram}); showing optical variability identified by \citet{Nascimbeni14}; showing variability in the \U\ filter (Fig.~\ref{f:m4_sigma_u_plot}, upper quartile). ABs may not show any of these, but the likelihood of a chance coincidence of an X-ray source with an unremarkable star is significant, while the likelihood of matching stars in one of these categories is small (Section~\ref{sec:chance_superpositions}), so we identify counterparts with 
any 
of the above characteristics as `AB' and potential counterparts without these as `AB?'.

The majority of AB counterparts lie on the binary sequence in the (\V, \vi) CMD. 
These objects include
CX3,
CX11,
CX20,
CX21, 
CX25,
CX26,
CX27,
CX28,
CX30,
CX72, 
CX83,
CX84,
CX86,
CX88,
CX94,
CX97,
CX102,
CX104,
CX105,
CX110,
CX124,
CX125, 
and CX168.
Most of these are verified proper motion members, though CX30, CX97, and CX105 
lack UVIS data 
(so have no proper motion measurement).   The colour-colour magnitude diagram comparing \hr\ and \vi\ identifies high-confidence ($>$90 per cent) \ha\ emission from  CX20, CX21, CX26, CX27, and CX97, and lower-confidence  ($>$75 per cent) \ha\ from  CX11,  CX28, CX72, CX82, CX83, CX84, CX102, CX104, CX125, and CX168. 

An additional four objects, CX8, CX10, CX24, 
and CX75, fall significantly further above the binary sequence region. CX8 and CX10 fall in the sub-subgiant region, while CX75 falls below this region. These three objects are secure proper-motion members, while the proper-motion membership status of CX24 is undetermined (likely a foreground star). 

{CX15, CX74, and CX75 are identified, via their light curves, as likely contact eclipsing binaries with orbital periods of 0.3~day by \citet{Nascimbeni14}. While their HUGS \vi\ colour offsets from the MS appear large -- CX15 and CX74 are to the blue of the MS and CX75 is to the red of the MS -- it appears likely that these offsets are a consequence of unrepresentative sampling of the light curves for these short-period binaries. \citet{Kaluzny13} and \citet{Nascimbeni14} have presented CMDs for M4, based on high-time-resolution sampling of the light curves that allows the determination of appropriate mean colours. In both the \bv\ CMD \citep{Kaluzny13} and the F467M\,$-$\,F775W CMD \citep{Nascimbeni14}, all three counterparts fall very close to the MS. The largest deviation is that of CX74, which is on the binary sequence just to the red of the MS in F467M\,$-$\,F775W. We thus classify all three objects as ABs, of the W~UMa type, based on their identification as contact eclipsing binaries with variability amplitudes of over 0.3 magnitudes, short 0.3~day orbital periods, light curve morphology, lack of flaring, and approximate agreement with the \citet{Rucinski95} relation between absolute magnitude, period, and colour for W~UMa variables. }

CX33 (formerly CX5/CX9), to the red of the giant branch in \vi\ but a clear proper motion member, has been associated with variable V56 by \citet{Kaluzny13}. They note that it shows clear evidence of luminosity and velocity variability, which supports its identification as a binary, and thus an (RS CVn-type) AB. 

CX13 is a proper motion nonmember, a contact eclipsing binary with a 1-mag variation amplitude \citep{Nascimbeni14}. Although it is on the MS in the (\V, \vi) CMD (Fig.~\ref{f:m4_cmd3_hugs}), it is well to the red of the MS in both \bv\ \citep{Kaluzny13} and F467M\,$-$\,F775W \citep{Nascimbeni14}. Since these two studies provide high-time-resolution sampling of the 0.3~day period light curve, they provide a more reliable measure of the mean colour of this star than does the HUGS database for which the \V\ and \I\ magnitudes are measured at different orbital phases. While this star is clearly quite red, the HUGS measurements of its \UV, \U, and \B\ magnitudes appear to be too faint, likely as a result of its large proper motion. Therefore we performed aperture photometry for this star, in the UVIS filters, choosing several surrounding stars to serve as photometric standards. The determined aperture UVIS magnitudes for CX13 are given in Table~\ref{t:positions_photometry} and plotted in Figs.~\ref{f:m4_cmd1_hugs} and \ref{f:m4_cmd2_hugs}.



In the bluer broad-band colours, \uvu\ and \ub, AB candidates may fall either to the right or left of the MS, presumably indicating the degree of UV emission, which increases with chromospheric activity. The \uvu\ colour is most sensitive to chromospheric UV emission. Examination of Fig.~\ref{f:m4_cmd2_hugs} indicates that in \ub, most of the AB candidates fall near the MS, with the fainter ones lying to the blue side of the MS. In Fig.~\ref{f:m4_cmd1_hugs}, a number of the AB candidates lie distinctly to the blue of the MS in \uvu, including 
CX21,
CX25,
CX83,
CX84,
CX87,
CX102,
CX104,
and CX110.
We take this as an indication of enhanced chromospheric activity.
Examples of AB candidates that fall on the \vi\ binary sequence or in the sub-subgiant region and do not show significant UV excesses are 
CX3, 
CX8,
CX10,
CX11, 
CX20,
CX26,
CX27,
CX28,
CX86,
CX94,
and CX168.


\subsection{Summary of source identifications}
\label{summary_IDs}
For clarity, we  summarise our 24 new confident optical identifications with \chandra\ X-ray sources from the current study. We identify 19 new confident ABs (CX11, CX21, CX30, CX72, CX74, CX75, CX83, CX84, CX86, CX88, CX94, CX97, CX102, CX104, CX105, CX110, CX124, CX125, and CX168). We identify one new confident CV (CX76), one new confident blue straggler (CX80), two new confident subgiants (CX82 and CX85), and one new confident foreground or background star (CX63). We also suggest one new candidate AB (CX87) and two new candidate CVs (CX81 and CX101). We also note two main sequence stars that are the only stars that fall in the X-ray error circles (CX70 and CX166); these are likely chance superpositions. 
CX70 has a radio counterpart (VLA1) and is therefore likely an AGN that is unrelated to the main sequence star in the X-ray error circle.

We also summarise the previous 16 confident optical identifications from \citet{Bassa04b,Bassa05}. These include 10 ABs (CX3, CX8, CX10, CX15, CX20, CX25, CX26, CX27, CX28, and CX33), one CV (CX4), one MSP (CX12), one possible MSP (CX1), two foreground or background stars (CX13 and CX24), and one AGN (CX2). While \citet{Bassa04b} classified CX22 as an AB, we now consider it to be a candidate AB, given its location in the blue straggler region of all CMDs. {CX8 is also a VLA source (VLA31), and is discussed further in \S~\ref{sec:individual_matches_with_VLA_sources}.}

\subsection{Evaluating the probability of chance superpositions}
\label{sec:chance_superpositions}

\begin{figure}
  \centering
  \includegraphics[width=\columnwidth]{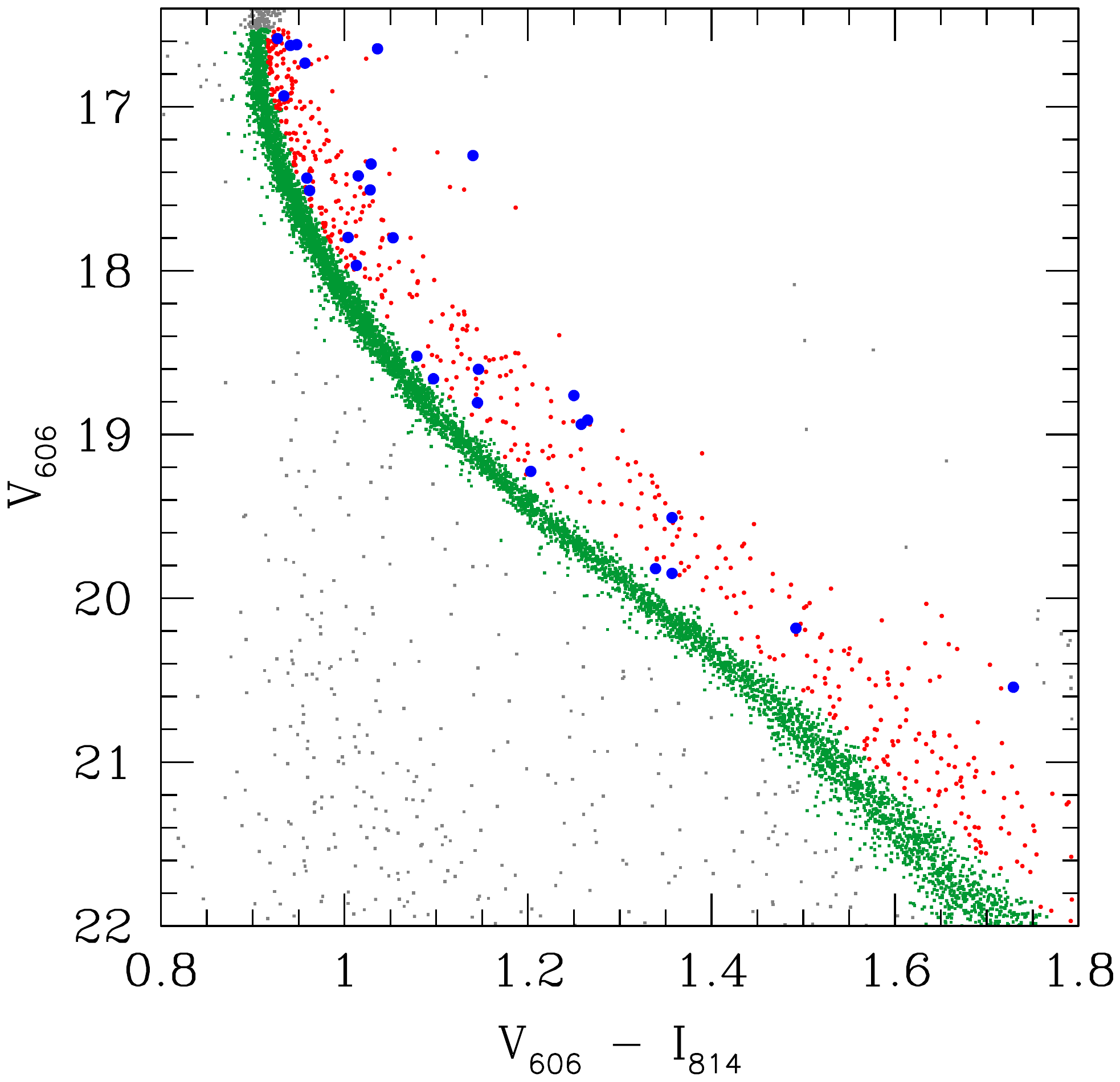}
  \caption{Stellar population selection using {\small GLUE} software. Colour key: green--MS; red--BSEQ; grey--all other stars. BSEQ stars within \chandra\ error circles are shown as blue dots.}
  \label{f:m4_glue_selection}
\end{figure}

Given the substantial number of apparent ABs in \chandra\ error circles, we have investigated the expected number of chance superpositions of such stars with error circles. We followed the approach of \citet{Cohn21}, who used the {\small GLUE} software package \citep{Beaumont15,Robitaille17} to define regions in the \vi\ CMD. This is illustrated in Fig.~\ref{f:m4_glue_selection}, where MS and binary sequence (BSEQ) groups are selected. The BSEQ region has been extended to include sub-subgiant stars. We then determined the number of stars in these groups inside X-ray error circles, and the radial density profiles of these groups. The latter were used to compute the local densities of group members at the locations of each source and thus to predict the number of chance superpositions in each error circle. The total numbers of observed and predicted objects in the error circles were calculated along with the significance of the excesses, based on \citet{Gehrels86}, using the formulae presented by \citet{Cohn21}. The results of this analysis are summarised in Table~\ref{t:chance_coincidences}. 
{There is an insignificant 1.7\,$\sigma$ excess of 7.0 MS stars in the error circles. Of these excess `MS' stars, three are actually other types. These stars, and their types from Table~\ref{t:counterparts}, are: CX4 (CV), CX13 (AB), and CX87 (AB?). Reducing the number of MS stars in error circles to 14 results in an excess of 4.0 stars over the expected background of 10.0 stars, which has a significance level of 0.9\,$\sigma$. In contrast, the excess of 27.2 BSEQ stars in the \chandra\ error circles registers at a 6.3\,$\sigma$ level, strongly indicating that the majority of the AB and AB? identifications are the likely counterparts to the \chandra\ sources.}


\begin{table}
\caption{Chance coincidence analysis.}\label{t:chance_coincidences}
\begin{center}
\begin{tabular}{lcccc} 
\hline
Population & $N_{\mathrm{obs}}\,^a$ & $N_{\mathrm{pred}}\,^b$ & Excess$^c$ & Significance $(\sigma)^d$ \\
\hline 
MS     &  17  & 10.0  & ~~7.0  &  1.7 \\ 
BSEQ   &  28  &  0.8  &  27.2  &  6.3 \\ 
\hline \\
\multicolumn{5}{p{0.9\columnwidth}}{\emph{Notes.}$^a$Observed number of population members in all error circles.} \\
\multicolumn{5}{p{0.9\columnwidth}}{$^b$Predicted number of population members in all error circles.} \\
\multicolumn{5}{p{0.9\columnwidth}}{$^c$Excess of observed versus predicted number in all error circles.} \\
\multicolumn{5}{p{0.9\columnwidth}}{$^d$Significance of excess expressed as a Gaussian-equivalent $\sigma$ level, based on \citet{Gehrels86} statistics.} \\
\end{tabular}
\end{center}
\end{table}

\subsection{X-ray/optical flux ratio}

Another piece of evidence that can help classify the nature of X-ray sources is 
comparison of X-ray luminosity vs.\ absolute magnitude. 
As discussed by \citet{Verbunt08}, quiescent LMXBs can reach the highest X-ray/optical ratios, followed by CVs, with ABs producing relatively little X-ray flux compared to their optical flux. The dividing line between CVs and ABs (or, rather, the ceiling for ABs, since CVs can be X-ray dim in quiescence) is not firmly settled; while the line $\log L_X(0.5\!-\!2.5\,\kev) = 34.0 - 0.4\,M_V$ seems to bound ABs empirically in globular clusters, the line $\log L_X (0.5\!-\!2.5\,\kev) = 32.3 - 0.27\,M_V$ seems to bound ABs in the solar neighbourhood \citep{Verbunt08}. 

\begin{figure*}
  \centering
  \includegraphics[width=0.8\textwidth]{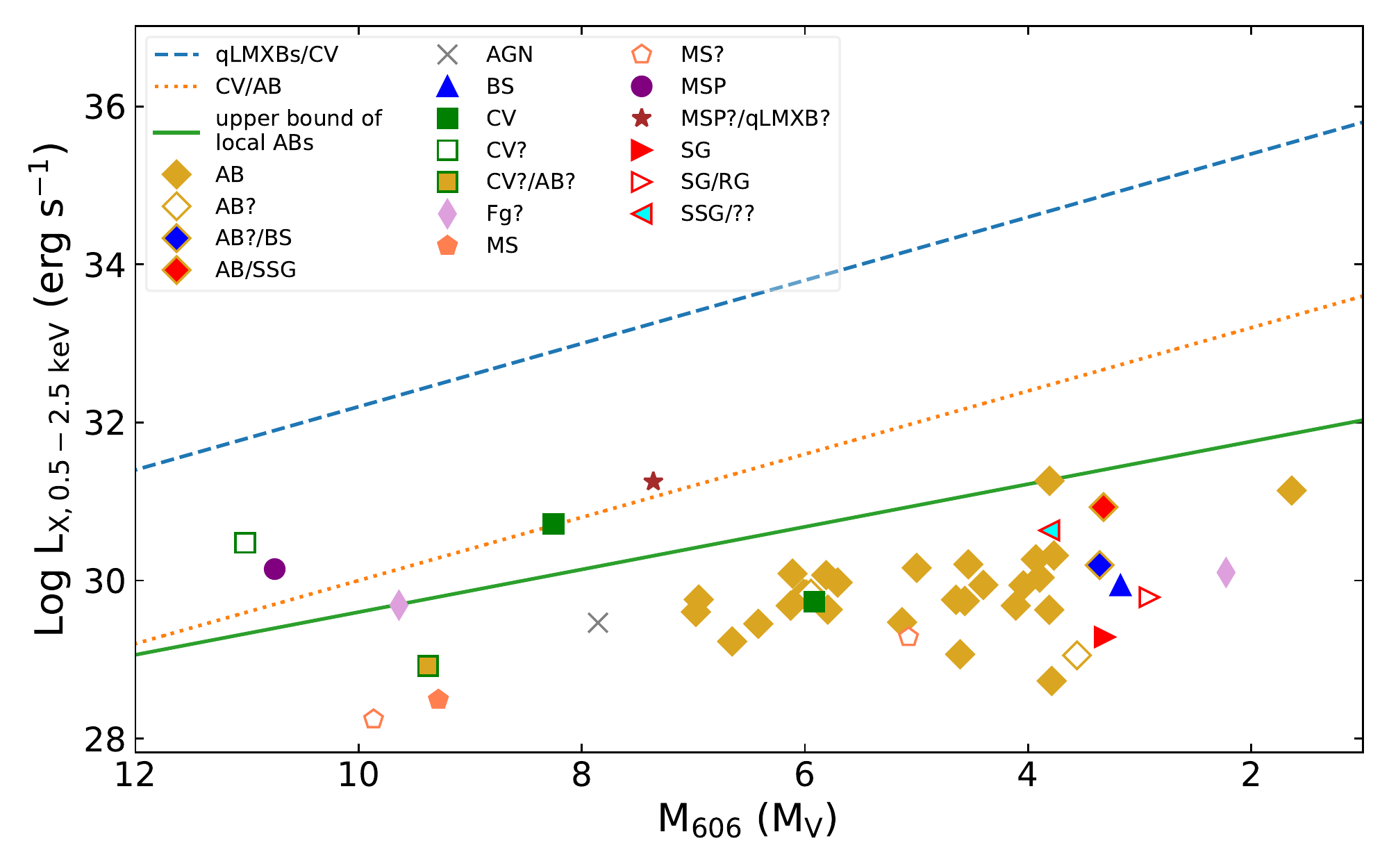}
  \caption{0.5--2.5 keV X-ray luminosity plotted against $M_V$, estimated from HUGS F606W magnitudes. The lines suggested by \citet{Verbunt08} are plotted; the highest separates X-ray bright quiescent LMXBs from CVs, while the others are different suggested separating lines between (X-ray brighter) CVs, and ABs. Most of our counterparts lie below all these lines, consistent with ABs, matching our classifications.}
  \label{f:x_opt_ratio}
\end{figure*}

We create a plot of 0.5--2.5 keV X-ray luminosity vs.\ $M_V$, for our M4 sources. We use the F606W HUGS magnitudes and {an extinction-corrected} distance modulus of 12.82 \citep[][2010 edition]{Harris96} to derive $M_V$. The small shift from using F606W is not important for our purpose. 
{
Similarly, small variations due to differential reddening are not important to the total extinction correction.} 
We extrapolate the 0.5--2.0 keV $L_X$ values of \citet{Bahramian20} to the 0.5--2.5 keV band {(note that reddening, in the form of $N_{\rm H}$, is included in their X-ray spectral fitting).} 
We find (Fig.~\ref{f:x_opt_ratio}) that all but four X-ray sources with HUGS F606W magnitudes lie below both empirical AB upper limit lines.  The exceptions were identified above as a CV (CX4), a CV candidate (CX81), an MSP (CX12 = pulsar M4 A), and a candidate MSP or qLMXB (CX1). All other objects are consistent with an AB nature. However, we do not take this as evidence that, e.g., the CV candidate CX76 is actually an AB, as CVs can have moderately bright optical counterparts without being X-ray bright. The X-ray/optical luminosity ratio plot generally supports our classifications of \chandra\ sources presented in Table~\ref{t:counterparts}, as these are primarily found to be ABs. 

\begin{landscape}
    
\begin{table}
    \centering
    \caption{{\it Gaia} counterparts to 17 X-ray sources}
    \begin{tabular}{lccccccccccl}
    \hline
ID$^a$	&	RA, Dec (J2000)$^b$	&	Dist$^c$	&	${L_X}^d$	& Type$^e$ &	\gaia\ ID$^f$	& Parallax &	PM$^g$	&	BPmag$^h$	&	RPmag$^h$	&	Offset$^i$ & Note$^j$	\\
 & (hhmmss.sss, ddmmss.ss) & (arcmin) & (0.5--10\,keV) & & & (mas) & (mas yr$^{-1}$) & (mag) & (mag) & (arcsec) \\
\hline
7	&	16:23:45.854	$-$26:28:55.04	&	3.55	&	4.71E+30	&	--	&	6045466468307366656		&	3.42	$\pm$	0.01	&	60.74	$\pm$	0.02	&	12.892	$\pm$	0.003	&	11.672	$\pm$	0.004	&	0.47 & Fg	\\
18	&	16:23:45.758	$-$26:31:16.89	&	2.38	&	1.13E+30	&	--	&	6045465506234526592		&	0.62	$\pm$	0.06	&	23.28	$\pm$	0.10	&	16.987	$\pm$	0.023	&	15.801	$\pm$	0.021	&	0.11 & Member	\\
30	&	16:23:28.378	$-$26:30:22.33	&	1.92	&	2.11E+30	&	AB	&	6045466159069806464		&		--		&		--		&	16.534	$\pm$	0.009	&	15.197	$\pm$	0.007	&	0.07 & --	\\
32	&	16:23:35.202	$-$26:35:26.08	&	3.89	&	4.86E+31	&	--	&	6045463169761296000		&	0.09	$\pm$	0.65	&	0.47	$\pm$	1.11	&	20.335	$\pm$	0.062	&	19.269	$\pm$	0.044	&	0.22 & Bkg?	\\
33	&	16:23:34.268	$-$26:29:56.22	&	1.62	&	2.24E+31	&	AB	&	6045466502667300096		&	0.48	$\pm$	0.02	&	22.01	$\pm$	0.05	&	14.967	$\pm$	0.005	&	13.374	$\pm$	0.005	&	0.04 & Member	\\
38	&	16:23:37.768	$-$26:35:19.13	&	3.81	&	9.11E+30	&	--	&	6045464647232984960		&		--		&		--		&		--		&		--		&	0.22 & --	\\
42	&	16:23:35.533	$-$26:27:08.41	&	4.42	&	7.95E+30	&	--	&	6045478459856719104		&	0.41	$\pm$	0.05	&	5.11	$\pm$	0.08	&	17.089	$\pm$	0.004	&	15.421	$\pm$	0.005	&	0.46 & Field	\\
63	&	16:23:42.332	$-$26:30:38.47	&	1.83	&	2.00E+30	&	Fg?	&	6045466262148847232		&	2.83	$\pm$	0.03	&	7.78	$\pm$	0.05	&	15.631	$\pm$	0.004	&	13.810	$\pm$	0.004	&	0.16 & Fg	\\
65	&	16:23:40.165	$-$26:29:25.61	&	2.40	&	1.70E+30	&	--	&	6045466330858826752		&	0.37	$\pm$	0.15	&	23.01	$\pm$	0.23	&	18.014	$\pm$	0.149	&	16.789	$\pm$	0.096	&	0.46 & Member	\\
69	&	16:23:17.264	$-$26:33:30.33	&	4.47	&	1.41E+30	&	--	&	6045464436777542784		&	1.23	$\pm$	0.19	&	6.51	$\pm$	0.30	&	19.784	$\pm$	0.136	&	17.187	$\pm$	0.010	&	0.15 & Fg	\\
70	&	16:23:29.214	$-$26:29:49.45	&	2.18	&	1.40E+30	&	MS?	&	6045466193420208256		&	1.80	$\pm$	1.47	&	23.66	$\pm$	2.40	&		--		&		--		&	0.26 & Member	\\
90	&	16:23:24.277	$-$26:30:13.79	&	2.77	&	7.26E+29	&	--	&	6045477910092876800		&	0.75	$\pm$	0.19	&	22.69	$\pm$	0.30	&	17.972	$\pm$	0.433	&	16.117	$\pm$	0.074	&	0.32 & Member	\\
91	&	16:23:33.777	$-$26:34:05.89	&	2.57	&	7.23E+29	&	--	&	6045464922118910720		&	0.57	$\pm$	0.07	&	23.18	$\pm$	0.12	&	17.465	$\pm$	0.011	&	16.075	$\pm$	0.006	&	0.05 & Member	\\
97	&	16:23:39.157	$-$26:29:55.42	&	1.85	&	5.64E+29	&	AB	&	6045466330859725184		&	0.46	$\pm$	0.48	&	23.27	$\pm$	0.70	&		--		&		--		&	0.22 & Member	\\
98	&	16:23:22.810	$-$26:31:55.58	&	2.80	&	4.91E+29	&	--	&	6045466021630961408		&	0.60	$\pm$	0.07	&	22.38	$\pm$	0.12	&	17.303	$\pm$	0.009	&	15.959	$\pm$	0.008	&	0.09 & Member	\\
105	&	16:23:26.986	$-$26:32:49.43	&	2.24	&	3.72E+29	&	AB	&	6045465746742763264		&	0.90	$\pm$	0.08	&	22.78	$\pm$	0.12	&	16.856	$\pm$	0.050	&	15.187	$\pm$	0.134	&	0.32 & Member?	\\
124	&	16:23:28.588	$-$26:30:56.40	&	1.60	&	1.47E+29	&	AB	&	6045466090350294784		&	0.40	$\pm$	0.07	&	22.47	$\pm$	0.10	&	16.850	$\pm$	0.018	&	15.597	$\pm$	0.012	&	0.47 & Member	\\
\hline
\multicolumn{9}{l}{{\it Notes}: $^a$Source numbering in this work. See Table~\ref{t:counterparts}.} \\
\multicolumn{9}{l}{$^b$Source position from \citet{Bahramian20}, advanced to 2015.0 using \emph{Gaia} mean cluster proper motion from \citet{Vasiliev19}.} \\
\multicolumn{9}{l}{$^c$Source distance from the cluster centre in arcmin, from \citet{Bahramian20}.}\\
\multicolumn{9}{l}{$^d$X-ray luminosity in \ergs, based on a BXA power-law fit from \citet{Bahramian20}.}\\
\multicolumn{9}{l}{{$^e$Optical counterpart type identified by HUGS photometry. See Table~\ref{t:counterparts}.}} \\
\multicolumn{9}{l}{$^f$\gaia\ unique source identifier in DR3.}\\
\multicolumn{9}{l}{$^g$Total proper motion, from \gaia\ DR3.}\\
\multicolumn{9}{l}{$^h$Mean magnitudes from \gaia\ DR3.}\\
\multicolumn{9}{l}{$^i$Offset of {\it Gaia-Chandra} match in arcsec.}\\
\multicolumn{12}{p{22cm}}{{$^j$Association with M4, based on parallax and proper motion. Member: considered true member of M4; Fg: foreground source; Bkg: background source; Field: Field star. ? indicates less certain classification.}} \\
    \end{tabular}
    \label{tab:gaia_matches}
\end{table}
\end{landscape}

\begin{figure}
    \centering
    \includegraphics[width=\columnwidth]{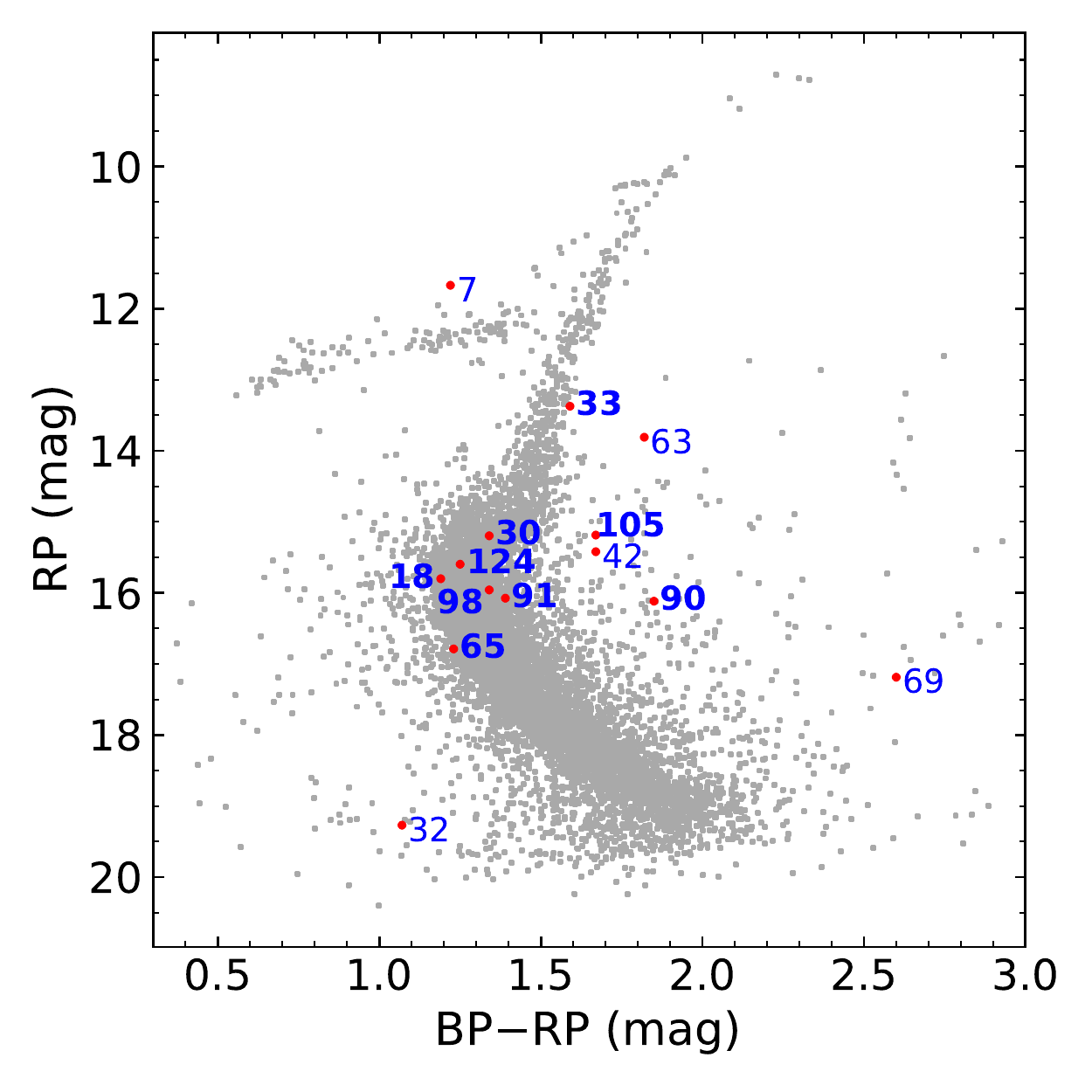}
    \caption{Colour-magnitude diagram of M4, based on {\it Gaia} DR3 photometry. The counterparts to 10 X-ray sources in the outer region of M4 are highlighted as red dots with CX IDs labelled, {while labels in boldface indicate true M4 members.}}
    \label{fig:gaia_cmd}
\end{figure}

\subsection{\emph{Gaia-Chandra} matches}
\label{sec:gaia-chandra}

Since the HUGS FoV only covers the central $\sim$1.7 arcmin radius region of M4, whereas the half-light radius of M4 is $\sim$4.3 arcmin \citep[][2010 edition]{Harris96}, many X-ray sources in \citet{Bahramian20} 
are not covered by HUGS. Therefore, we used {\it Gaia} Data Release 3 \citep[DR3;][]{GaiaCollaboration16,GaiaCollaboration22} to seek potential optical counterparts 
in the outer regions of the cluster. 

We first selected X-ray sources with distances from the cluster centre $>$1.5 arcmin, 
and then cross-matched those X-ray sources with DR3 sources {(positions at 2016.0 epoch)} in {\sc TOPCAT}~\citep{Taylor05}. Given that the typical centroiding uncertainties of those X-ray sources are less than 0.5 arcsec \citep[see][]{Bahramian20}, we thus limited the matching offsets to less than 0.5 arcsec, and found 
{17} {\it Gaia-Chandra} matches (Table~\ref{tab:gaia_matches}, Figure~\ref{fig:gaia_cmd}). 
{ The small proper motion components for M4 of $\overline{\mu_\alpha} = -12.490\,\mathrm{mas\,yr}^{-1}$ and $\overline{\mu_\delta} = -19.001\,\mathrm{mas\,yr}^{-1}$ \citep{Vasiliev19} produce only small shifts ($\Delta \alpha \cos\delta = -0.012$ arcsec and $\Delta \delta = -0.019$ arcsec) over the one-year interval between the epochs of X-ray positions (advanced to 2015.0 epoch; see Table~\ref{t:counterparts}) and \gaia\ DR3 positions.  
Boresight corrections between these epochs are not needed for finding \gaia\ counterparts.}
To check for chance coincidences, we shifted the X-ray sources by 5 arcseconds in different directions and redid the cross-matching process correspondingly. 
We find 3 matches on average per shift, indicating of order 
three of our matches 
are likely spurious, while 
roughly 14 
of them are real. 
Ten of these matches have proper motions (see Table~\ref{tab:gaia_matches}) within 2 sigma of the proper motion of M4 (22.7 mas yr$^{-1}$, with a central proper motion dispersion of $\sim$0.5 mas yr$^{-1}$), and we therefore suggest these stars {(CX 18, 33, 65, 70, 90, 91, 97, 98, 105, and 124)} are members of M4.
In addition, we checked parallaxes in \citet{GaiaCollaboration22} and distances in \citet{Bailer-Jones21} for those counterparts, and found that all 
of these likely matches have parallaxes that are consistent within 2 sigma of the distance to M4 (1.85$\pm$0.02 kpc, \citealt{Baumgardt21}; but cf. \citealt[][2010 revision]{Harris96} which gives 2.2 kpc, which would not change this result),
{except for the counterpart to CX105, which has a parallax of 0.90$\pm$0.08 mas. Considering both the total proper motion, and proper motions in RA and Dec, of the CX105 counterpart are consistent with M4, we suggest it is probably an  M4 member. The relatively high parallax might be a consequence of intrinsic motion of the binary stars, as suggested by its moderately high Gaia EDR3 RUWE value of 1.225 \citep{Stassun21}.} 
On the other hand, the proper motions and (large) parallaxes of {CX7}, CX63, and CX69 suggest they are foreground stars. CX32 is likely a background AGN (its proper motion is consistent with zero). CX42 appears to be a field star; its parallax distance is 2.2$^{+0.3}_{-0.2}$ kpc, but its proper motion is completely inconsistent with M4. 
We note that CXs 30, 33, 63, 70, 97, 105, and 124 also have HST magnitudes 
(see column 
5 in Table~\ref{tab:gaia_matches}). 

CX33 lies on the red giant branch of the \gaia\ CMD,  consistent with its \hst\ identification as an RS CVn type AB (Sec.~\ref{sec:chromospherically_AB_candidates}). 
{CX90 and CX105 have interestingly red colours for M4 members.}

\subsection{\emph{VLA-Chandra} matches}
\label{sec:vla-chandra-matches}
\cory{We match the \citet[][S20]{Shishkovsky20} 
VLA catalogue to the \citet[][B20]{Bahramian20} 
{\it Chandra} catalogue 
to find positional matches (Table~\ref{tab:chandra_vla_match}). We only consider sources that have confident or marginal detections in \citetalias{Bahramian20} ({\tt detection\_quality\_flag=0} or {\tt 1}),  excluding  poor detections. To determine search radii, we first calculate source-specific composite positional errors ($\sigma_{r,x}$; $r,x$ stands for `radio-X-ray') by combining \chandra\, and VLA positional uncertainties in quadrature; this includes \chandra\, positional uncertainty characterised by the 95 per cent error radius \citep{Hong05}, \chandra\, boresight offsets, and VLA positional error. The VLA positional errors in RA and Dec are set to the larger of the positional uncertainties reported in \citetalias{Shishkovsky20}, and $0.1$ of the projected beam sizes; the VLA error radius (denoted by $\sigma_\mathrm{VLA}$) {is set to the maximum of VLA positional errors in RA and Dec,} of which the median is $\approx 0.10\arcsec$ for sources in M4. With search radii of 2.0$\sigma_{r,x}$, we found 6 \chandra\, counterparts to VLA sources, one of which 
is a 
 known MSP (CX12 = VLA9). Considering the low spatial densities of 
 \chandra\, and VLA sources, it is very unlikely that these matches are 
 coincidental 
 {(all are within 1.0 $\sigma_{r,x}$)}. 
}

{What are radio sources likely to be? We summarize possibilities.  For the radio spectral index $\alpha$, 
`steeper' sources have more negative $\alpha$ values, $\alpha=0$ corresponds to a `flat' source, and $\alpha > 0$ is termed `inverted.' MSPs generally have very `steep' radio spectra ($\alpha=-1.6\pm0.2$, \citealt{Kramer98}), while background AGN and starbursts  typically have moderately steep spectra ($\alpha=-0.7$, \citealt{Gordon21}), and X-ray binaries typically have `flat' spectra ($\alpha\sim0$, \citealt{Espinasse18}). 
Only two MSPs with X-ray studies have $L_X<3\times10^{29}$\,\ergs\ \citep{PichardoMarcano23}, our X-ray limit, indicating that MSPs should be X-ray detected in M4.  Redback MSPs (with main-sequence companions) are X-ray brighter, $L_X>7\times10^{30}$\,\ergs\ \citep[e.g.][]{Zhao22}.   
Coronal activity in close binaries can produce radio emission, following $L_X/L_R = 10^{15\pm1}$ \citep{GudelBenz93}, with typically flat spectra \citep{Gudel02}. This is potentially detectable for X-ray active binaries in M4, given the deep VLA radio limits of 10 $\mu$Jy, which would imply $L_X=4\times10^{31\pm1}$ following the G\"udel-Benz relation. \citet{Gudel02} refer to BY Dra stars up to $10^{15}$\,\ergsh\ (0.2 $\mu$Jy at M4), and RS CVn stars (coronally active binaries including giants) up to $10^{18}$\,\ergsh\ (200 $\mu$Jy at M4), so evolved stars with coronal $L_X\sim10^{31}$\,\ergs\ may plausibly be radio detected in M4.
}

\subsection{\emph{VLA-HUGS} matches}
\cory{We also search VLA error circles for positional matches with HUGS sources. Since HUGS astrometry is matched to {\it Gaia} (within a few $\mathrm{mas}$), we only use the VLA positional uncertainty. We use search radii of 3 times the VLA positional uncertainties (Section~\ref{sec:vla-chandra-matches}) and found a total of 7 matches including the known MSP (VLA9 or CX12); three of the matches have separations more than $2\sigma_\mathrm{VLA}$. These matches are summarised in Table~\ref{tab:vla_hugs_match}.

A more complete VLA-HUGS cross-match using the full VLA catalogue estimates $\approx 0.6$ chance coincidences with MS sources per VLA source per arcsec$^2$ (Zhao, Y. et al., in prep); if a HUGS source marginally matches a VLA source, i.e., it has an offset of $3\sigma_\mathrm{VLA}\approx 0.3\arcsec$, the false alarm rate is around 0.2.

}


\subsection{Individual matches with VLA sources}
\label{sec:individual_matches_with_VLA_sources}
\cory{
We assume a core radius of 1.16 arcmin  \citep[][2010 online version]{Harris96}, 
and a distance of $1.85~\mathrm{kpc}$  \citep{Baumgardt21}. 

{\bf VLA1/CX70} is the brightest among the 37 VLA sources in M4; its radio spectral index 
has 
an intermediate steepness ($\alpha = -0.71\pm 0.01$). Its position coincides with 
CX70, the X-ray spectrum of which can be modelled by a power-law with a (relatively hard) photon index of $1.2^{+1.6}_{-1.1}$ (\citetalias{Bahramian20}). We found no HUGS counterparts within the search region, but there is a faint red object barely visible in $V$ and $I$ (the area is not covered by the WFC3 UV filters). 
VLA1's radio and X-ray properties allow for the possibility of it being an MSP, though the optical non-detection would imply a very low-mass companion, likely a black widow system. 
However, these features could also be explained by a background AGN. Indeed, VLA1's radial offset ($2.2\arcmin$) from the cluster centre is about 
two core radii; 
using the normalised radio source counts in \citetalias{Shishkovsky20}, we estimate $\approx 10$ AGN with $\slow > 10\,\mujy$ within a radius of $2.2\arcmin$, so an AGN 
is likely. 

{\bf VLA4} has a HUGS counterpart (R0009986) with a slight \uvu\ excess in the (\UV, \uvu) CMD, but is consistent with the MS in the other broad-band CMDs. In 
the colour-colour diagram (Fig.~\ref{f:color-color_diagram}) the counterpart shows a significant \ha\ deficit. 
R0009986 is a 
cluster member 
(PM member probability = 96.9 per cent). The optical position is only $0.04\arcsec$ from the VLA position, translating to $0.003$ chance coincidences within a search circle at a radius of this offset. 
The radio spectrum is marginally steep enough to be consistent with a typical MSP ($\alpha=-0.4\pm 0.1$), {though MSPs, and especially redback MSPs (given the main sequence star identification), should be detected in X-rays (see above).   
 Other interpretations also seem unlikely--an AGN is inconsistent with the high-confidence optical counterpart, an X-ray binary should show \ha\ emission and $L_X \gtrsim 10^{30}$ \ergs, while the (relatively) bright radio emission, with strong limits on X-rays, makes a coronally active binary very unlikely. This leaves the nature of this system unclear.}

{\bf VLA5} {is positionally consistent (with a chance coincidence number of 0.2) with HUGS R0004648, which is a cluster member on the main sequence.}
The projected radial offset from the cluster centre is less than 1\,arcmin, within which we expect $\approx 1$ AGN.
The steep radio spectrum ($\alpha = -1.4 \pm 0.1$) 
may suggest an 
MSP identification, {though some AGN have such steep radio spectra}. 
The MS photometry, $\V=19.8$ and $\vi=1.3$, correspond to a K3-K4 dwarf of $0.3~\Msun$, suggesting a redback MSP; however, the X-ray non-detection pushes the X-ray luminosity to $<3\times 10^{29}~\ergs$ -- again, unusually low for a redback MSP. 
{An AGN appears the most likely scenario, in which case the HUGS counterpart would be spurious.}

{\bf VLA9/CX12} is 
{the known MSP \citep{Lyne88,McKenna88} PSR B1620$-$26 (or PSR M4 A, hereafter M4A). 
It is detected in the X-rays (CX12) and radio (M4-VLA9), with a quite steep radio spectrum ($\alpha<-2.60$). Its X-ray emission is consistent with thermal, blackbody-like radiation (see \S~\ref{sec:M4A}), as seen in most MSPs in other clusters \citep{Bogdanov06}. }

{\bf VLA13/CX77} is another steep-spectrum source ($\alpha = -1.2\pm 0.6$) located at a relatively large projected distance (2.4 arcmin) from the cluster centre, which makes it more likely to be a background 
AGN. 
The absence of HUGS counterparts, if an MSP,  would suggest a highly-ablated low-mass companion (a black widow),
though an AGN is more likely.

{\bf VLA19/CX14} has 
an upper limit on $\alpha$ ($<1.3$). The \chandra\ ID CX14 is within $1\,\sigma_\mathrm{r,x}$, so the match is relatively confident. \citetalias{Bahramian20} found that a power-law model fits the X-ray spectrum best, giving a photon index of $2.5^{+1.0}_{-1.2}$ and  $L_X$ (0.5-10~$\kev$) 
$=2.7_{-1.5}^{+8.5}\times 10^{30}~\ergs$. 
VLA19's large projected distance from the cluster centre (3.1 arcmin) 
{indicates it is} 
a background AGN.

{\bf VLA20} 
has a poorly determined radio spectral index 
($\alpha < -0.4$). A HUGS ID (R0005914) consistent with the WD sequence is found at $2.1 \sigma_\mathrm{VLA}$ from the radio position. Despite this relatively large offset, this match is very unlikely to be a chance coincidence,  
due to 
the paltry number of WDs compared to MS stars. 
While R0005914 does not have a HUGS membership probability, its proximity to the cluster centre $\approx 1$ arcmin 
argues against a background nature. 
{The lack of X-rays argues against an MSP nature, but as the likely companion star is a white dwarf (rather than a main-sequence star, which would require a redback nature), this is more plausible than for VLA4 or VLA5 above.}


{\bf VLA30/CX57} is a 
$5\,\sigma$ 
detection 
in both the low and high-frequency sub-bands, 
producing a spectral index of 
$\alpha=-0.2\pm 0.8$. Its \chandra\, ID, CX57, is $0.6\sigma_{r,x}$ away from the radio position, so the match is relatively confident. CX57's X-ray spectrum can be 
fit by a 
power-law with a best-fit photon index of $1.0_{-1.0}^{+1.7}$. 
Similar to VLA1 and VLA13, a black widow scenario is 
possible, 
where the optical companion is 
very low-mass and faint.
A distant AGN also fits the data, 
{and seems most likely.} 

{\bf VLA31/CX8} is a flat-to-inverted ($\alpha=0.4_{-0.8}^{+0.6}$) radio source that has both a HUGS (R0001877) and a \chandra\ ID (CX8). R0001877 is at a relatively large offset ($2.9\sigma_\mathrm{VLA}$) from the radio position; however, it is a confident cluster member (PM = 97.5 per cent) that belongs to a rare population with unusual photometry -- in all HUGS CMDs, it is located below the sub-giant branch and redder than the main sequence. In the (\hr, \vi) colour-colour diagram, R0001877 shows strong signs of \ha\ emission. Objects in this region of CMDs are 
referred to as `sub-subgiants' (SSGs; \citealt{Leiner22}). 
SSGs in GCs are typically binary stars and X-ray sources, 
suggesting an evolutionary path distinct from that for single stars \citep{Geller17}. So far, there are three cases of SSG-radio associations in GCs: two MSPs, PSR J1740-5340A and B, in NGC 6397 \citep{Zhao20a, Zhang22} match with steep radio sources, 
and an SSG matches a flat-to-inverted source in M10 (M10-VLA1) \citep{Shishkovsky18} that is likely a BH binary or an unusual RS CVn-type AB. The \chandra\ ID can be best modelled by a power-law, with a photon index of $2.5\pm 0.4$, consistent with typical accreting stellar-mass BHs in quiescence \citep[e.g.,][]{Reynolds14}. 
VLA31's X-ray and radio luminosities make VLA31 consistent with the {black hole} $L_X-L_R$ correlation (Figure \ref{f:lrlx-bh-plot}). {The flat spectrum argues against an MSP nature, but does not completely rule it out.}
{The radio and X-ray fluxes are consistent with the G\"udel-Benz relation, within the (order of magnitude) scatter, and the sub-subgiant star appears large enough to sustain substantial coronal activity, so this object may be an RS CVn. 
}
}

\begin{table*}
    \centering
    \caption{Six VLA counterparts to \chandra \, sources}
    \begin{tabular}{llcccccl}
    \hline
    M4-VLA & CX & \multicolumn{2}{c}{Offset} & $\slow$  & $\shigh$ & $\alpha$ & Notes \\
       &        &  ($\arcsec$) & ($\sigma_{r,x}$)  & ($\mujy$) & ($\mujy$) & ($S_\nu \propto \nu^\alpha$) \\
    \hline
            1  & 70 & 0.6 & 0.7 & $1289.40 \pm 2.90$ & $995.50\pm 3.50$ & $-0.71^{+0.01}_{-0.01}$ & no PM info \\
		9  & 12 & 0.2 & 0.3 & $40.50   \pm 2.40$ & $<6.50$ & $<-2.60$ & Known MSP \\
		13 & 77 & 0.4 & 0.2 & $26.10   \pm 2.80$ & $16.90\pm 2.90$ & $-1.24^{+0.55}_{-0.61}$ & no PM info \\
		19 & 14 & 0.3 & 0.6 & $19.10   \pm 3.70$ & $<14.70$ & $<1.30$ & no PM info \\
		30 & 57 & 0.4 & 0.6 & $11.10   \pm 2.10$ & $10.60\pm 2.20$ & $-0.20^{+0.75}_{-0.82}$ & no PM info \\
		31 & 8  & 0.4 & 0.8 & $9.70    \pm 2.70$ & $12.20\pm 2.20$ & $0.37^{+0.59}_{-0.77}$ & PM member \\
    \hline
    \end{tabular}
    \label{tab:chandra_vla_match}
\end{table*}

\begin{table*}
    \centering
    \caption{UV/Optical counterparts to VLA sources.}
    \begin{tabular}{llccccccl}
    \hline
    M4-VLA & HUGS \# & \multicolumn{2}{c}{Offset} & $\slow$ & $\shigh$ & $\alpha$ & PM & Notes \\
           &         & ($\arcsec$) & ($\sigma_\mathrm{VLA}$) & ($\mujy$) & ($\mujy$) & ($S_\nu \propto \nu^\alpha$) & (\%) & \\
    \hline
    	4  & R0009986 & 0.04 & 0.36 & $106.20  \pm 2.50$ & $90.50\pm 2.50$ & $-0.44^{+0.10}_{-0.10}$ & $96.9$ & slight UV excess \\
		5  & R0004648 & 0.25 & 2.42 & $82.70   \pm 2.20$ & $49.20\pm 2.10$ & $-1.43^{+0.14}_{-0.14}$ & $92.3$ &  \\
		9  & R0006434 & 0.02 & 0.23 & $40.50   \pm 2.40$ & $<6.50$ & $<-2.60$ & -- & Known MSP \\
		20 & R0005914 & 0.22 & 2.10 & $18.80   \pm 2.60$ & $<7.30$ & $<-0.40$ & $96.4$ & WD \\
		31 & R0001877 & 0.30 & 2.89 & $9.70    \pm 2.70$ & $12.20\pm 2.20$ & $0.37^{+0.59}_{-0.77}$ & $97.5$ & H$\alpha$ excess; SSG \\
    \hline
    \end{tabular}
    \label{tab:vla_hugs_match}
\end{table*}

\begin{figure*}
    \centering
    \includegraphics[width=0.7\textwidth]{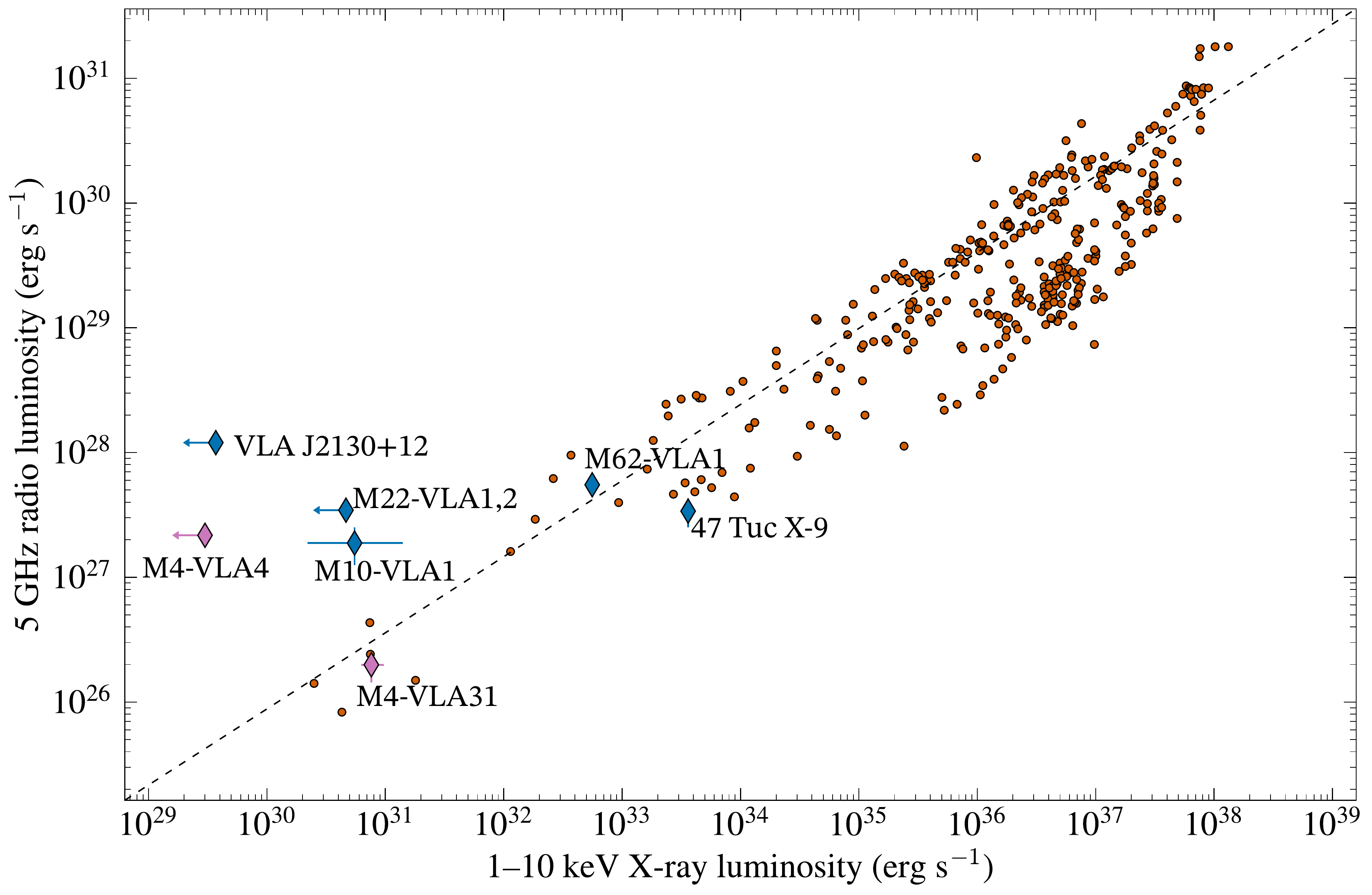}
    \caption{5 GHz radio and 1--10 keV X-ray luminosity of quiescent/hard-state BHs (filled orange circles) and BH candidates found in GCs (filled blue diamonds) from the database compiled by \citet{Bahramian22}. 
    {The two radio sources in M4 which could possibly be black hole X-ray binaries, M4-VLA4 and M4-VLA31,} 
    are in pink, both of which are consistent with or above the {quiescent/hard-state BH} correlation \citep[dashed line;][]{Gallo14}.}
    \label{f:lrlx-bh-plot}
\end{figure*}

\section{MSP population in M4} 
\label{sec:msp-population-in-m4}
\label{sec:X-ray_CMD}

%
{M4A (CX12/VLA9) is the only confirmed MSP in M4 (Sec.~\ref{sec:individual_matches_with_VLA_sources}).
\citet{Kaluzny12} provided strong evidence  that CX1 is likely to be a MSP, and likely a redback, as well.
}
Using the correlation between stellar encounter rate and the number of MSPs in a GC, \citet{Zhao22} predicted $\sim$10 MSPs 
in M4, suggesting a few MSPs may remain undiscovered. 
In this section, we 
attempt to constrain the number of MSPs that might be hidden among the uncategorized X-ray sources.

\begin{figure*}
    \centering
    \includegraphics[width=0.8\textwidth]{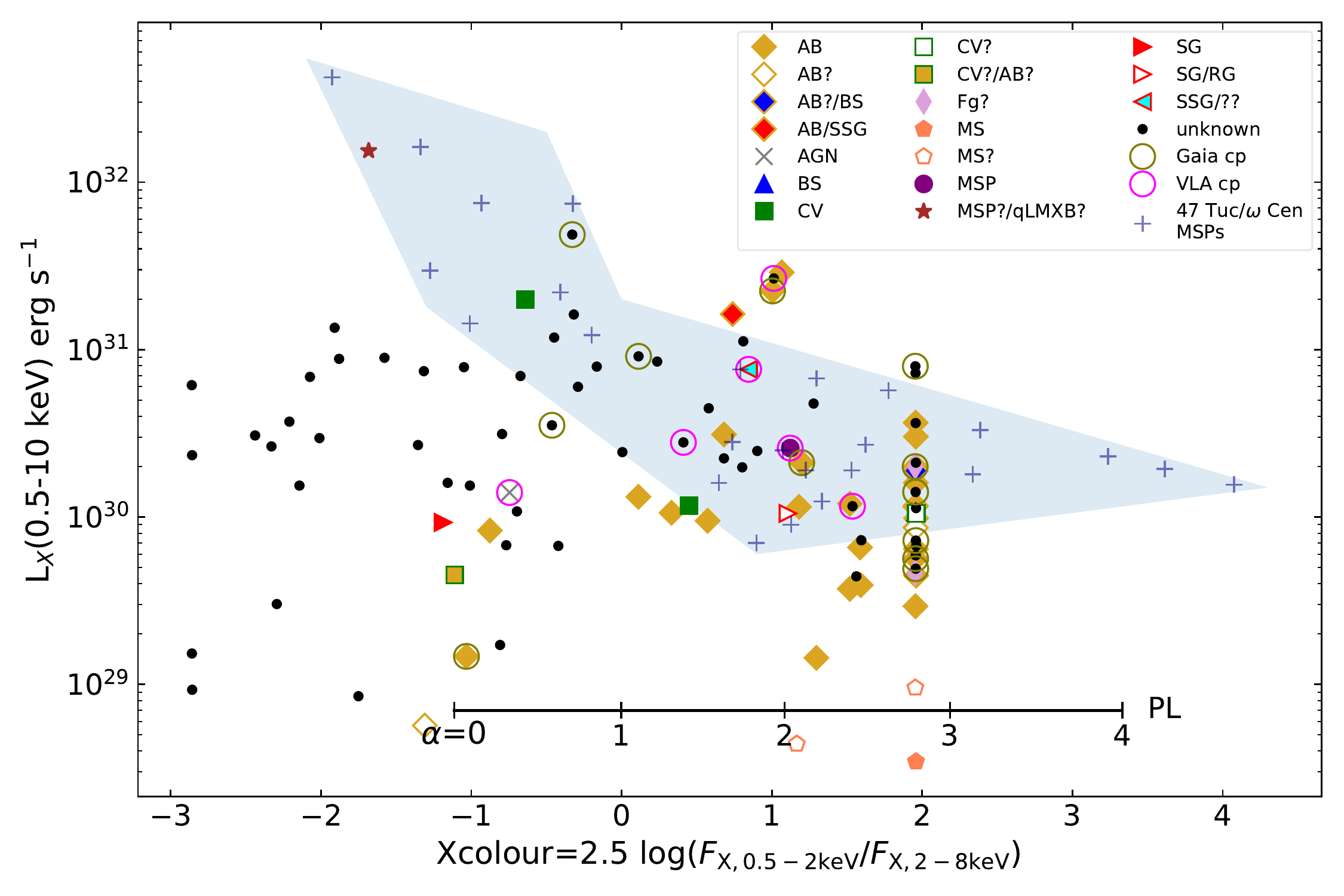}
    \caption{X-ray colour-magnitude diagram of M4, plotting  X-ray luminosities in the band 0.5--10 keV vs. X-ray colours 
    from source fluxes in 0.5--2 keV and 2--8 keV \citep{Bahramian20}. 
    Optical identifications 
    from Table~\ref{t:counterparts}. We 
    circle 12 sources with \gaia\ counterparts (olive circles; see Table~\ref{tab:gaia_matches}), and six sources with VLA counterparts (magenta circles; see Table~\ref{tab:chandra_vla_match}). Navy pluses represent 
    25 MSPs from other clusters (see text),  
    while the shaded area indicates the region where we find these MSPs in the X-ray CMD. 
    The X-ray colours of power-law models of different indices are also indicated.
    }
    \label{fig:x_ray_cmd}
\end{figure*}

Using the 
X-ray luminosities and 
\textcolor{black}{fluxes in the 0.5-2 and 2-8 keV ranges} 
from \citet{Bahramian20}, and optical identifications in Table~\ref{t:counterparts}, we 
plotted 
the X-ray CMD of M4 (Figure~\ref{fig:x_ray_cmd}), comparing X-ray hardness with X-ray luminosity. 
\textcolor{black}{Note that the X-ray flux ratios are calculated from the fitted power-law models in \citet{Bahramian20}, which gives a flux ratio limit at the assumed parameter boundaries.}
Most of the  sources of unknown nature are outside the core region, with distances to the centre $>$1.5 arcmin, which {\it HST} observations do not cover. 
These objects are likely to be dominated by background AGN,
and indeed we find that most
have relatively hard X-ray colours, typical of background AGN (which often have substantial intrinsic absorption). 
We plotted the positions of {
25 
known cluster 
MSPs with 
$7 \times 10^{29}$ 
$< L_X <$
$4 \times 10^{32}$ \ergs\ 
(18  in 47 Tuc \citep{Heinke05,Bhattacharya17}, 
2 in $\omega$\,Cen \citep{Zhao22}, 
2 in NGC 6397 \citep{Bogdanov10},
2 in Terzan 5 \citep{Bogdanov21},
and one in M92 \citep{Zhao22}), 
}
on the same X-ray CMD (navy-blue plus signs in Figure~\ref{fig:x_ray_cmd}).
We used the spectral fits in the listed sources, and their luminosities, to plot their positions, as the spectral response of {\it Chandra} changes with time. 
By comparing the colours and luminosities of those known MSPs with the positions of unknown sources in M4, we 
identify a region in the X-ray CMD (shaded area in Figure~\ref{fig:x_ray_cmd}) 
where 
unidentified MSPs are likely. 
{While most MSPs' X-ray emission is dominated by thermal radiation from the surface, redback MSPs are dominated by harder shock emission, so we include a significant redback MSP sample from several clusters. These produce two overlapping loci on the X-ray CMD.}

\textcolor{black}{Within 
this region are 45 X-ray sources in M4,
including 
15 confident ABs (including 2 eclipsing AB candidates, CX25 and CX28), 
2 likely ABs, 
1 sub-subgiant/red giant, 
1 blue straggler,
2 confident and 1 likely CV,
1 certain MSP (M4A/VLA9) and one good candidate (CX1), 
1 likely foreground source (with a {\it Gaia} counterpart), 
and three unknown sources with {\it Gaia} counterparts. 
These objects we consider to have relatively confident identifications that are unlikely to be additional MSPs. We also {note three objects discussed in \S~\ref{sec:individual_matches_with_VLA_sources} which could possibly be candidate MSPs; CX8/VLA31 (a sub-subgiant), CX77/VLA13  and CX57/VLA30 (no HUGS counterparts).} 
{We also see 
another 16 }
unknown sources 
without \gaia\ counterparts.
However, only 
three
(CXs 17, 19, 
and 62) of those 16 unknown sources are covered by HST observations, showing empty error circles. 
As MSPs in a GC tend to be concentrated in the 
core, 
the 12 unknown sources located at distances $>$2 core radii (2.2 arcmin) from the centre, without additional information, are unlikely to be MSPs.}
We thus identify 
seven objects without optical counterparts, or with uncertain optical counterparts, in Table~\ref{tab:MSP_candidate}, which we consider as possible candidate MSPs.
We 
estimate that $\lesssim$10 MSPs remain undetected in M4, 
consistent with the estimate made from the correlation between stellar encounter rates and numbers of MSPs in GCs \citep{Zhao22}.

\begin{table*}
    \centering
    \caption{Candidate MSPs in M4}
    \begin{tabular}{lcccccc}
    \hline
Source$^a$	&	RA, Dec (J2000)$^b$	&	${L_X}^c$	& Xcolour$^d$ &	Dist$^e$	&	Identification$^f$ & VLA ID$^g$	\\
 & & (erg s$^{-1}$) & & (arcmin) & from optical & \\
\hline
1   &   16:23:34.125$-$26:31:34.90	&	1.54E+32   & $-$1.68   & 0.25  &   MSP?/qLMXB? &   - \\
8	&	16:23:31.478$-$26:30:57.91	&	7.62E+30	&	0.85	&	1.02	&	SSG/??	& 31 \\
17 & 16:23:35.984$-$26:31:01.71	&	2.24E+30 & 0.68 & 0.54 & empty error & - \\
19	&	16:23:28.923$-$26:29:50.95 &	2.48E+30 & 0.90 & 2.20 & empty error & - \\
57	&	16:23:30.245$-$26:30:45.12	&	2.79E+30 & 0.41 & 1.36 & empty error & 30 \\
62	&	16:23:30.055$-$26:32:05.40	&	2.11E+30 & 1.96 & 1.23 & empty error & - \\
77	&	16:23:41.592$-$26:29:37.81	&	1.16E+30 & 1.54 & 2.39 & outside HUGS & 13 \\
\hline
\multicolumn{7}{p{1.5\columnwidth}}{{\it Notes}: $^a$Source numbering in this work (Table~\ref{t:counterparts}). $^b$Source position \citep{Bahramian20}. $^c$X-ray luminosity in \ergs, based on a BXA power-law fit \citep{Bahramian20}. $^d$X-ray colours 
from source fluxes in 0.5--2 keV and 2--8 keV. $^e$Source distance from  cluster centre in arcmin \citep{Bahramian20}. $^f$Optical identifications by {\it HST} observations (Table~\ref{t:counterparts}). $^g$\textcolor{black}{VLA counterparts to the X-ray sources listed (Table~\ref{tab:chandra_vla_match}).}} \\
    \end{tabular}
    \label{tab:MSP_candidate}
\end{table*}

\section{Radial Distributions}
\label{sec:radial-distributions}
\begin{figure}
  \centering
  \includegraphics[width=\columnwidth]{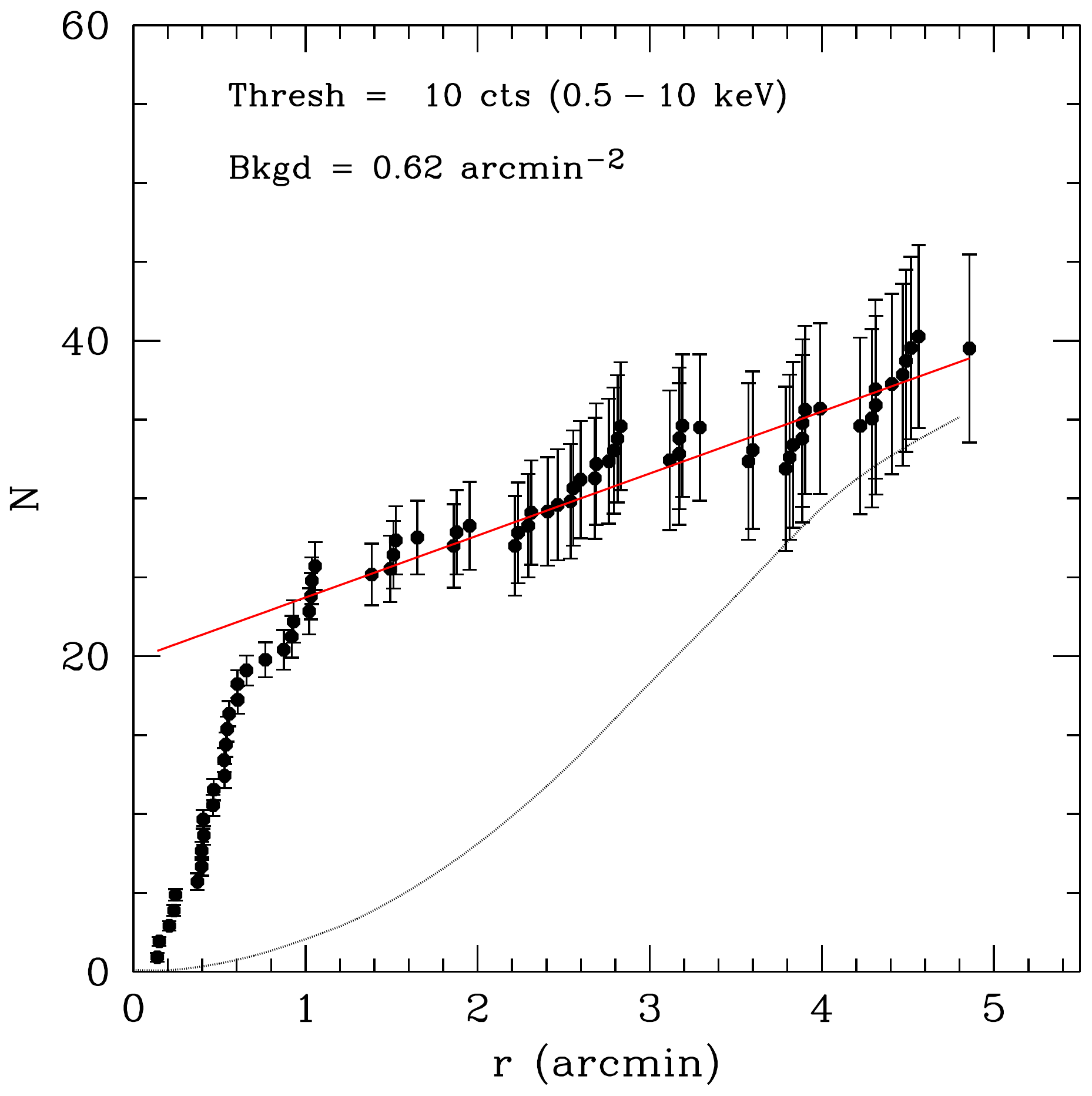}
  \caption{Cumulative radial distribution of excess Chandra source counts over the background predicted from the \citet{Giacconi01} extragalactic source counts. Error bars represent the statistical uncertainty of the background correction. The dashed curve is the cumulative background distribution for the expected background density of 0.62 arcmin$^{-2}$.}
  \label{f:m4_bkgd_corr_1}
\end{figure}


The radial distribution of a population of objects in a cluster depends on the characteristic mass of the objects. Populations of interest include all \chandra\ sources, bright CVs, faint CVs, ABs, and blue stragglers. Analysis of population radial distributions to determine object masses relative to the main sequence turnoff (MSTO) mass have been performed for a number of clusters. These include 47~Tuc \citep{Grindlay02,Heinke05,Cheng19}, M30 \citep{Lugger07,Mansfield22}, M71 \citep{Elsner08}, NGC~6397 \citep{Cohn10},  NGC~6752 \citep{Lugger17,Cohn21}, Terzan 5 \citep{Cheng19b}, M28 \citep{Cheng20a}, and $\omega$\,Cen \citep{Cheng20b}. Estimated object masses provide an important information on their nature. For CV candidates in particular, binary system mass estimates provide a measure of the masses of the white dwarf primaries, which provides an important constraint on CV evolution \citep{Pala22}. 

The method of characteristic mass determination in these studies is based on the assumption that the populations with members more massive than the main-sequence turnoff (MSTO) mass are in thermal equilibrium. In this case, the greater the characteristic mass of a population, the higher its degree of central concentration. Our analysis algorithm has most recently been described by \citet{Cohn21}. The essential component is a maximum-likelihood fitting of a generalised King model to the radial distribution of each population, including a MSTO group. The surface density profile of the generalised King model takes the form,
\begin{equation}
\label{eqn:Generalised_King_Model} 
S(r) = S_0 \left[1 + \left({\frac{r}{r_0}}\right)^2 \right]^{\alpha/2}.
\end{equation}
The core radius $r_c$, defined as the radius where the surface density drops to half of its central value, is related to the scale parameter $r_0$ by,
\begin{equation}
r_c = \left(2^{-2/\alpha} -1 \right)^{1/2} r_0\,.
\end{equation}

For the MSTO sample, $\alpha \approx -2$ and $r_c \approx r_0$. These parameter values may either be determined from a fit of Eqn.~\ref{eqn:Generalised_King_Model} to the MSTO sample or by the adoption of a King-model profile and the determination of the optical core radius. While we attempted to determine both $\alpha$ and $r_c$ for the MSTO sample directly from the HUGS \V-band star counts, this approach is complicated by the large core of M4, which fills much of the \hst\ ACS/WFC field. We also explored adopting a value of $\alpha = -2$ and determining $r_c$, which results in a value of $r_c = 65.4 \pm 3.3$~arcsec. Two studies have determined the core radius of M4 by fitting King models to the ground-based surface brightness profile presented by \citet{Trager95}, producing different results. These are 50.1 arcsec \citep{Trager95} and 69.7 arcsec \citep{McLaughlin05}.
Additionally, \citet{Baumgardt21} obtained a value of 43.5 arcsec by fitting an N-body model to a set of ground and space-based data. We conclude that the determination of $r_c$ for M4 is somewhat uncertain. 

A refinement of our determination of the characteristic \chandra\ source mass in the present study is that we now incorporate an X-ray exposure map, which allows us to select those pixels in the \chandra\ image that satisfy a minimum exposure criterion. This allows the removal from consideration of those regions that received either no or inadequate exposure. A second refinement concerns the application of the background source correction for the \chandra\ source sample. We now generate 1000 bootstrap resamplings of the original source sample and for each of these we generate 1000 Monte-Carlo removals of $N_{\mathrm{bkgd}}$ uniformly distributed objects, where $N_{\mathrm{bkgd}}$ is chosen as a random Poisson deviate with an expectation value equal to the predicted extragalactic source number calculated from \citet{Giacconi01}. 

This procedure results in a determination of the value of the mass ratio $q_X = m_\mathrm{X}/m_\mathrm{MSTO}$. We note that in our recent analyses of the entire \chandra\ source distributions for NGC~6752 in \citet{Lugger17} and \citet{Cohn21}, we neglected to apply a background correction. This resulted in a moderate underestimate of the mass of the typical \chandra\ source in these studies. For comparison, \citet{Cohn21} obtained a value of $q_X = 1.25 \pm 0.10$ for NGC~6752, while \citet{Heinke05} obtained $q_X = 1.63 \pm 0.11$ for 47~Tuc; these values differ at the $2.6\,\sigma$ level.

Fig.~\ref{f:m4_bkgd_corr_1} shows the cumulative excess number of sources (for a 10-count detection threshold) over the expected background number, calculated for a background density of 0.62 arcmin$^{-2}$, based on \citet{Giacconi01}. The red dashed line shows a linear regression fit to this profile for $r>3~\mathrm{arcmin}$. The expectation is that the profile should be asymptotically flat at large $r$, as the incremental contribution of the cluster to the total source counts dwindles. The continuous rise of the red line suggests the presence of an excess number of about 11 cluster sources in the halo, between 2 and 5\,arcmin from the cluster centre. Such an extended-halo X-ray source population has been observed in 47~Tuc by \citet{Cheng19}, who interpret it as having descended from a primordial binary population in the cluster halo. 

For the adopted background density of 0.62 arcmin$^{-2}$ and a core radius value of 50.1 arcsec \citep{Trager95}, the maximum-likelihood fit to a limiting radius of 5~arcmin of a generalised King model produces a mass ratio value of $q_X = 1.53^{+0.25}_{-0.22}.$ The greater uncertainty limits relative to 47~Tuc are likely due to the larger background correction for M4 as a result of its greater proximity and thus larger angular size on the sky. For the 74 10-count sources within 5~arcmin of the cluster centre that meet the minimum exposure criterion, 35 are likely background objects. For the same background density of 0.62 arcmin$^{-2}$ and a core radius of 69.7 arcsec \citep{McLaughlin05}, $q_X = 1.84^{+0.38}_{-0.31}.$ These two values for $q_X$ are consistent to within the large uncertainty range. The corresponding range in typical source mass is $m_x \sim 1.2-1.5\,\msun$. 


We next determine the $q$ value for the AB group. In this case, no background correction is necessary, since all of the AB candidates are likely cluster members, as indicated by their proper motions. The result depends somewhat sensitively on the value adopted for the core radius. For $r_c = 50.1$\,arcsec, $q_\mathrm{AB} = 1.79^{+0.32}_{-0.27}$, while for $r_c = 69.7$\,arcsec, $q_\mathrm{AB} = 2.49^{+0.49}_{-0.43}$. Although these two values are formally consistent with each other, given the large uncertainty, the $q_\mathrm{AB}$ value for the smaller core radius is much more similar to the range of $q_X$ values for all of the X-ray sources. Since the ABs are by far the largest subgroup of X-ray sources, this suggests that the smaller core radius value is the more appropriate one for this analysis. For a MSTO mass of 0.80\,\msun\ and a core radius of 50.1 arcsec, the inferred characteristic mass of the ABs is $1.43^{+0.26}_{-0.22}\,\msun$. For comparison, \citet{Cohn10} obtained a smaller characteristic mass of $1.06\pm0.08\,\msun$ for a sample of 36 ABs in NGC~6397. These two values do not differ significantly, given the large uncertainty of the M4 AB characteristic mass estimate. 

\section{Conclusions}

Previous Chandra and HST data of M4 revealed 20 optical counterparts to 31 X-ray sources \citep{Bassa04b,Bassa05,Kaluzny12}. Here we have used an X-ray source list \citep{Bahramian20} from significantly deeper Chandra observations (119 ks, vs. 26 ks used by \citealt{Bassa04b}), and analyzed HST broad-band optical and ultraviolet data using the ACS and WFC3 filters, largely from the HUGS survey.\footnote{A caveat is that our HST data ($3.4 \times 3.4$\,arcmin FoV) do not fully cover the 4.3\,arcmin half-light radius, only 20 per cent of it. However, since the HST/ACS data cover 53 of the 88 high-quality \chandra\ detections within the half-light radius, and we anticipate $\sim$30 background sources in the remaining area (see Section~\ref{sec:radial-distributions}, and cf. \citealt{Heinke05}), this is not a strong caveat.}
We also discuss MAVERIC radio imaging data using the Jansky VLA, which finds 38 radio sources within M4 \citep{Shishkovsky20}.  

We consider 88 high-quality \chandra\ sources within the 4.3\,arcmin half-light radius of M4, of which 53 are covered by some HST data, and 16 were previously identified. 
 We add 24 
 new HST optical/UV confident counterparts. 
In total, we find 28 confident ABs, and 6 other likely coronal sources (2 AB candidates, 1 blue straggler, 2 subgiants, 1 red giant), for 34 likely coronally-emitting sources in the HST/ACS field. 
We also find two confident CVs (CX4 and CX76, which is new) and two CV candidates (CX81 and CX101), along with a known MSP (CX12) and a previously suggested candidate MSP (CX1). 
We find one new foreground source (CX63), and 8 empty error circles,  likely dominated by distant AGN (as  $\sim$7 background AGN are expected in this region, and 1 empty error circle also shows a radio source).
Using \gaia\ data, we identify seven brighter likely optical counterparts outside the HST coverage.

We also identify 9 potential matches between VLA radio sources and X-ray or optical sources (6 X-ray, 5 optical). 
Some are almost certainly AGN (e.g. VLA1/CX70), 
but some appear highly likely to be cluster sources (VLA9/CX12 is a known MSP; VLA31/CX8, a sub-subgiant, and VLA20, a WD, appear to be cluster members). 
Our knowledge of the nature of faint cluster radio sources is quite limited so far. 

We attempt to identify potential candidate MSPs in M4 without detected radio pulsations so far. Among X-ray sources, we agree with \citet{Kaluzny12} that CX1 is a likely MSP (probably a redback); its lack of detected radio emission may be due to radio eclipses, faintness, or a misaligned beam geometry. CX8/VLA31 has a sub-subgiant optical counterpart. While its X-ray properties are consistent with MSPs, its flat radio spectrum would be unusual; dedicated spectroscopy of this object {will be reported elsewhere}. 
We identify another five objects with X-ray properties similar to MSPs but without optical counterparts, two of which have radio counterparts.  
A few other VLA-detected radio sources (VLA5 and VLA20) with possible optical counterparts have steep spectra, suggestive of MSPs, {but 
their X-ray upper limits make an MSP nature unlikely.}
Assuming that MSPs in M4 resemble those in other globular clusters, we conservatively constrain the number of MSPs in M4 to $<$10. 

The background-corrected radial distribution of 10-count \chandra\ sources suggests a relaxed population of 27 sources within the central 2 arcmin, with average mass $\sim1.2 - 1.5\,\Msun$, as well as a cluster halo population of $\sim10$ sources. The majority of these cluster sources, within both the region covered by the HUGS field and the cluster halo, are likely to be ABs. 
 
We can compare the numbers of different object types in M4 to other well-studied clusters, at similar $L_X$ limits (Tables \ref{t:Cluster_properties_Chandra_obs}, \ref{t:CVs_in_clusters}). 
We now have two confirmed, plus two candidate CVs, in M4, down to $\log L_X=29.5$; this agrees with the predicted number of four CVs produced per unit stellar mass in the field. NGC 6397, with a similar mass as M4, has almost four times as many CVs as M4 (this is also true for $L_X>10^{30}~\ergs$, or for $L_X>10^{31}~\ergs$), suggesting dynamical production of NGC 6397's CVs.  Although the number of CVs in M4 is  small, their optical and X-ray luminosity functions seem to be consistent with those in other clusters \citep{Cohn10, RiveraSandoval18, Lugger17,Belloni19}. 
Interestingly, which CVs are `bright' or `faint' differs depending on whether one is referring to X-rays or $M_V$; CX4 has $L_X=2\times10^{31}~\ergs$ but $M_V\sim8$, while CX76 has $L_X=1.2\times10^{30}~\ergs$ but $M_V\sim6$. 

 MSP progenitors are dynamically produced in clusters \citep{Clark75}, and indeed the numbers of MSPs known\footnote{\url{http://www.naic.edu/~pfreire/GCpsr.html}, as of January 2023} in the nearby globular clusters NGC 6397 (2), NGC 6752 (9), 47 Tucanae (29), and M22 (4) are similar to their predicted stellar encounter rates compared to M4 (3, 15, 37, 3 respectively; \citealt{Bahramian13}), although $\omega$ Cen has significantly more MSPs (18) than predicted \citep[3,][]{Chen23}. 
 

 Finally, M4 and NGC 6397 have similar numbers of ABs with $L_X>3\times10^{29}~\ergs$ (a limit to which both clusters are fairly complete in identifications), 28 vs. 26. This is consistent with the similarity in their masses. However, it is somewhat surprising when compared to the dramatically different binary fractions in the two clusters (10 per cent at the half-mass radius for M4, 2.4 per cent for NGC 6397; \citealt{Milone12}). We resolve this apparent contradiction by noting that X-ray detectable ABs are generally in much tighter orbits than average binaries in clusters (e.g. \citealt{Heinke05} detect X-rays from most short-period binaries found by \citealt{Albrow01} in 47 Tuc). Thus, this indicates that although the total binary fractions differ significantly, the short-orbit (or `hard') binary fraction is similar between M4 and NGC 6397. This seems to be in agreement with simulations that indicate that hard binaries are not substantially destroyed by cluster dynamics (e.g. \citealt{Hurley07}), but that the binary fraction differences may be explained by destruction of `soft' (wide-orbit) binaries in denser clusters, perhaps especially during a higher-density initial cluster formation phase \citep{Fregeau09,Leigh15}.

We note promising future directions regarding this cluster. First, the radio sources associated with cluster members, apart from the known millisecond pulsar, are mysterious; optical spectroscopic studies are in progress. A second epoch of HST $B$ data would permit proper motion membership to be ascertained for the faintest objects. In particular, the faintest CV candidate, CX101, has very limited photometry, with its apparent strong \ha\ excess being very low signal-to-noise.







\section*{Acknowledgements}

We thank the anonymous referee for thoughtful comments that improved the manuscript. We acknowledge useful discussions with M. van den Berg, D. Pooley, and J. Strader. M. Legnardi generously provided differential-reddening-corrected photometry for M4. CH acknowledges support from NSERC Discovery Grant RGPIN-2016-04602. 
JZ is supported by a China Scholarship Council scholarship No.\ 202108180023. 
This work is based on observations made with the NASA/ESA \emph{Hubble Space Telescope} obtained from the Space Telescope Science Institute, which is operated by the Association of Universities for Research in Astronomy, Inc., under NASA contract NAS 5-26555. The results reported in this article are based in part on observations made by the \emph{Chandra X-ray Observatory.}
Funding for the {\it Gaia} Data Processing and Analysis Consortium
has been provided by national institutions, in particular the institutions
participating in the {\it Gaia} Multilateral Agreement.

\section*{Data Availability}

The \chandra\ data used in this paper are available in the \chandra\ Data Archive (\url{https://cxc.harvard.edu/cda/}) by searching the Observation ID listed in Section~\ref{sec:M4A} in the Search and Retrieval interface, ChaSeR (\url{https://cda.harvard.edu/chaser/}). The \hst\ data used in this work can be retrieved from the Mikulski Archive for Space Telescope (MAST) Portal (\url{https://mast.stsci.edu/search/ui/#/hst}) by searching the proposal IDs listed in Table~\ref{t:UV_optical_data}. The VLA observations are available at 
\url{https://data.nrao.edu/portal/#/} under the proposal IDs VLA/13B-014 and VLA/15A-100.
This work has made use of data from the European Space Agency (ESA) mission
{\it Gaia} (\url{https://www.cosmos.esa.int/gaia}), processed by the {\it Gaia}
Data Processing and Analysis Consortium (DPAC,
\url{https://www.cosmos.esa.int/web/gaia/dpac/consortium}).


\bibliographystyle{mnras}
\bibliography{M4}

\begin{thebibliography}{}
\makeatletter
\relax
\def\mn@urlcharsother{\let\do\@makeother \do\$\do\&\do\#\do\^\do\_\do\%\do\~}
\def\mn@doi{\begingroup\mn@urlcharsother \@ifnextchar [ {\mn@doi@}
  {\mn@doi@[]}}
\def\mn@doi@[#1]#2{\def\@tempa{#1}\ifx\@tempa\@empty \href
  {http://dx.doi.org/#2} {doi:#2}\else \href {http://dx.doi.org/#2} {#1}\fi
  \endgroup}
\def\mn@eprint#1#2{\mn@eprint@#1:#2::\@nil}
\def\mn@eprint@arXiv#1{\href {http://arxiv.org/abs/#1} {{\tt arXiv:#1}}}
\def\mn@eprint@dblp#1{\href {http://dblp.uni-trier.de/rec/bibtex/#1.xml}
  {dblp:#1}}
\def\mn@eprint@#1:#2:#3:#4\@nil{\def\@tempa {#1}\def\@tempb {#2}\def\@tempc
  {#3}\ifx \@tempc \@empty \let \@tempc \@tempb \let \@tempb \@tempa \fi \ifx
  \@tempb \@empty \def\@tempb {arXiv}\fi \@ifundefined
  {mn@eprint@\@tempb}{\@tempb:\@tempc}{\expandafter \expandafter \csname
  mn@eprint@\@tempb\endcsname \expandafter{\@tempc}}}

\bibitem[\protect\citeauthoryear{{Albrow} et~al.,}{{Albrow}
  et~al.}{2001}]{Albrow01}
{Albrow} M.~D.,  et~al., 2001, \mn@doi [\apj] {10.1086/322353}, \href
  {http://adsabs.harvard.edu/abs/2001ApJ...559.1060A} {559, 1060}

\bibitem[\protect\citeauthoryear{{Anderson} et~al.,}{{Anderson}
  et~al.}{2008}]{Anderson08}
{Anderson} J.,  et~al., 2008, \mn@doi [\aj] {10.1088/0004-6256/135/6/2055},
  \href {http://adsabs.harvard.edu/abs/2008AJ....135.2055A} {135, 2055}

\bibitem[\protect\citeauthoryear{{Bagchi} \& {Lorimer}}{{Bagchi} \&
  {Lorimer}}{2011}]{Bagchi11}
{Bagchi} M.,  {Lorimer} D.~R.,  2011, in {Burgay} M.,  et~al., eds,  American
  Institute of Physics Conference Series Vol. 1357, Radio Pulsars: An
  Astrophysical Key to Unlock the Secrets of the Universe. pp 173--176
  (\mn@eprint {arXiv} {1012.4705}), \mn@doi{10.1063/1.3615109}

\bibitem[\protect\citeauthoryear{{Bahramian} \& {Rushton}}{{Bahramian} \&
  {Rushton}}{2022}]{Bahramian22}
{Bahramian} A.,  {Rushton} A.,  2022, {bersavosh/XRB-LrLx\_pub: update 220808},
  Zenodo, \mn@doi{10.5281/zenodo.6972578}

\bibitem[\protect\citeauthoryear{{Bahramian}, {Heinke}, {Sivakoff}  \&
  {Gladstone}}{{Bahramian} et~al.}{2013}]{Bahramian13}
{Bahramian} A.,  {Heinke} C.~O.,  {Sivakoff} G.~R.,   {Gladstone} J.~C.,  2013,
  \mn@doi [\apj] {10.1088/0004-637X/766/2/136}, \href
  {http://adsabs.harvard.edu/abs/2013ApJ...766..136B} {766, 136}

\bibitem[\protect\citeauthoryear{{Bahramian} et~al.,}{{Bahramian}
  et~al.}{2015}]{Bahramian15}
{Bahramian} A.,  et~al., 2015, \mn@doi [\mnras] {10.1093/mnras/stv1585}, \href
  {https://ui.adsabs.harvard.edu/abs/2015MNRAS.452.3475B} {452, 3475}

\bibitem[\protect\citeauthoryear{{Bahramian} et~al.,}{{Bahramian}
  et~al.}{2020}]{Bahramian20}
{Bahramian} A.,  et~al., 2020, \mn@doi [\apj] {10.3847/1538-4357/aba51d}, \href
  {https://ui.adsabs.harvard.edu/abs/2020ApJ...901...57B} {901, 57}

\bibitem[\protect\citeauthoryear{{Bailer-Jones} et~al.,}{{Bailer-Jones}
  et~al.}{2021}]{Bailer-Jones21}
{Bailer-Jones} C.~A.~L.,  et~al., 2021, \mn@doi [\aj]
  {10.3847/1538-3881/abd806}, \href
  {https://ui.adsabs.harvard.edu/abs/2021AJ....161..147B} {161, 147}

\bibitem[\protect\citeauthoryear{{Bassa} et~al.,}{{Bassa}
  et~al.}{2004}]{Bassa04b}
{Bassa} C.,  et~al., 2004, \mn@doi [\apj] {10.1086/421259}, \href
  {http://adsabs.harvard.edu/abs/2004ApJ...609..755B} {609, 755}

\bibitem[\protect\citeauthoryear{{Bassa} et~al.,}{{Bassa}
  et~al.}{2005}]{Bassa05}
{Bassa} C.,  et~al., 2005, \mn@doi [\apj] {10.1086/426683}, \href
  {https://ui.adsabs.harvard.edu/abs/2005ApJ...619.1189B} {619, 1189}

\bibitem[\protect\citeauthoryear{{Baumgardt} et~al.,}{{Baumgardt}
  et~al.}{2021}]{Baumgardt21}
{Baumgardt} H.,  et~al., 2021, Fundamental parameters of Galactic globular
  clusters (as of May 2021),
  \urlwofont{https://people.smp.uq.edu.au/HolgerBaumgardt/globular/}

\bibitem[\protect\citeauthoryear{{Beaumont}, {Goodman}  \&
  {Greenfield}}{{Beaumont} et~al.}{2015}]{Beaumont15}
{Beaumont} C.,  {Goodman} A.,   {Greenfield} P.,  2015, in {Taylor} A.~R.,
  {Rosolowsky} E.,  eds,  Astronomical Society of the Pacific Conference Series
  Vol. 495, Astronomical Data Analysis Software an Systems XXIV (ADASS XXIV).
  p.~101

\bibitem[\protect\citeauthoryear{{Beccari}, {De Marchi}, {Panagia}  \&
  {Pasquini}}{{Beccari} et~al.}{2014}]{Beccari14}
{Beccari} G.,  {De Marchi} G.,  {Panagia} N.,   {Pasquini} L.,  2014, \mn@doi
  [\mnras] {10.1093/mnras/stt2074}, \href
  {http://adsabs.harvard.edu/abs/2014MNRAS.437.2621B} {437, 2621}

\bibitem[\protect\citeauthoryear{{Belloni} et~al.,}{{Belloni}
  et~al.}{2019}]{Belloni19}
{Belloni} D.,  et~al., 2019, \mn@doi [\mnras] {10.1093/mnras/sty3097}, \href
  {https://ui.adsabs.harvard.edu/abs/2019MNRAS.483..315B} {483, 315}

\bibitem[\protect\citeauthoryear{{Bhattacharya} et~al.,}{{Bhattacharya}
  et~al.}{2017}]{Bhattacharya17}
{Bhattacharya} S.,  et~al., 2017, \mn@doi [\mnras] {10.1093/mnras/stx2241},
  \href {https://ui.adsabs.harvard.edu/abs/2017MNRAS.472.3706B} {472, 3706}

\bibitem[\protect\citeauthoryear{{Bogdanov} et~al.,}{{Bogdanov}
  et~al.}{2006}]{Bogdanov06}
{Bogdanov} S.,  et~al., 2006, \mn@doi [\apj] {10.1086/505133}, \href
  {https://ui.adsabs.harvard.edu/abs/2006ApJ...646.1104B} {646, 1104}

\bibitem[\protect\citeauthoryear{{Bogdanov} et~al.,}{{Bogdanov}
  et~al.}{2010}]{Bogdanov10}
{Bogdanov} S.,  et~al., 2010, \mn@doi [\apj] {10.1088/0004-637X/709/1/241},
  \href {http://adsabs.harvard.edu/abs/2010ApJ...709..241B} {709, 241}

\bibitem[\protect\citeauthoryear{{Bogdanov} et~al.,}{{Bogdanov}
  et~al.}{2021}]{Bogdanov21}
{Bogdanov} S.,  et~al., 2021, \mn@doi [\apj] {10.3847/1538-4357/abee78}, \href
  {https://ui.adsabs.harvard.edu/abs/2021ApJ...912..124B} {912, 124}

\bibitem[\protect\citeauthoryear{{Chabrier}}{{Chabrier}}{2001}]{Chabrier01}
{Chabrier} G.,  2001, \mn@doi [\apj] {10.1086/321401}, \href
  {https://ui.adsabs.harvard.edu/abs/2001ApJ...554.1274C} {554, 1274}

\bibitem[\protect\citeauthoryear{{Chen} et~al.,}{{Chen} et~al.}{2023}]{Chen23}
{Chen} W.,  et~al., 2023, \mn@doi [\mnras] {10.1093/mnras/stad029}, \href
  {https://ui.adsabs.harvard.edu/abs/2023MNRAS.tmp...59C} {}

\bibitem[\protect\citeauthoryear{{Cheng}, {Li}, {Xu}  \& {Li}}{{Cheng}
  et~al.}{2018}]{Cheng18}
{Cheng} Z.,  {Li} Z.,  {Xu} X.,   {Li} X.,  2018, \mn@doi [\apj]
  {10.3847/1538-4357/aaba16}, \href
  {https://ui.adsabs.harvard.edu/abs/2018ApJ...858...33C} {858, 33}

\bibitem[\protect\citeauthoryear{{Cheng} et~al.,}{{Cheng}
  et~al.}{2019a}]{Cheng19}
{Cheng} Z.,  et~al., 2019a, \mn@doi [\apj] {10.3847/1538-4357/ab1593}, \href
  {https://ui.adsabs.harvard.edu/abs/2019ApJ...876...59C} {876, 59}

\bibitem[\protect\citeauthoryear{{Cheng} et~al.,}{{Cheng}
  et~al.}{2019b}]{Cheng19b}
{Cheng} Z.,  et~al., 2019b, \mn@doi [\apj] {10.3847/1538-4357/ab3c6d}, \href
  {https://ui.adsabs.harvard.edu/abs/2019ApJ...883...90C} {883, 90}

\bibitem[\protect\citeauthoryear{{Cheng} et~al.,}{{Cheng}
  et~al.}{2020a}]{Cheng20a}
{Cheng} Z.,  et~al., 2020a, \mn@doi [\apj] {10.3847/1538-4357/ab7933}, \href
  {https://ui.adsabs.harvard.edu/abs/2020ApJ...892...16C} {892, 16}

\bibitem[\protect\citeauthoryear{{Cheng} et~al.,}{{Cheng}
  et~al.}{2020b}]{Cheng20b}
{Cheng} Z.,  et~al., 2020b, \mn@doi [\apj] {10.3847/1538-4357/abbdfc}, \href
  {https://ui.adsabs.harvard.edu/abs/2020ApJ...904..198C} {904, 198}

\bibitem[\protect\citeauthoryear{{Clark}}{{Clark}}{1975}]{Clark75}
{Clark} G.~W.,  1975, \apjl, \href
  {http://adsabs.harvard.edu/cgi-bin/nph-bib_query?bibcode=1975ApJ...199L.143C&db_key=AST}
  {199, L143}

\bibitem[\protect\citeauthoryear{{Cohn} et~al.,}{{Cohn} et~al.}{2010}]{Cohn10}
{Cohn} H.~N.,  et~al., 2010, \mn@doi [\apj] {10.1088/0004-637X/722/1/20}, \href
  {http://adsabs.harvard.edu/abs/2010ApJ...722...20C} {722, 20}

\bibitem[\protect\citeauthoryear{{Cohn} et~al.,}{{Cohn} et~al.}{2021}]{Cohn21}
{Cohn} H.~N.,  et~al., 2021, \mn@doi [\mnras] {10.1093/mnras/stab2636}, \href
  {https://ui.adsabs.harvard.edu/abs/2021MNRAS.508.2823C} {508, 2823}

\bibitem[\protect\citeauthoryear{{Cool} et~al.,}{{Cool} et~al.}{1995}]{Cool95}
{Cool} A.~M.,  et~al., 1995, \mn@doi [\apj] {10.1086/175209}, \href
  {https://ui.adsabs.harvard.edu/abs/1995ApJ...439..695C} {439, 695}

\bibitem[\protect\citeauthoryear{{Cool} et~al.,}{{Cool} et~al.}{2013}]{Cool13}
{Cool} A.~M.,  et~al., 2013, \mn@doi [\apj] {10.1088/0004-637X/763/2/126},
  \href {http://adsabs.harvard.edu/abs/2013ApJ...763..126C} {763, 126}

\bibitem[\protect\citeauthoryear{{Davies}}{{Davies}}{1997}]{Davies97}
{Davies} M.~B.,  1997, \mn@doi [\mnras] {10.1093/mnras/288.1.117}, \href
  {https://ui.adsabs.harvard.edu/abs/1997MNRAS.288..117D} {288, 117}

\bibitem[\protect\citeauthoryear{{De Marchi}, {Panagia}  \& {Romaniello}}{{De
  Marchi} et~al.}{2010}]{DeMarchi10}
{De Marchi} G.,  {Panagia} N.,   {Romaniello} M.,  2010, \mn@doi [\apj]
  {10.1088/0004-637X/715/1/1}, \href
  {https://ui.adsabs.harvard.edu/abs/2010ApJ...715....1D} {715, 1}

\bibitem[\protect\citeauthoryear{{Dotter} et~al.,}{{Dotter}
  et~al.}{2007}]{Dotter07}
{Dotter} A.,  et~al., 2007, \mn@doi [\aj] {10.1086/517915}, \href
  {https://ui.adsabs.harvard.edu/abs/2007AJ....134..376D} {134, 376}

\bibitem[\protect\citeauthoryear{{Edmonds}, {Gilliland}, {Heinke}  \&
  {Grindlay}}{{Edmonds} et~al.}{2003a}]{Edmonds03b}
{Edmonds} P.~D.,  {Gilliland} R.~L.,  {Heinke} C.~O.,   {Grindlay} J.~E.,
  2003a, \mn@doi [\apj] {10.1086/378194}, \href
  {http://adsabs.harvard.edu/abs/2003ApJ...596.1197E} {596, 1197}

\bibitem[\protect\citeauthoryear{{Edmonds}, {Gilliland}, {Heinke}  \&
  {Grindlay}}{{Edmonds} et~al.}{2003b}]{Edmonds03a}
{Edmonds} P.~D.,  {Gilliland} R.~L.,  {Heinke} C.~O.,   {Grindlay} J.~E.,
  2003b, \mn@doi [\apj] {10.1086/378193}, \href
  {https://ui.adsabs.harvard.edu/abs/2003ApJ...596.1177E} {596, 1177}

\bibitem[\protect\citeauthoryear{{Elsner} et~al.,}{{Elsner}
  et~al.}{2008}]{Elsner08}
{Elsner} R.~F.,  et~al., 2008, \mn@doi [\apj] {10.1086/591899}, \href
  {https://ui.adsabs.harvard.edu/abs/2008ApJ...687.1019E} {687, 1019}

\bibitem[\protect\citeauthoryear{{Espinasse} \& {Fender}}{{Espinasse} \&
  {Fender}}{2018}]{Espinasse18}
{Espinasse} M.,  {Fender} R.,  2018, \mn@doi [\mnras] {10.1093/mnras/stx2467},
  \href {https://ui.adsabs.harvard.edu/abs/2018MNRAS.473.4122E} {473, 4122}

\bibitem[\protect\citeauthoryear{{Fregeau}, {Ivanova}  \& {Rasio}}{{Fregeau}
  et~al.}{2009}]{Fregeau09}
{Fregeau} J.~M.,  {Ivanova} N.,   {Rasio} F.~A.,  2009, \mn@doi [\apj]
  {10.1088/0004-637X/707/2/1533}, \href
  {https://ui.adsabs.harvard.edu/abs/2009ApJ...707.1533F} {707, 1533}

\bibitem[\protect\citeauthoryear{{Fruscione} et~al.,}{{Fruscione}
  et~al.}{2006}]{Fruscione06}
{Fruscione} A.,  et~al., 2006, in {Silva} D.~R.,  {Doxsey} R.~E.,  eds,
  Society of Photo-Optical Instrumentation Engineers (SPIE) Conference Series
  Vol. 6270, Society of Photo-Optical Instrumentation Engineers (SPIE)
  Conference Series. p. 62701V, \mn@doi{10.1117/12.671760}

\bibitem[\protect\citeauthoryear{{Gaia Collaboration} et~al.,}{{Gaia
  Collaboration} et~al.}{2016}]{GaiaCollaboration16}
{Gaia Collaboration} et~al., 2016, \mn@doi [\aap]
  {10.1051/0004-6361/201629272}, \href
  {https://ui.adsabs.harvard.edu/abs/2016A&A...595A...1G} {595, A1}

\bibitem[\protect\citeauthoryear{{Gaia Collaboration} et~al.,}{{Gaia
  Collaboration} et~al.}{2022}]{GaiaCollaboration22}
{Gaia Collaboration} et~al., 2022, arXiv e-prints, \href
  {https://ui.adsabs.harvard.edu/abs/2022arXiv220800211G} {p. arXiv:2208.00211}

\bibitem[\protect\citeauthoryear{{Gallo} et~al.,}{{Gallo}
  et~al.}{2014}]{Gallo14}
{Gallo} E.,  et~al., 2014, \mn@doi [\mnras] {10.1093/mnras/stu1599}, \href
  {https://ui.adsabs.harvard.edu/abs/2014MNRAS.445..290G} {445, 290}

\bibitem[\protect\citeauthoryear{{Ge} et~al.,}{{Ge} et~al.}{2015}]{Ge15}
{Ge} C.,  et~al., 2015, \mn@doi [\apj] {10.1088/0004-637X/812/2/130}, \href
  {https://ui.adsabs.harvard.edu/abs/2015ApJ...812..130G} {812, 130}

\bibitem[\protect\citeauthoryear{{Gehrels}}{{Gehrels}}{1986}]{Gehrels86}
{Gehrels} N.,  1986, \mn@doi [\apj] {10.1086/164079}, \href
  {https://ui.adsabs.harvard.edu/abs/1986ApJ...303..336G} {303, 336}

\bibitem[\protect\citeauthoryear{{Geller} et~al.,}{{Geller}
  et~al.}{2017}]{Geller17}
{Geller} A.~M.,  et~al., 2017, \mn@doi [\apj] {10.3847/1538-4357/aa6af3}, \href
  {https://ui.adsabs.harvard.edu/abs/2017ApJ...840...66G} {840, 66}

\bibitem[\protect\citeauthoryear{{Giacconi} et~al.,}{{Giacconi}
  et~al.}{2001}]{Giacconi01}
{Giacconi} R.,  et~al., 2001, \mn@doi [\apj] {10.1086/320222}, \href
  {https://ui.adsabs.harvard.edu/abs/2001ApJ...551..624G} {551, 624}

\bibitem[\protect\citeauthoryear{{Gordon} et~al.,}{{Gordon}
  et~al.}{2021}]{Gordon21}
{Gordon} Y.~A.,  et~al., 2021, \mn@doi [\apjs] {10.3847/1538-4365/ac05c0},
  \href {https://ui.adsabs.harvard.edu/abs/2021ApJS..255...30G} {255, 30}

\bibitem[\protect\citeauthoryear{{Grindlay}, {Heinke}, {Edmonds}  \&
  {Murray}}{{Grindlay} et~al.}{2001a}]{Grindlay01b}
{Grindlay} J.~E.,  {Heinke} C.,  {Edmonds} P.~D.,   {Murray} S.~S.,  2001a,
  \mn@doi [Science] {10.1126/science.1061135}, \href
  {https://ui.adsabs.harvard.edu/abs/2001Sci...292.2290G} {292, 2290}

\bibitem[\protect\citeauthoryear{{Grindlay} et~al.,}{{Grindlay}
  et~al.}{2001b}]{Grindlay01}
{Grindlay} J.~E.,  et~al., 2001b, \mn@doi [\apjl] {10.1086/338499}, \href
  {http://adsabs.harvard.edu/cgi-bin/nph-bib_query?bibcode=2001ApJ...563L..53G&db_key=AST}
  {563, L53}

\bibitem[\protect\citeauthoryear{{Grindlay} et~al.,}{{Grindlay}
  et~al.}{2002}]{Grindlay02}
{Grindlay} J.~E.,  et~al., 2002, \mn@doi [\apj] {10.1086/344150}, \href
  {http://adsabs.harvard.edu/cgi-bin/nph-bib_query?bibcode=2002ApJ...581..470G&db_key=AST}
  {581, 470}

\bibitem[\protect\citeauthoryear{{G{\"u}del}}{{G{\"u}del}}{2002}]{Gudel02}
{G{\"u}del} M.,  2002, \mn@doi [\araa]
  {10.1146/annurev.astro.40.060401.093806}, \href
  {https://ui.adsabs.harvard.edu/abs/2002ARA&A..40..217G} {40, 217}

\bibitem[\protect\citeauthoryear{{Guedel} \& {Benz}}{{Guedel} \&
  {Benz}}{1993}]{GudelBenz93}
{Guedel} M.,  {Benz} A.~O.,  1993, \mn@doi [\apjl] {10.1086/186766}, \href
  {https://ui.adsabs.harvard.edu/abs/1993ApJ...405L..63G} {405, L63}

\bibitem[\protect\citeauthoryear{{Haggard}, {Cool}  \& {Davies}}{{Haggard}
  et~al.}{2009}]{Haggard09}
{Haggard} D.,  {Cool} A.~M.,   {Davies} M.~B.,  2009, \mn@doi [\apj]
  {10.1088/0004-637X/697/1/224}, \href
  {https://ui.adsabs.harvard.edu/abs/2009ApJ...697..224H} {697, 224}

\bibitem[\protect\citeauthoryear{{Hansen} et~al.,}{{Hansen}
  et~al.}{2004}]{Hansen04}
{Hansen} B. M.~S.,  et~al., 2004, \mn@doi [\apjs] {10.1086/424832}, \href
  {https://ui.adsabs.harvard.edu/abs/2004ApJS..155..551H} {155, 551}

\bibitem[\protect\citeauthoryear{{Harding} \& {Muslimov}}{{Harding} \&
  {Muslimov}}{2002}]{Harding02}
{Harding} A.~K.,  {Muslimov} A.~G.,  2002, \mn@doi [\apj] {10.1086/338985},
  \href {https://ui.adsabs.harvard.edu/abs/2002ApJ...568..862H} {568, 862}

\bibitem[\protect\citeauthoryear{{Harris}}{{Harris}}{1996}]{Harris96}
{Harris} W.~E.,  1996, \aj, \href
  {http://adsabs.harvard.edu/cgi-bin/nph-bib_query?bibcode=1996AJ....112.1487H&amp;db_key=AST}
  {112, 1487}

\bibitem[\protect\citeauthoryear{{Heinke} et~al.,}{{Heinke}
  et~al.}{2003}]{Heinke03}
{Heinke} C.~O.,  et~al., 2003, \mn@doi [\apj] {10.1086/378885}, \href
  {http://adsabs.harvard.edu/abs/2003ApJ...598..501H} {598, 501}

\bibitem[\protect\citeauthoryear{{Heinke} et~al.,}{{Heinke}
  et~al.}{2005}]{Heinke05}
{Heinke} C.~O.,  et~al., 2005, \mn@doi [\apj] {10.1086/429899}, \href
  {http://adsabs.harvard.edu/abs/2005ApJ...625..796H} {625, 796}

\bibitem[\protect\citeauthoryear{{Heinke}, {Rybicki}, {Narayan}  \&
  {Grindlay}}{{Heinke} et~al.}{2006}]{Heinke06}
{Heinke} C.~O.,  {Rybicki} G.~B.,  {Narayan} R.,   {Grindlay} J.~E.,  2006,
  \mn@doi [\apj] {10.1086/503701}, \href
  {http://adsabs.harvard.edu/abs/2006ApJ...644.1090H} {644, 1090}

\bibitem[\protect\citeauthoryear{{Heinke} et~al.,}{{Heinke}
  et~al.}{2020}]{Heinke20}
{Heinke} C.~O.,  et~al., 2020, \mn@doi [\mnras] {10.1093/mnras/staa194}, \href
  {https://ui.adsabs.harvard.edu/abs/2020MNRAS.492.5684H} {492, 5684}

\bibitem[\protect\citeauthoryear{{Hellier}}{{Hellier}}{2001}]{Hellier01}
{Hellier} C.,  2001, {Cataclysmic Variable Stars: How and Why They Vary}.
Springer-Praxis

\bibitem[\protect\citeauthoryear{{Henleywillis} et~al.,}{{Henleywillis}
  et~al.}{2018}]{Henleywillis18}
{Henleywillis} S.,  et~al., 2018, \mn@doi [\mnras] {10.1093/mnras/sty675},
  \href {https://ui.adsabs.harvard.edu/abs/2018MNRAS.479.2834H} {479, 2834}

\bibitem[\protect\citeauthoryear{{Hong} et~al.,}{{Hong} et~al.}{2005}]{Hong05}
{Hong} J.,  et~al., 2005, \mn@doi [\apj] {10.1086/496966}, \href
  {http://adsabs.harvard.edu/abs/2005ApJ...635..907H} {635, 907}

\bibitem[\protect\citeauthoryear{{Howell}, {Nelson}  \& {Rappaport}}{{Howell}
  et~al.}{2001}]{Howell01}
{Howell} S.~B.,  {Nelson} L.~A.,   {Rappaport} S.,  2001, \mn@doi [\apj]
  {10.1086/319776}, \href {http://adsabs.harvard.edu/abs/2001ApJ...550..897H}
  {550, 897}

\bibitem[\protect\citeauthoryear{{Hurley}, {Aarseth}  \& {Shara}}{{Hurley}
  et~al.}{2007}]{Hurley07}
{Hurley} J.~R.,  {Aarseth} S.~J.,   {Shara} M.~M.,  2007, \mn@doi [\apj]
  {10.1086/517879}, \href
  {https://ui.adsabs.harvard.edu/abs/2007ApJ...665..707H} {665, 707}

\bibitem[\protect\citeauthoryear{{Hut} et~al.,}{{Hut} et~al.}{1992}]{Hut92}
{Hut} P.,  et~al., 1992, \pasp, \href
  {http://adsabs.harvard.edu/cgi-bin/nph-bib_query?bibcode=1992PASP..104..981H&db_key=AST}
  {104, 981}

\bibitem[\protect\citeauthoryear{{Ivanova}, {Belczynski}, {Fregeau}  \&
  {Rasio}}{{Ivanova} et~al.}{2005}]{Ivanova05}
{Ivanova} N.,  {Belczynski} K.,  {Fregeau} J.~M.,   {Rasio} F.~A.,  2005,
  \mn@doi [\mnras] {10.1111/j.1365-2966.2005.08804.x}, \href
  {https://ui.adsabs.harvard.edu/abs/2005MNRAS.358..572I} {358, 572}

\bibitem[\protect\citeauthoryear{{Ivanova} et~al.,}{{Ivanova}
  et~al.}{2006}]{Ivanova06}
{Ivanova} N.,  et~al., 2006, \mn@doi [\mnras]
  {10.1111/j.1365-2966.2006.10876.x}, \href
  {http://adsabs.harvard.edu/abs/2006MNRAS.372.1043I} {372, 1043}

\bibitem[\protect\citeauthoryear{{Johnston} \& {Verbunt}}{{Johnston} \&
  {Verbunt}}{1996}]{Johnston96}
{Johnston} H.~M.,  {Verbunt} F.,  1996, \aap, \href
  {https://ui.adsabs.harvard.edu/abs/1996A&A...312...80J} {312, 80}

\bibitem[\protect\citeauthoryear{{Kaluzny}, {Thompson}  \&
  {Krzeminski}}{{Kaluzny} et~al.}{1997}]{Kaluzny97}
{Kaluzny} J.,  {Thompson} I.~B.,   {Krzeminski} W.,  1997, \mn@doi [\aj]
  {10.1086/118432}, \href
  {https://ui.adsabs.harvard.edu/abs/1997AJ....113.2219K} {113, 2219}

\bibitem[\protect\citeauthoryear{{Kaluzny} et~al.,}{{Kaluzny}
  et~al.}{2012}]{Kaluzny12}
{Kaluzny} J.,  et~al., 2012, \mn@doi [\apjl] {10.1088/2041-8205/750/1/L3},
  \href {https://ui.adsabs.harvard.edu/abs/2012ApJ...750L...3K} {750, L3}

\bibitem[\protect\citeauthoryear{{Kaluzny}, {Thompson}, {Rozyczka}  \&
  {Krzeminski}}{{Kaluzny} et~al.}{2013}]{Kaluzny13}
{Kaluzny} J.,  {Thompson} I.~B.,  {Rozyczka} M.,   {Krzeminski} W.,  2013,
  \actaa, \href {https://ui.adsabs.harvard.edu/abs/2013AcA....63..181K} {63,
  181}

\bibitem[\protect\citeauthoryear{{Knigge}}{{Knigge}}{2012}]{Knigge12}
{Knigge} C.,  2012, \memsai, \href
  {https://ui.adsabs.harvard.edu/abs/2012MmSAI..83..549K} {83, 549}

\bibitem[\protect\citeauthoryear{{Knigge} et~al.,}{{Knigge}
  et~al.}{2003}]{Knigge03}
{Knigge} C.,  et~al., 2003, \mn@doi [\apj] {10.1086/379609}, \href
  {https://ui.adsabs.harvard.edu/abs/2003ApJ...599.1320K} {599, 1320}

\bibitem[\protect\citeauthoryear{{Kong} et~al.,}{{Kong} et~al.}{2006}]{Kong06}
{Kong} A. K.~H.,  et~al., 2006, \mn@doi [\apj] {10.1086/505485}, \href
  {https://ui.adsabs.harvard.edu/abs/2006ApJ...647.1065K} {647, 1065}

\bibitem[\protect\citeauthoryear{{Kramer} et~al.,}{{Kramer}
  et~al.}{1998}]{Kramer98}
{Kramer} M.,  et~al., 1998, \mn@doi [\apj] {10.1086/305790}, \href
  {https://ui.adsabs.harvard.edu/abs/1998ApJ...501..270K} {501, 270}

\bibitem[\protect\citeauthoryear{{Kremer} et~al.,}{{Kremer}
  et~al.}{2020}]{Kremer20}
{Kremer} K.,  et~al., 2020, in {Bragaglia} A.,  {Davies} M.,  {Sills} A.,
  {Vesperini} E.,  eds, ~ Vol. 351, Star Clusters: From the Milky Way to the
  Early Universe. pp 357--366 (\mn@eprint {arXiv} {1907.12564}),
  \mn@doi{10.1017/S1743921319007269}

\bibitem[\protect\citeauthoryear{{Legnardi} et~al.,}{{Legnardi}
  et~al.}{2023}]{Legnardi23}
{Legnardi} M.~V.,  et~al., 2023, \mn@doi [\mnras] {10.1093/mnras/stad1056},
  \href {https://ui.adsabs.harvard.edu/abs/2023MNRAS.522..367L} {522, 367}

\bibitem[\protect\citeauthoryear{{Leigh} et~al.,}{{Leigh}
  et~al.}{2015}]{Leigh15}
{Leigh} N. W.~C.,  et~al., 2015, \mn@doi [\mnras] {10.1093/mnras/stu2110},
  \href {https://ui.adsabs.harvard.edu/abs/2015MNRAS.446..226L} {446, 226}

\bibitem[\protect\citeauthoryear{{Leiner} et~al.,}{{Leiner}
  et~al.}{2022}]{Leiner22}
{Leiner} E.~M.,  et~al., 2022, \mn@doi [\apj] {10.3847/1538-4357/ac53b1}, \href
  {https://ui.adsabs.harvard.edu/abs/2022ApJ...927..222L} {927, 222}

\bibitem[\protect\citeauthoryear{{Lugger} et~al.,}{{Lugger}
  et~al.}{2007}]{Lugger07}
{Lugger} P.~M.,  et~al., 2007, \mn@doi [\apj] {10.1086/507572}, \href
  {http://adsabs.harvard.edu/abs/2007ApJ...657..286L} {657, 286}

\bibitem[\protect\citeauthoryear{{Lugger} et~al.,}{{Lugger}
  et~al.}{2017}]{Lugger17}
{Lugger} P.~M.,  et~al., 2017, \mn@doi [\apj] {10.3847/1538-4357/aa6c56}, \href
  {https://ui.adsabs.harvard.edu/\#abs/2017ApJ...841...53L} {841, 53}

\bibitem[\protect\citeauthoryear{{Lyne} et~al.,}{{Lyne} et~al.}{1988}]{Lyne88}
{Lyne} A.~G.,  et~al., 1988, \mn@doi [\nat] {10.1038/332045a0}, \href
  {https://ui.adsabs.harvard.edu/abs/1988Natur.332...45L} {332, 45}

\bibitem[\protect\citeauthoryear{{Mansfield} et~al.,}{{Mansfield}
  et~al.}{2022}]{Mansfield22}
{Mansfield} S.,  et~al., 2022, \mn@doi [\mnras] {10.1093/mnras/stac242}, \href
  {https://ui.adsabs.harvard.edu/abs/2022MNRAS.511.3785M} {511, 3785}

\bibitem[\protect\citeauthoryear{{McKenna} \& {Lyne}}{{McKenna} \&
  {Lyne}}{1988}]{McKenna88}
{McKenna} J.,  {Lyne} A.~G.,  1988, \mn@doi [\nat] {10.1038/336226a0}, \href
  {https://ui.adsabs.harvard.edu/abs/1988Natur.336..226M} {336, 226}

\bibitem[\protect\citeauthoryear{{McLaughlin} \& {van der Marel}}{{McLaughlin}
  \& {van der Marel}}{2005}]{McLaughlin05}
{McLaughlin} D.~E.,  {van der Marel} R.~P.,  2005, \mn@doi [\apjs]
  {10.1086/497429}, \href
  {https://ui.adsabs.harvard.edu/abs/2005ApJS..161..304M} {161, 304}

\bibitem[\protect\citeauthoryear{{Miller-Jones} et~al.,}{{Miller-Jones}
  et~al.}{2015}]{Miller-Jones15}
{Miller-Jones} J.~C.~A.,  et~al., 2015, \mn@doi [\mnras]
  {10.1093/mnras/stv1869}, \href
  {https://ui.adsabs.harvard.edu/abs/2015MNRAS.453.3918M} {453, 3918}

\bibitem[\protect\citeauthoryear{{Milone} et~al.,}{{Milone}
  et~al.}{2012}]{Milone12}
{Milone} A.~P.,  et~al., 2012, \mn@doi [\aap] {10.1051/0004-6361/201016384},
  \href {https://ui.adsabs.harvard.edu/abs/2012A&A...540A..16M} {540, A16}

\bibitem[\protect\citeauthoryear{{Mochejska}, {Kaluzny}, {Thompson}  \&
  {Pych}}{{Mochejska} et~al.}{2002}]{Mochejska02}
{Mochejska} B.~J.,  {Kaluzny} J.,  {Thompson} I.,   {Pych} W.,  2002, \mn@doi
  [\aj] {10.1086/342015}, \href
  {https://ui.adsabs.harvard.edu/abs/2002AJ....124.1486M} {124, 1486}

\bibitem[\protect\citeauthoryear{{Nardiello} et~al.,}{{Nardiello}
  et~al.}{2018}]{Nardiello18}
{Nardiello} D.,  et~al., 2018, \mn@doi [\mnras] {10.1093/mnras/sty2515}, \href
  {http://adsabs.harvard.edu/abs/2018MNRAS.481.3382N} {481, 3382}

\bibitem[\protect\citeauthoryear{{Nascimbeni} et~al.,}{{Nascimbeni}
  et~al.}{2014}]{Nascimbeni14}
{Nascimbeni} V.,  et~al., 2014, \mn@doi [\mnras] {10.1093/mnras/stu930}, \href
  {https://ui.adsabs.harvard.edu/abs/2014MNRAS.442.2381N} {442, 2381}

\bibitem[\protect\citeauthoryear{{Pala} et~al.,}{{Pala} et~al.}{2020}]{Pala20}
{Pala} A.~F.,  et~al., 2020, \mn@doi [\mnras] {10.1093/mnras/staa764}, \href
  {https://ui.adsabs.harvard.edu/abs/2020MNRAS.494.3799P} {494, 3799}

\bibitem[\protect\citeauthoryear{{Pala} et~al.,}{{Pala} et~al.}{2022}]{Pala22}
{Pala} A.~F.,  et~al., 2022, \mn@doi [\mnras] {10.1093/mnras/stab3449}, \href
  {https://ui.adsabs.harvard.edu/abs/2022MNRAS.510.6110P} {510, 6110}

\bibitem[\protect\citeauthoryear{{Pallanca} et~al.,}{{Pallanca}
  et~al.}{2017}]{Pallanca17}
{Pallanca} C.,  et~al., 2017, \mn@doi [\apj] {10.3847/1538-4357/aa7ca6}, \href
  {https://ui.adsabs.harvard.edu/abs/2017ApJ...845....4P} {845, 4}

\bibitem[\protect\citeauthoryear{{Pichardo Marcano} et~al.,}{{Pichardo Marcano}
  et~al.}{2023}]{PichardoMarcano23}
{Pichardo Marcano} M.,  et~al., 2023, \mn@doi [\mnras] {10.1093/mnras/stad722},
  \href {https://ui.adsabs.harvard.edu/abs/2023MNRAS.521.5026P} {521, 5026}

\bibitem[\protect\citeauthoryear{{Piotto} et~al.,}{{Piotto}
  et~al.}{2015}]{Piotto15}
{Piotto} G.,  et~al., 2015, \mn@doi [\aj] {10.1088/0004-6256/149/3/91}, \href
  {http://adsabs.harvard.edu/abs/2015AJ....149...91P} {149, 91}

\bibitem[\protect\citeauthoryear{{Pooley}}{{Pooley}}{2015}]{Pooley15}
{Pooley} D.,  2015, in IAU General Assembly. p. 2257028

\bibitem[\protect\citeauthoryear{{Pooley}}{{Pooley}}{2016}]{Pooley16}
{Pooley} D.,  2016, \memsai, \href
  {https://ui.adsabs.harvard.edu/abs/2016MmSAI..87..547P} {87, 547}

\bibitem[\protect\citeauthoryear{{Pooley} \& {Hut}}{{Pooley} \&
  {Hut}}{2006}]{Pooley06}
{Pooley} D.,  {Hut} P.,  2006, \mn@doi [\apjl] {10.1086/507027}, \href
  {http://adsabs.harvard.edu/abs/2006ApJ...646L.143P} {646, L143}

\bibitem[\protect\citeauthoryear{{Pooley} et~al.,}{{Pooley}
  et~al.}{2002}]{Pooley02}
{Pooley} D.,  et~al., 2002, \mn@doi [\apj] {10.1086/339210}, \href
  {http://adsabs.harvard.edu/abs/2002ApJ...569..405P} {569, 405}

\bibitem[\protect\citeauthoryear{{Pooley} et~al.,}{{Pooley}
  et~al.}{2003}]{Pooley03}
{Pooley} D.,  et~al., 2003, \mn@doi [\apjl] {10.1086/377074}, \href
  {http://adsabs.harvard.edu/cgi-bin/nph-bib_query?bibcode=2003ApJ...591L.131P&db_key=AST}
  {591, L131}

\bibitem[\protect\citeauthoryear{{Pretorius} \& {Knigge}}{{Pretorius} \&
  {Knigge}}{2012}]{Pretorius12}
{Pretorius} M.~L.,  {Knigge} C.,  2012, \mn@doi [\mnras]
  {10.1111/j.1365-2966.2011.19801.x}, \href
  {https://ui.adsabs.harvard.edu/abs/2012MNRAS.419.1442P} {419, 1442}

\bibitem[\protect\citeauthoryear{{Reynolds} et~al.,}{{Reynolds}
  et~al.}{2014}]{Reynolds14}
{Reynolds} M.~T.,  et~al., 2014, \mn@doi [\mnras] {10.1093/mnras/stu832}, \href
  {https://ui.adsabs.harvard.edu/abs/2014MNRAS.441.3656R} {441, 3656}

\bibitem[\protect\citeauthoryear{{Ridolfi} et~al.,}{{Ridolfi}
  et~al.}{2021}]{Ridolfi21}
{Ridolfi} A.,  et~al., 2021, arXiv e-prints, \href
  {https://ui.adsabs.harvard.edu/abs/2021arXiv210304800R} {p. arXiv:2103.04800}

\bibitem[\protect\citeauthoryear{{Rivera Sandoval} et~al.,}{{Rivera Sandoval}
  et~al.}{2018}]{RiveraSandoval18}
{Rivera Sandoval} L.~E.,  et~al., 2018, \mn@doi [\mnras]
  {10.1093/mnras/sty058}, \href
  {https://ui.adsabs.harvard.edu/abs/2018MNRAS.475.4841R} {475, 4841}

\bibitem[\protect\citeauthoryear{{Robitaille} et~al.,}{{Robitaille}
  et~al.}{2017}]{Robitaille17}
{Robitaille} T.,  et~al., 2017, {glueviz v0.13.1: multidimensional data
  exploration}, \mn@doi{10.5281/zenodo.1237692}

\bibitem[\protect\citeauthoryear{{Rucinski}}{{Rucinski}}{1995}]{Rucinski95}
{Rucinski} S.,  1995, \mn@doi [\pasp] {10.1086/133603}, \href
  {https://ui.adsabs.harvard.edu/abs/1995PASP..107..648R} {107, 648}

\bibitem[\protect\citeauthoryear{{Shara} \& {Hurley}}{{Shara} \&
  {Hurley}}{2006}]{Shara06}
{Shara} M.~M.,  {Hurley} J.~R.,  2006, \mn@doi [\apj] {10.1086/504679}, \href
  {https://ui.adsabs.harvard.edu/abs/2006ApJ...646..464S} {646, 464}

\bibitem[\protect\citeauthoryear{{Shishkovsky} et~al.,}{{Shishkovsky}
  et~al.}{2018}]{Shishkovsky18}
{Shishkovsky} L.,  et~al., 2018, \mn@doi [\apj] {10.3847/1538-4357/aaadb1},
  \href {https://ui.adsabs.harvard.edu/abs/2018ApJ...855...55S} {855, 55}

\bibitem[\protect\citeauthoryear{{Shishkovsky} et~al.,}{{Shishkovsky}
  et~al.}{2020}]{Shishkovsky20}
{Shishkovsky} L.,  et~al., 2020, \mn@doi [\apj] {10.3847/1538-4357/abb880},
  \href {https://ui.adsabs.harvard.edu/abs/2020ApJ...903...73S} {903, 73}

\bibitem[\protect\citeauthoryear{{Sigurdsson} et~al.,}{{Sigurdsson}
  et~al.}{2003}]{Sigurdsson03}
{Sigurdsson} S.,  et~al., 2003, \mn@doi [Science] {10.1126/science.1086326},
  \href {https://ui.adsabs.harvard.edu/abs/2003Sci...301..193S} {301, 193}

\bibitem[\protect\citeauthoryear{{Stassun} \& {Torres}}{{Stassun} \&
  {Torres}}{2021}]{Stassun21}
{Stassun} K.~G.,  {Torres} G.,  2021, \mn@doi [\apjl]
  {10.3847/2041-8213/abdaad}, \href
  {https://ui.adsabs.harvard.edu/abs/2021ApJ...907L..33S} {907, L33}

\bibitem[\protect\citeauthoryear{{Taylor}}{{Taylor}}{2005}]{Taylor05}
{Taylor} M.~B.,  2005, in {Shopbell} P.,  {Britton} M.,   {Ebert} R.,  eds,
  Astronomical Society of the Pacific Conference Series Vol. 347, Astronomical
  Data Analysis Software and Systems XIV. p.~29

\bibitem[\protect\citeauthoryear{{Taylor}, {Grindlay}, {Edmonds}  \&
  {Cool}}{{Taylor} et~al.}{2001}]{Taylor01}
{Taylor} J.~M.,  {Grindlay} J.~E.,  {Edmonds} P.~D.,   {Cool} A.~M.,  2001,
  \mn@doi [\apjl] {10.1086/320676}, \href
  {http://adsabs.harvard.edu/abs/2001ApJ...553L.169T} {553, L169}

\bibitem[\protect\citeauthoryear{{Thomson} et~al.,}{{Thomson}
  et~al.}{2012}]{Thomson12}
{Thomson} G.~S.,  et~al., 2012, \mn@doi [\mnras]
  {10.1111/j.1365-2966.2012.21104.x}, \href
  {http://adsabs.harvard.edu/abs/2012MNRAS.423.2901T} {423, 2901}

\bibitem[\protect\citeauthoryear{{Thorsett}, {Arzoumanian}, {Camilo}  \&
  {Lyne}}{{Thorsett} et~al.}{1999}]{Thorsett99}
{Thorsett} S.~E.,  {Arzoumanian} Z.,  {Camilo} F.,   {Lyne} A.~G.,  1999,
  \mn@doi [\apj] {10.1086/307771}, \href
  {https://ui.adsabs.harvard.edu/abs/1999ApJ...523..763T} {523, 763}

\bibitem[\protect\citeauthoryear{{Trager}, {King}  \& {Djorgovski}}{{Trager}
  et~al.}{1995}]{Trager95}
{Trager} S.~C.,  {King} I.~R.,   {Djorgovski} S.,  1995, \mn@doi [\aj]
  {10.1086/117268}, \href
  {https://ui.adsabs.harvard.edu/abs/1995AJ....109..218T} {109, 218}

\bibitem[\protect\citeauthoryear{{Vasiliev}}{{Vasiliev}}{2019}]{Vasiliev19}
{Vasiliev} E.,  2019, \mn@doi [\mnras] {10.1093/mnras/stz171}, \href
  {https://ui.adsabs.harvard.edu/abs/2019MNRAS.484.2832V} {484, 2832}

\bibitem[\protect\citeauthoryear{{Verbunt}}{{Verbunt}}{2000}]{Verbunt00}
{Verbunt} F.,  2000, in {Pallavicini} R.,  {Micela} G.,   {Sciortino} S.,  eds,
   Astronomical Society of the Pacific Conference Series Vol. 198, Stellar
  Clusters and Associations: Convection, Rotation, and Dynamos. p.~421
  (\mn@eprint {arXiv} {astro-ph/9907202})

\bibitem[\protect\citeauthoryear{{Verbunt}}{{Verbunt}}{2001}]{Verbunt01}
{Verbunt} F.,  2001, \mn@doi [\aap] {10.1051/0004-6361:20000469}, \href
  {https://ui.adsabs.harvard.edu/abs/2001A&A...368..137V} {368, 137}

\bibitem[\protect\citeauthoryear{{Verbunt}}{{Verbunt}}{2002}]{Verbunt02}
{Verbunt} F.,  2002, in {van Leeuwen} F.,  {Hughes} J.~D.,   {Piotto} G.,  eds,
   Astronomical Society of the Pacific Conference Series Vol. 265, Omega
  Centauri, A Unique Window into Astrophysics. p.~289 (\mn@eprint {arXiv}
  {astro-ph/0111441}), \mn@doi{10.48550/arXiv.astro-ph/0111441}

\bibitem[\protect\citeauthoryear{{Verbunt}, {Pooley}  \& {Bassa}}{{Verbunt}
  et~al.}{2008}]{Verbunt08}
{Verbunt} F.,  {Pooley} D.,   {Bassa} C.,  2008, in {Vesperini} E.,  {Giersz}
  M.,   {Sills} A.,  eds,  Proceedings of the International Astronomical Union
  Vol. 246, Dynamical Evolution of Dense Stellar Systems. pp 301--310
  (\mn@eprint {arXiv} {0710.1804}), \mn@doi{10.1017/S1743921308015822}

\bibitem[\protect\citeauthoryear{{Vesperini}}{{Vesperini}}{2010}]{Vesperini10}
{Vesperini} E.,  2010, \mn@doi [Philosophical Transactions of the Royal Society
  of London Series A] {10.1098/rsta.2009.0260}, \href
  {https://ui.adsabs.harvard.edu/abs/2010RSPTA.368..829V} {368, 829}

\bibitem[\protect\citeauthoryear{{Wilms}, {Allen}  \& {McCray}}{{Wilms}
  et~al.}{2000}]{Wilms2000}
{Wilms} J.,  {Allen} A.,   {McCray} R.,  2000, \mn@doi [\apj] {10.1086/317016},
  \href {https://ui.adsabs.harvard.edu/abs/2000ApJ...542..914W} {542, 914}

\bibitem[\protect\citeauthoryear{{Zhang} et~al.,}{{Zhang}
  et~al.}{2022}]{Zhang22}
{Zhang} L.,  et~al., 2022, \mn@doi [\apjl] {10.3847/2041-8213/ac81c3}, \href
  {https://ui.adsabs.harvard.edu/abs/2022ApJ...934L..21Z} {934, L21}

\bibitem[\protect\citeauthoryear{{Zhao} \& {Heinke}}{{Zhao} \&
  {Heinke}}{2022}]{Zhao22}
{Zhao} J.,  {Heinke} C.~O.,  2022, \mn@doi [\mnras] {10.1093/mnras/stac442},
  \href {https://ui.adsabs.harvard.edu/abs/2022MNRAS.511.5964Z} {511, 5964}

\bibitem[\protect\citeauthoryear{{Zhao} et~al.,}{{Zhao} et~al.}{2020}]{Zhao20a}
{Zhao} Y.,  et~al., 2020, \mn@doi [\mnras] {10.1093/mnras/staa631}, \href
  {https://ui.adsabs.harvard.edu/abs/2020MNRAS.493.6033Z} {493, 6033}

\bibitem[\protect\citeauthoryear{{Zhao} et~al.,}{{Zhao} et~al.}{2021}]{Zhao21}
{Zhao} Y.,  et~al., 2021, \mn@doi [\apj] {10.3847/1538-4357/abfc58}, \href
  {https://ui.adsabs.harvard.edu/abs/2021ApJ...914...77Z} {914, 77}

\bibitem[\protect\citeauthoryear{{van den Berg}}{{van den
  Berg}}{2020}]{van_den_Berg20}
{van den Berg} M.,  2020, in {Bragaglia} A.,  {Davies} M.,  {Sills} A.,
  {Vesperini} E.,  eds, ~ Vol. 351, Star Clusters: From the Milky Way to the
  Early Universe. pp 367--376 (\mn@eprint {arXiv} {1910.07595}),
  \mn@doi{10.1017/S1743921319007981}

\makeatother
\end{thebibliography}

\appendix

\section{X-ray analysis of the MSP M4A}
\label{sec:M4A}
We used archived CXO ACIS-S X-ray observations of M4 (observation IDs 946, 7446, and 7447) to analyze the X-ray properties of M4A, with a total exposure time of 119.2 ks. 
We reduced and extracted data using {\sc ciao}\footnote{\url{https://cxc.cfa.harvard.edu/ciao/}} \citep[][]{Fruscione06}, version 4.13, {\sc caldb} 4.9.4. All the {\it Chandra} data were first reprocessed to generate new level 2 files for further analysis through the {\tt chandra\_repro} script. No background flares were found for those observations. 

We searched for an X-ray counterpart to M4A using its radio timing position. We applied {\tt wavdetect}, a Mexican-Hat Wavelet source detection script, to identify sources and generate regions for spectral extraction using {\tt specextract}. The X-ray spectra of M4A were 
combined for spectral fitting using the {\tt combine\_spectra} script. The corresponding background spectra were extracted from source-free regions around M4A.

We performed X-ray spectral fitting in {\sc sherpa}, the modelling and fitting package of {\sc ciao}. We first filtered the energy range to 0.5--6 keV.
(The full spectrum suggests hard-to-model flux above 6 keV; but a merged image of the region does not show evidence of a source above 6 keV. We chose to ignore data above 6 keV for this work.)
Data were grouped to five counts per bin. We applied {\tt wstat} statistics in {\sc sherpa} to estimate fitting uncertainties and goodness. The X-ray absorption by the interstellar medium towards M4 was modelled by {\tt xstbabs} with {\it wilm} abundances \citep{Wilms2000}. The hydrogen column number density ($N_{\rm H}$) was fixed to 3.05$\times$10$^{21}$ cm$^{-2}$, which was obtained by converting the interstellar reddening to the cluster, $E(B\!-\!V) = 0.35$ \citep[][2010 edition]{Harris96}, to $N_{\rm H}$ using the conversion of \citet{Bahramian15}. 

We considered three spectral models to fit the X-ray emission from M4A: a blackbody (BB; {\tt xsbbodyrad}), a neutron star hydrogen atmosphere model \citep[NSA; {\tt xsnsatmos}, see][]{Heinke06}, and a power-law model (PL; {\tt xspegpwrlw}). For the NSA model, we fixed the NS mass and radius parameters to 1.4 M$_{\sun}$ and 10 km, respectively, and the distance to M4 was set to 1.85 kpc \citep{Baumgardt21}. The fitting results are given in Table~\ref{tab:spec_fits_M4A}.

\begin{table}
    \caption{Spectral fits for PSR B1620$-$26 in M4.}
    \centering
    \begin{tabular}{lccc}
      \hline
       & & Spectral Model & \\
       & BB & NSA & PL  \\
      \hline
      {$kT_{\rm BB}$/$\log T_{\rm eff}$}/$\Gamma$\,$^a$ & $0.30 \pm 0.03$ & $6.34 \pm 0.08$ & $2.8 \pm 0.3$ \\
      Reduced Stat. & 1.06 & 0.99 & 1.36 \\
      Q-value & 0.39 & 0.44 & 0.21 \\
      {$F_X$(0.5--6 keV)}$^b$ & $4.4^{+0.8}_{-1.0} $ & $4.6^{+2.5}_{-1.3}$ & $9.5^{+2.0}_{-2.0}$ \\
      \hline
    \multicolumn{4}{p{0.9\linewidth}}{{\it Notes}: $N_{\rm H}$ was fixed for all fits at $3.05 \times 10^{21}~$cm$^{-2}$.} \\
    \multicolumn{4}{p{0.9\linewidth}}{$^a$ $kT_{\rm BB}$: blackbody temperature in units of keV; $\log T_{\rm eff}$: unredshifted effective temperature of the NS surface in units of log Kelvin; $\Gamma$: photon index.} \\
    \multicolumn{4}{l}{$^b$ Unabsorbed flux in units of $10^{-15}$ erg cm$^{-2}$~s$^{-1}$.} \\
    \end{tabular}
    \label{tab:spec_fits_M4A}
\end{table}

\begin{figure}
    \centering
    \includegraphics[width=\linewidth]{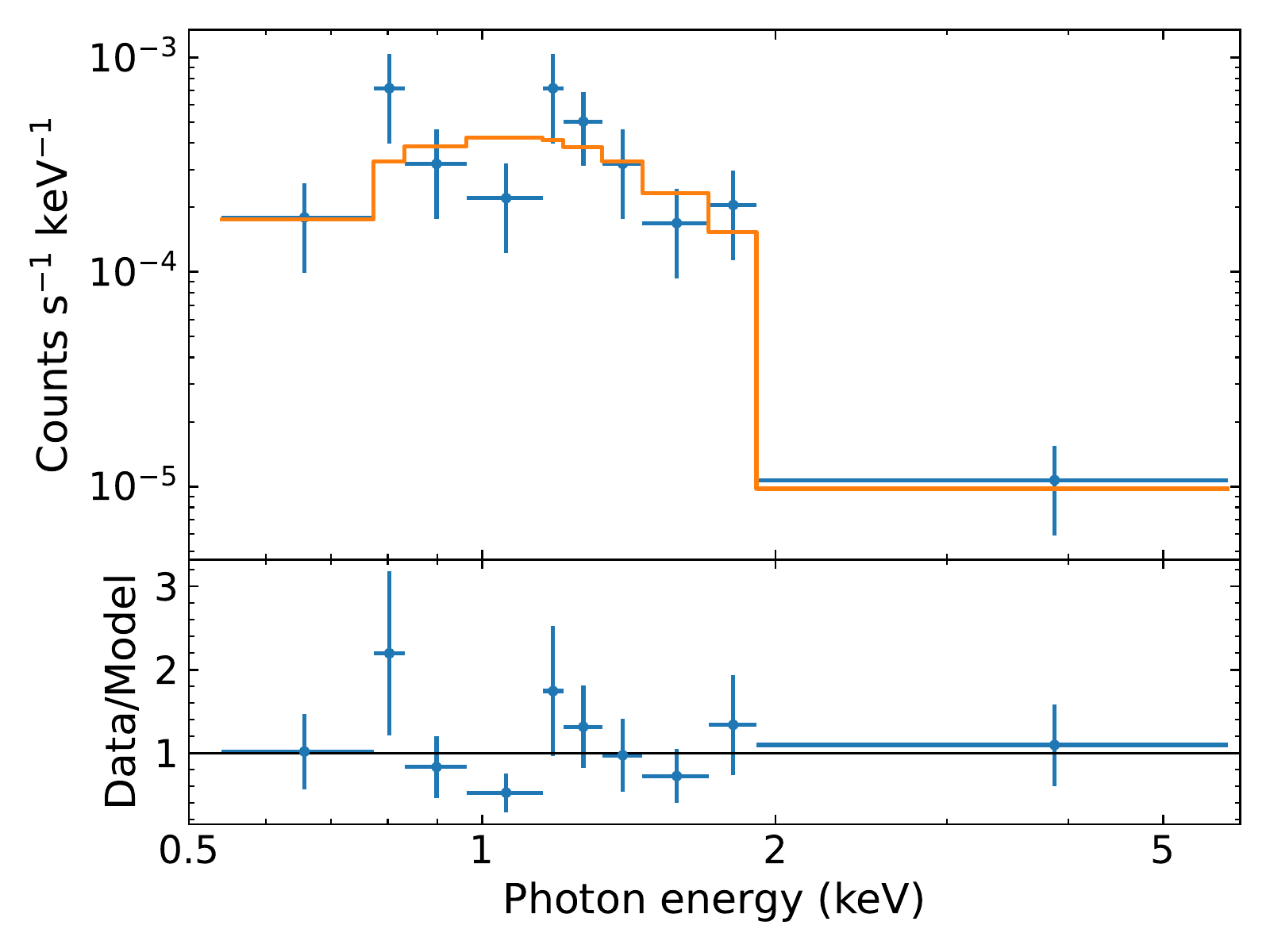}
    \caption{The X-ray spectrum and best fit of M4A. Data are filtered into the energy band of 0.5--6 keV and grouped to five counts per bin. The spectrum is well fitted by a single NSA model, using {\tt wstat} statistics. }
    \label{fig:M4A_spec}
\end{figure}

The X-ray spectrum of M4A in the band 0.5--6 keV is well described by a single BB or NSA model, 
consistent with 
thermal emission from this source. 
For 
a single PL model, 
the best-fit photon index is 2.8$\pm$0.3. 
Typically, nonthermal emission from MSPs (e.g. synchrotron) can be represented by a power-law of photon index 1-2, while thermal blackbody-like surface emission can be fit by 
a blackbody, or (for relatively low-count spectra) by a power-law of index $\gg2$. The fit parameters for M4A suggest that its emission is largely thermal surface emission. 
We also tested spectral fitting with combinations of PL and BB/NSA models, 
but the additional component does not improve the fit. 
The spectral fitting results of M4A indicate that the X-ray emission is principally thermal and hence likely originates from hotspots near the magnetic poles on the NS surface, heated by returning particles from the magnetosphere \citep{Harding02}. 
The 0.5--6 keV X-ray luminosity inferred from the NSA model is $L_{\rm X}$=1.9$^{+1.0}_{-0.5}$$\times$10$^{30}$ erg s$^{-1}$, for a distance of 1.85$\pm$0.02 kpc \citep{Baumgardt21}. The X-ray spectrum and best-fit NSA model are shown in Figure~\ref{fig:M4A_spec}.

\section{Finding Charts}

(This material is intended for online distribution only.)


                                                 %
                                                 %
\bsp	
\label{lastpage}                                 %
\end{document}